# (Sub-)stellar companions shape the winds of evolved stars

**Authors:** L. Decin[1,2,*], M. Montargès[1], A. M. S. Richards[3], C. A. Gottlieb[4], W. Homan[1], I. McDonald[3,5], I. El Mellah[1,6], T. Danilovich[1], S. H. J. Wallström[1], A. Zijlstra[3,7], A. Baudry[8], J. Bolte[1], E. Cannon[1], E. De Beck[9], F. De Ceuster[1,10], A. de Koter[1,11], J. De Ridder[1], S. Etoka[3], D. Gobrecht[1], M. Gray[3,12], F. Herpin[8], M. Jeste[13], E. Lagadec[14], P. Kervella[15], T. Khouri[9], K. Menten[13], T. J. Millar[16], H. S. P. Müller[17], J. M. C. Plane[2], R. Sahai[18], H. Sana[1], M. Van de Sande[1], L. B. F. M. Waters[7,19], K. T. Wong[20], J. Yates[10]

**Affiliations:**

[1]KU Leuven, Institute of Astronomy, 3001 Leuven, Belgium.

[2]University of Leeds, School of Chemistry, Leeds LS2 9JT, United Kingdom.

[3]The University of Manchester, Jodrell Bank Centre for Astrophysics, Manchester M13 9PL, United Kingdom.

[4]Harvard-Smithsonian Center for Astrophysics, Cambridge MA 02138, USA.

[5]Open University, Walton Hall, Milton Keynes MK7 6AA, United Kingdom.

[6]KU Leuven, Center for mathematical Plasma Astrophysics, 3001 Leuven, Belgium.

[7]University of Hong Kong, Laboratory for Space Research, Pokfulam, Hong Kong.

[8]Université de Bordeaux, Laboratoire d'Astrophysique de Bordeaux, 33615 Pessac, France.

[9]Chalmers University of Technology, Onsala Space Observatory, 43992 Onsala, Sweden.

[10]University College London, Department of Physics and Astronomy, London WC1E 6BT, United Kingdom.

[11]University of Amsterdam, Anton Pannekoek Institute for Astronomy, 1090 GE Amsterdam, The Netherlands.

[12]National Astronomical Research Institute of Thailand, Chiangmai 50180, Thailand.

[13]Max-Planck-Institut für Radioastronomie, 53121 Bonn, Germany.

[14]Université Côte d'Azur, Laboratoire Lagrange, Observatoire de la Côte d'Azur, F-06304 Nice Cedex 4, France.

[15]Laboratoire d'Etudes Spatiales et d'Instrumentation en Astrophysique, Observatoire de Paris, Université Paris Sciences et Lettres, Centre National de la Recherche Scientifique, Sorbonne Université, Université de Paris, 92195 Meudon, France.

[16]Queen's University Belfast, Astrophysics Research Centre, Belfast BT7 1NN, United Kingdom.

[17]Universität zu Köln, I. Physikalisches Institut, 50937 Köln, Germany.

[18]California Institute of Technology, Jet Propulsion Laboratory, Pasadena CA 91109, USA.

[19]SRON Netherlands Institute for Space Research, NL-3584 CA Utrecht, The Netherlands.

[20]Institut de Radioastronomie Millimétrique, 38406 Saint Martin d'Hères, France.

*Corresponding author. Email: leen.decin@kuleuven.be.

**Abstract:** Binary interactions dominate the evolution of massive stars, but their role is less clear for low and intermediate mass stars. The evolution of a spherical wind from an Asymptotic Giant Branch (AGB) star into a non-spherical planetary nebula (PN) could be due to binary interactions. We observe a sample of AGB stars with the Atacama Large Millimeter/submillimeter Array (ALMA), finding that their winds exhibit distinct non-spherical geometries with morphological similarities to PNe. We infer that the same physics shapes both AGB winds and PNe. The morphology and AGB mass-loss rate are correlated. These characteristics can be explained by binary interaction. We propose an evolutionary scenario for AGB morphologies which is consistent with observed phenomena in AGB stars and PNe.

**Main Text:**

At the end of their life, low and intermediate mass (0.8 to 8 solar masses, $M_\odot$) stars turn into luminous cool red giant stars when ascending the AGB. Our Sun will reach that phase in ~7.7 Gyr from now *(1)*. During the AGB phase, the star's radius may become as large as one astronomical unit (au) and its luminosity may reach thousands of times that of the Sun. The AGB phase lasts between ~$0.1 - 20$ Myr, the more massive stars are short-lived *(2)*. At the start of the AGB phase, stars are oxygen-rich with a carbon-to-oxygen (C/O) ratio lower than 1, and are referred to as M-type stars. During the AGB phase, carbon is fused in the stellar core and brought to the surface by convection. Eventually, the C/O ratio gets larger than 1, leading to a carbon star. The AGB phase is characterized by a stellar wind with mass-loss rate greater than $\sim 10^{-8}\ M_\odot yr^{-1}$. The increase in luminosity while ascending the AGB induces an increase in mass-loss rate, of up to ~$10^{-4}\ M_\odot\ yr^{-1}$ *(3)*. For stars with mass-loss rate greater than $10^{-7}\ M_\odot yr^{-1}$, the mass-loss rate exceeds the envelope's hydrogen nuclear burning rate, and mass loss determines the further stellar evolution *(4)*. The wind strips away the star's outer envelope. At the moment that the envelope is less than about 1% of the stellar mass, the star becomes a post-AGB star *(5)*. During this short evolutionary phase which takes a few 1000 yr, the temperature of the star increases at constant luminosity and it becomes a planetary nebula (PN), characterized by a hot central star which ionizes the gas ejected during the previous red giant phase. The lifetime of PNe is roughly 20 000 yr. The PN nebula then disperses quickly, leaving an inert white dwarf which slowly cools *(6)*.

One puzzling aspect about PNe formation concerns the mechanism that shapes the nebulae into a wide range of morphologies, including elliptical, bipolar, and `butterfly'-shaped geometries *(7)*. While ~80% of the AGB stars have a wind with overall spherical symmetry *(8)*, less than 20% of PNe are circularly symmetric *(9,10)*. Various hypotheses - including rapidly spinning or strongly magnetic single stars *(11)* - have been proposed to explain this morphological metamorphosis, but they have been questioned because strong asymmetries are not formed efficiently (*12)*. More recently, short-period (orbital period $P_{orb} \lesssim 10$ days) binary systems (orbital separation $a \lesssim 0.2\ au$) surrounded by a common gaseous envelope - referred to as the common-envelope phase - have become the favored hypothesis (*13*). The proposed PN shaping mechanisms operate over a short time, either during the final few hundred years of the AGB or during the early post-AGB phase *(14)*. Identifying the shaping mechanism and its time of occurrence are observationally challenging owing to the short lifetime of the post-AGB and PN stages; the strong observational bias towards detecting binary post-AGB stars and PNe with short orbital periods *(15)*; and the high mass-loss rates at the end of the AGB phase, which surrounds the star with high optical depth material and obscures the inner workings.

During the last few years, observations at high spatial resolution have shown that AGB winds may exhibit small-scale structural complexity - including arcs, shells, bipolar structures, clumps, spirals, tori, and rotating disks *(16, 17)* - embedded in a smooth, radially outflowing wind. Only about a dozen AGB winds have been studied in detail *(18)*. It has not been possible to determine any systematic morphological change during the AGB evolution, and the transition from the smaller scale structures observed during the AGB to the PN morphologies is not understood.

In the ALMA ATOMIUM - ALMA Tracing the Origins of Molecules In dUst-forming oxygen-rich M-type stars - Program *(18)*, we have observed a sample of oxygen-rich AGB stars spanning a range of (circum)stellar parameters and AGB evolutionary stages (Table S1). We study the wind morphology at a spatial resolution of ~0.24″ and ~1″ using the rotational lines of $^{12}$CO $J$=2→1, $^{28}$SiO $J$=5→4, and $^{28}$SiO $J$=6→5 in the ground vibrational state, with $J$ the rotational angular momentum quantum number. These two molecules have large fractional abundances with respect to molecular hydrogen and yield complementary information on the density (CO) and on the morphological and dynamical properties close to the stellar surface (SiO).

Figure 1 shows a gallery of the CO observations. None of the sources has a smooth, spherical geometry. The images exhibit various structures in common with post-AGB stars and PNe: bipolar morphologies with a central waist, equatorial density enhancements (EDE) and disk-like geometries, eye-like shapes, spiral-like structures, and arcs at regularly spaced intervals *(18)*. We infer from these images that the same physical mechanism shapes both AGB winds and PNe. These data constrain the mechanism shaping the winds while it is in operation in a sample of stars with a range of AGB properties, i.e. cover the moment in time when AGB morphologies are being transformed into aspherical geometries.

Combining the CO and SiO data provides an observational criterion (Fig. S2) for classifying the prevailing wind morphologies (Table S2). We find a correlation between the AGB mass-loss rate ($\dot{M}$) and the prevailing geometry (Table 1), with a Kendall's rank correlation coefficient $\tau_b$ of 0.79 (Fig. S3-S4, *(18)*). A dynamically complex EDE is often observed for oxygen-rich AGB stars with low mass-loss rates (which we refer to as 'Class 1'), a bipolar structure tends to be dominant for stars with medium mass-loss rates ('Class 2'), while the winds of high mass-loss rate stars preferentially exhibit a spiral-like structures ('Class 3'). Other oxygen-rich AGB stars whose geometry was deduced from previous observations follow this same schematic order (Table S3). This correlation suggests that a common mechanism controls the wind morphology throughout the AGB phase, and that it depends on the mass-loss rate.

Among the mechanisms proposed to explain asphericity, binary models including long-period systems ($P_{orb} \gtrsim 1$ yr, $a \gtrsim 2\ au$) *(19-21)* can explain both the morphologies and the correlation with mass-loss rate *(18)*. Stellar evolution models *(22)* show that the majority of AGB stars with a mass-loss rate above $10^{-7}\mathrm{M}_{\odot}\ yr^{-1}$ - including all those in the ATOMIUM sample - have masses above $\sim$1.5 $\mathrm{M}_{\odot}$. Planet and stellar binary population statistics *(23,24)* indicate that stars with these masses have an average number of companions (with mass above ~5 Jupiter mass) ≳ 1 *(18)*. Binary interaction is known to dominate the evolution of more massive stars *(25)*. We conjecture that (sub-)stellar binary interaction is the dominant wind shaping agent for the majority of AGB stars with mass-loss rate exceeding the nuclear burning rate. Our conjecture is supported by the growing number of aspherical PNe detected whose binary central stars have a long-period orbit ($P_{orb} \gtrsim 1$ yr) not undergoing a common-envelope evolution *(18)*.

On the assumption that binary interaction dominates, we derive *(18)* an analytical relation

$$Q^1 = 10^{-6}\, Q^p = 8.32\, \frac{1}{(1-e)^2}\, \frac{1}{f_w} \left(\frac{m_{comp}}{M_\odot}\right)^{1/3} \left(\frac{M_*}{M_\odot}\right)^{7/6} \left(\frac{a}{1\,au}\right)^{-3/2} \left(\frac{\dot{M}}{10^{-6} \mathrm{M}_\odot\, yr^{-1}}\right)^{-1}$$

that estimates the probability of a binary system forming a (possibly rotating) EDE structure (large value of $Q^1$) or being dominated by a spiral-like structure (low value of $Q^1$) *(18)*. Here $e$ denotes the eccentricity of the orbit, $f_w$ the fraction of the stellar wind mass present at a distance $r = a$, and $M_*$ and $m_{comp}$ are the mass of the primary star and companion, respectively. This relation holds for a wind velocity at $r = a$ that is lower than the orbital velocity, and can be easily reformulated for the case of a high wind velocity *(18)*. Our analytical relation supports the correlation observed in the ATOMIUM data. Higher mass-loss rate or orbital separation leads to lower injection of angular momentum into the initially spherical AGB wind by interactions with the orbiting companion, and weaker shaping of the material along the orbital plane into an EDE, a circumbinary disk or an accretion disk *(18)*. Wide binaries, with a separation of up to several tens of astronomical units, produce a spiral-like structure *(19)*.

The transition of the wind morphology during the AGB phase we observe applies to oxygen-rich AGB stars, whereas carbon-rich winds most often display a (broken) spiral/arc-like structure *(18)*. We attribute this differentiation to stronger wind acceleration for carbon-rich than oxygen-rich stars, due to the different dust composition (Fig. S5). Stronger acceleration results in a smaller geometrical region in which the velocity field is non-radial, in a lower probability of forming an EDE, and in a smaller radius beyond which the wind shows a self-similar morphology *(18)*. This implies that carbon-rich AGB stars are more commonly surrounded by an expanding, self-similar, spiral structure, which is consistent with past observations *(18)*. Although EDEs may form in carbon-rich winds, EDEs will be most commonly found around low mass-loss rate oxygen-rich AGB stars with slowly accelerating winds *(18)*.

Observations of binary companions around AGB progenitor stars indicate the highest fraction of binary companions are at an orbital distance greater than ~20 au *(23)*. We calculate *(18)* that those orbits will widen during the AGB evolution as the mass-loss rate increases (Fig. S6-S7). This implies that early-type low mass-loss rate AGB stars will often have an EDE, with complex flow patterns, and the wind of late-type high mass-loss rate AGB stars are mainly shaped by spiral structures (Fig. 2). Our results also imply that the effects of planets around evolved stars are more easily detected in early-type oxygen-rich AGB stars *(18)*.

Our proposed evolutionary and chemical scheme for AGB wind morphologies can explain multiple AGB, post-AGB and PN phenomena *(18),* including why (i) circular detached shells are only detected around carbon-rich AGB stars *(26)*; (ii) disks are mainly found around oxygen-rich post-AGB and PN binaries (*15*); (iii) carbon-rich stars can be surrounded by silicate dust (*27*); (iv) PNe in the bulge of the Milky Way can have a mixed carbon/oxygen chemistry (*28*); (v) post-AGB envelopes can be classified according to two distinct morphological types (*29*); (vi) post-AGB binaries can have non-zero eccentricities with values as high as 0.3 (*30*); and (vii) the low fraction of round PNe (*9,10*).

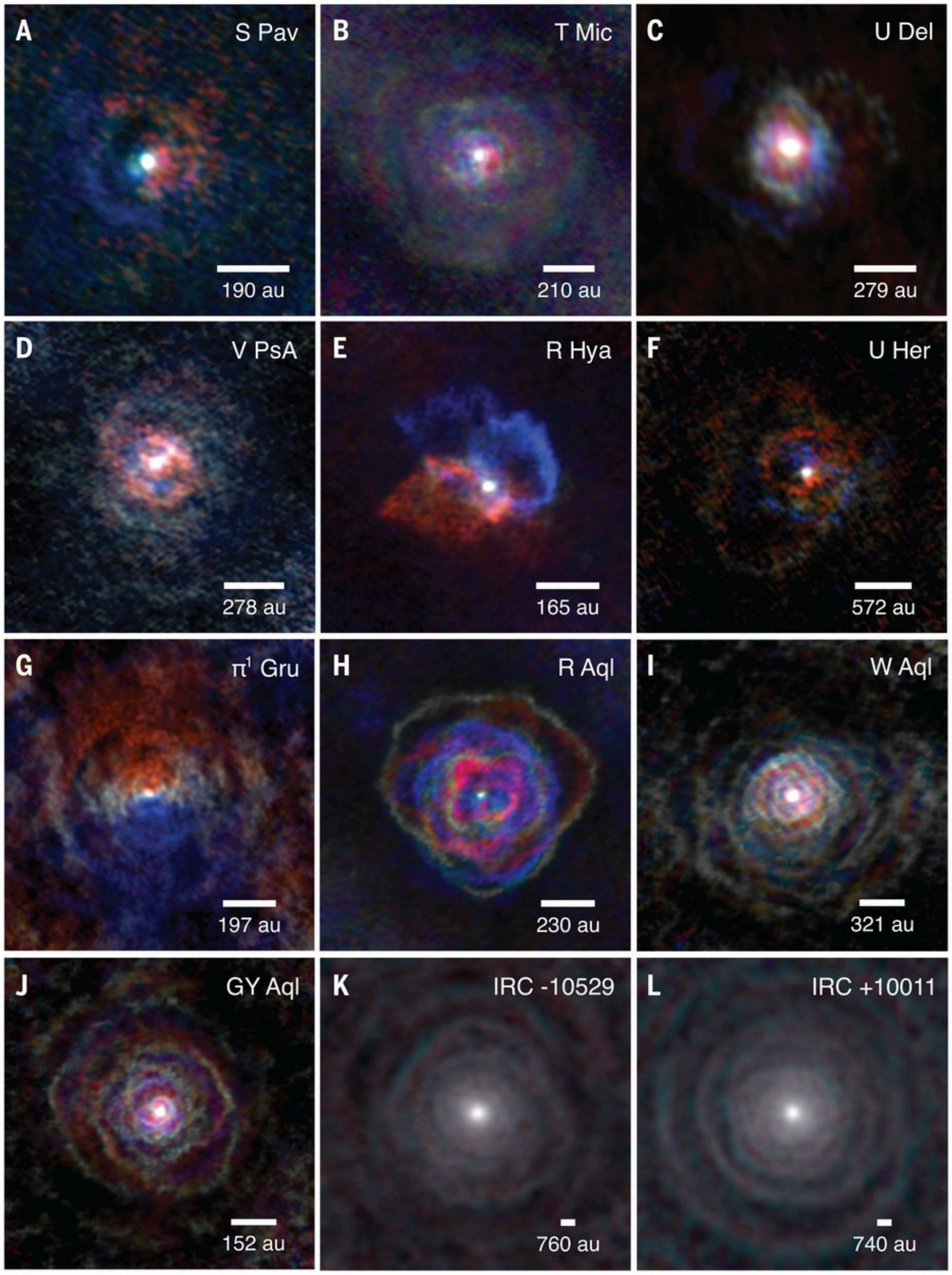

**Fig. 1. Gallery of AGB winds.** Emission maps of twelve stars are shown, derived from the ATOMIUM $^{12}$CO $J$=2→1 data. For each star, the emission which is red-shifted with respect to the local standard of rest velocity is shown in red, blue-shifted in blue, and rest velocity in white. The white scale bars are of length 1″. Full channel maps and position-velocity diagrams for each source are shown in Figures S8-S65. Fig. 1A: S Pav, Fig. 1B: T Mic, Fig. 1C: U Del, Fig 1D: V PsA, Fig. 1E: R Hya, Fig. 1F: U Her, Fig. 1G: $\pi^1$ Gru, Fig. 1H: R Aql, Fig 1I: W Aql, Fig. 1J: GY Aql, Fig. 1K: IRC -10529, and Fig. 1L: IRC +10011. For two stars (RW Sco and SV Aqr) the signal-to-noise of the data was too low to produce a three-colour map, although the individual channels show clear signs of asymmetry (Fig. S20, Fig. S28 *(18)*).

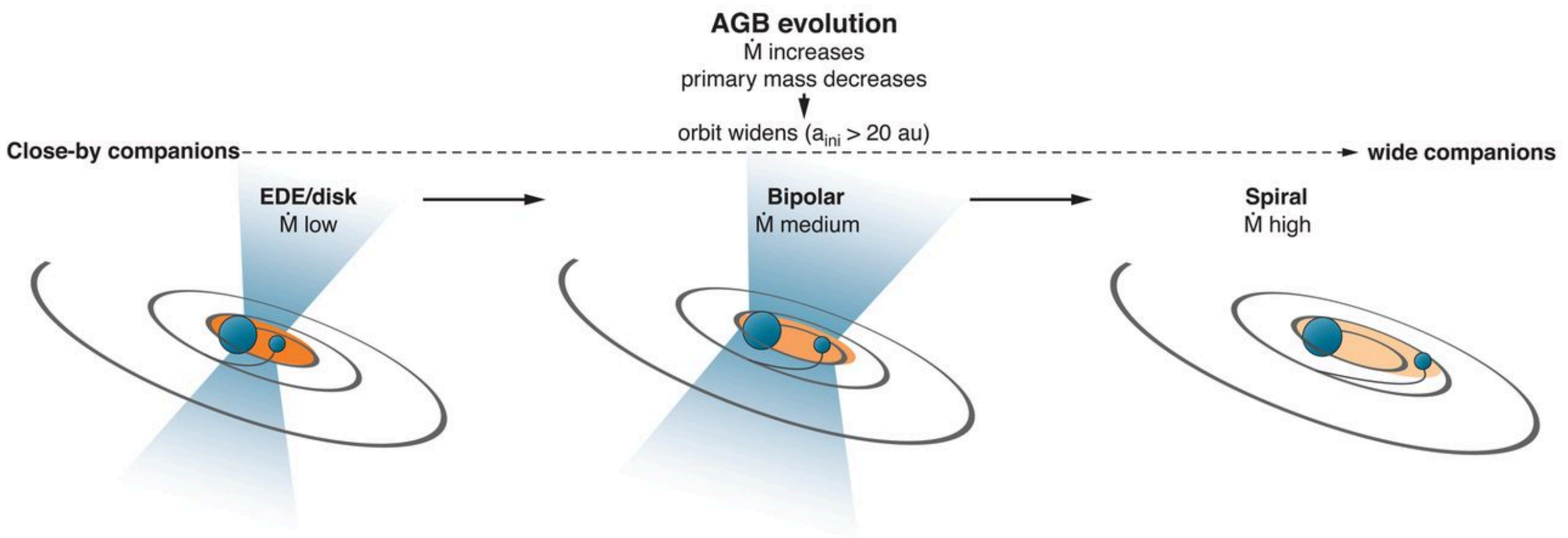

**Fig. 2. Schematic illustration of our inferred evolution of wind morphology during the AGB phase**. Most (sub-)stellar companions have initial orbits ($a_{ini}$) greater than 20 au *(24)*. These orbits widen during AGB evolution because the stellar mass decreases. Binary systems with close-by companions often have a high-density EDE and accretion disk (shown in orange) and complex inner wind dynamics. For increasingly wider orbits and higher mass-loss rates, the prevailing outflow morphology first transitions to a bipolar structure (in blue) and then to a regularly spaced spiral structure (in black). EDEs or accretion disks can be present at these later stages, but at lower density.

**Table 1. Wind characteristics of the AGB stars in the ATOMIUM sample.** Columns 1-6 contain the source name, luminosity in units of solar luminosity $L_\odot$, mass-loss rate, wind velocity based on the $^{12}$CO $J$=2→1 line ($v_{\rm wind}$), the identification of arc morphologies in the CO $J$=2→1 channel map, and the SiO wind dynamics characterizing the velocity field ($\boldsymbol{v}$) in the vicinity of the AGB star as derived from our ALMA data (Fig. S8-S65, Table S2 *(18)*). The stars are ordered by increasing mass-loss rate. The last column indicates objects with similar wind characteristics. Class 1 designates sources with multiple density arcs and dynamically complex inner wind structures, with signs of a biconical outflow and/or rotation, shaping the wind in an equatorial density enhancement (EDE). Class 2 indicates a bipolar structure, some with additional hourglass morphology in the CO channel maps. Class 3 has large density arc(s) often with a recognizable spiral-like structure.

| Name | Luminosity ($\boldsymbol{L_\odot}$) | Mass-loss rate ($\boldsymbol{M_\odot}$ yr$^{-1}$) | $\boldsymbol{v_{wind}}$ (km s$^{-1}$) | CO morphology arcs[(a)] | SiO inner wind dynamics[(b)] | ATOMIUM classification |
|---|---|---|---|---|---|---|
| S Pav | 4859 | $8.0 \times 10^{-8}$ | 14 | (x) | Skewed rotating $\boldsymbol{v}$-field | Class 1 |
| T Mic | 4654 | $8.0 \times 10^{-8}$ | 14 | x | Skewed rotating $\boldsymbol{v}$-field | Class 1 |
| U Del | 4092 | $1.5 \times 10^{-7}$ | 17 | c-xx | Bipolar/rotating flow | Class 2 |
| RW Sco | 7714 | $2.1 \times 10^{-7}$ | 19 | c-xx | - (low S/N) | Class 2 |
| V PsA | 4092 | $3.0 \times 10^{-7}$ | 20 | c-xx | Bipolar flow | Class 2 |
| SV Aqr | 4000 | $3.0 \times 10^{-7}$ | 16 | c-xx | - (low S/N) | Class 2 |
| R Hya | 7375 | $4.0 \times 10^{-7}$ | 22 | o-xx | Skewed rotating **v**-field | Class 2 |
| U Her | 8026 | $5.9 \times 10^{-7}$ | 20 | a-xx | Complex dynamics | Class 3 |
| **π**[1] Gru | 4683 | $7.7 \times 10^{-7}$ | 65 | o-xx | Bipolar/rotating flow | Class 2 |
| R Aql | 4937 | $1.1 \times 10^{-6}$ | 16 | xxx | - | Class 3 |
| W Aql | 9742 | $3.0 \times 10^{-6}$ | 25 | xxx | Complex dynamics | Class 3 |
| GY Aql | 9637 | $4.1 \times 10^{-6}$ | 18 | xxx | Complex dynamics | Class 3 |
| IRC -10529 | 14421 | $4.5 \times 10^{-6}$ | 20 | xxx | Bipolar/rotating flow | Class 3 |
| IRC +10011 | 13914 | $1.9 \times 10^{-5}$ | 23 | xxx | Complex dynamics | Class 3 |

(a) (x): faint arc, x: several arcs with extent $< 180°$ , c-xx: circular/elliptical arc centered around the star, o-xx: arcs symmetrically offset from the central star, a-xx: pronounced asymmetric arcs, xxx: more than 1 arc with extent $> 270°$, linked to a (complex) spiral structure.

(b) 'Bipolar/rotating flow' indicates a directed bipolar flow or an EDE/disk-like structure, sometimes with Keplerian rotation; 'skewed rotating **v**-field' denotes systematic, but complex, signs of rotation and the $v = 0$ signature in the map of the intensity weighted velocity field (moment1-map) is skewed; 'complex dynamics' refers to a clear blue-shifted and red-shifted velocity structure in the moment1-map, but no obvious systematic rotation can be deduced; '-' denotes that no conclusion could be drawn, sometimes owing to too low a signal-to-noise ratio of the SiO data ('low S/N').

**Acknowledgments:** ALMA is a partnership of ESO (representing its member states), NSF (USA) and NINS (Japan), together with NRC (Canada) and NSC and ASIAA (Taiwan), in cooperation with the Republic of Chile. The Joint ALMA Observatory is operated by ESO, AUI/NRAO and NAOJ. The CASA data reduction package was developed by an international consortium of scientists based at the National Radio Astronomical Observatory (NRAO), the European Southern Observatory (ESO), the National Astronomical Observatory of Japan (NAOJ), the CSIRO Australia Telescope National Facility (CSIRO/ATNF), and the Netherlands Institute for Radio Astronomy (ASTRON) under the guidance of NRAO. We thank the Data Reduction team at ESO for customizing the imaging pipeline. We thank the UK Science and Technology Facilities Council (STFC) IRIS for provision of high-performance computing facilities allowing UK's Radio and mm/sub-mm Interferometry Services to improve the data quality and much more data to be processed. We thank the ALMA Archive scientist Felix Stoehr for providing the technical information; **Funding:** L.D., D.G., W.H., J.B., J.M.C.P., and S.H.J.W. acknowledge support from the ERC consolidator grant 646758 AEROSOL, L.D., H.S., and E.C. acknowledge support from the KU Leuven under the C1 MAESTRO grant C16/17/007, H.S. acknowledges support from the European Research Council (ERC) under the European Union's DLV-772225-MULTIPLES Horizon 2020 research and innovation programme, W.H. acknowledges support from the FWO Flemish Fund of Scientific Research under grant G086217N, F. H. acknowledges support from the "Programme National de Physique Stellaire" (PNPS) of CNRS/INSU co-funded by CEA and CNES, F.D.C. is supported by the EPSRC iCASE studentship programme, Intel Corporation and Cray Inc., J.M.C.P. acknowledges support from the UK STFC grant ST/P00041X/1, J.Y. acknowledges support from the UK STFC grant ST/R001049/1, M.V.d.S. acknowledges support from the FWO through grant 12X6419N, T.D. acknowledges support from the FWO through grants 12N9917N & 12N9920N, M.M. acknowledges support from the European Union's Horizon 2020 research and innovation program under the Marie Skłodowska-Curie Grant agreement No. 665501 with the FWO ([PEGASUS]$^2$ Marie Curie fellowship 12U2717N, A.B. acknowledges support from the "Programme National de Physique Stellaire" (PNPS), I.M. acknowledges funding by the UK STFC grant ST/P000649/1, EDB acknowledges financial support from the Swedish National Space Agency, I.E.M. acknowledges support from the FWO and the European Union's Horizon 2020 research and innovation program under the Marie Skłodowska-Curie grant agreement No 665501, S.E. acknowledges funding from the UK STFC as part of the consolidated grant ST/P000649/1 to the Jodrell Bank Centre for Astrophysics at the University of Manchester, P.K. acknowledges support from the French PNPS of CNRS/INSU, C.A.G. acknowledges support from NSF grant AST-1615847, T.J.M. is grateful to the STFC for support under grant ST/P000321/1, A.A.Z. was supported by the STFC under grants ST/T000414/1 and ST/P000649/1, M.D.G. thanks the STFC for support under consolidated grant ST/P000649/1 to the JBCA; **Author contributions:** L.D. is principal investigator (PI) of the ALMA Large Program ATOMIUM, leads the ATOMIUM team and supervises the ATOMIUM project, developed the methodology, analyzed the results, wrote the draft manuscript, contributed to Fig. 1, made Fig. 2, Fig. S1-S2, wrote the software for Fig. S5-S7 and Fig. S8-S65; M.M. developed the observational strategy and contributed to Fig. 1; A.M.S.R. led the ATOMIUM data reduction team and wrote software used in the data reduction; C.A.G. is co-PI of the ATOMIUM project; W.H., I.E.M., and J.B. performed hydrodynamical simulations of binary systems; I.McD. and H.S. investigated the (sub-)stellar binary population statistics; M.M, T.D., W.H., S.H.J.W., A.B., S.E., F.H., K.T.W. contributed to the data reduction; J.D.R. performed the statistical analysis and made Fig. S3-S4, J. Y. provided hardware for the ALMA data reduction, P.K. and F.D.C. contributed to Fig. 1. All authors discussed the interpretation of the data, contributed scientific results, and helped prepare the paper; **Competing interests:** We declare no competing interests; **Data and materials availability:** The ALMA data are available from the ALMA data archive at http://almascience.eso.org/aq/ under project code 2018.1.00659.L. Scripts for producing Figure S5-S7 are available at https://github.com/LeenDecin/Supplementary_material_for_Science_paper_abb1229.

**Supplementary Materials:**

Materials and Methods

Supplementary Text

Figures S1-S65

Tables S1-S7

Data S1-S2

References (*31-168*)

Supplementary Materials for

# (Sub-)stellar companions shape the winds of evolved stars

L. Decin*, M. Montargès, A. M. S. Richards, C. A. Gottlieb, W. Homan, I. McDonald, I. El Mellah, T. Danilovich, S. H. J. Wallström, A. Zijlstra, A. Baudry, J. Bolte, E. Cannon, E. De Beck, F. De Ceuster, A. de Koter, J. De Ridder, S. Etoka, D. Gobrecht, M. Gray, F. Herpin, M. Jeste, E. Lagadec, P. Kervella, T. Khouri, K. Menten, T. Millar, H. S. P. Müller, J. Plane, R. Sahai, H. Sana, M. Van de Sande, L. B. F. M. Waters, K. T. Wong, J. Yates

*Corresponding author. Email: leen.decin@kuleuven.be.

**This PDF file includes:**

# Materials and Methods

## S1 ALMA ATOMIUM Large Program and target selection

The ALMA ATOMIUM Large Program aims to establish the dominant physical and chemical processes in the winds of oxygen-rich evolved stars over a range of initial stellar masses, pulsation behaviours, mass-loss rates, and evolutionary phases. The ATOMIUM observing proposal was selected in Cycle 6 as Large Program for a total of 113.2 hr (2018.1.00659.L, PI L. Decin). To disentangle the impact of various (circum)stellar properties, a sample of seventeen targets was selected that cover i) a range in initial stellar mass — hence Asymptotic Giant Branch (AGB) stars versus the more massive counterparts, the red supergiants (RSG); ii) various pulsation characteristics (regular versus semi-regular) and AGB versus RSG (large versus small amplitude); iii) various evolutionary phases — hence inclusion of sources with different mass-loss rates and S-type AGB stars (with C/O almost equal to 1). All selected targets have an apparent stellar angular size above 3 mas and an $R$-band magnitude $<$11 to allow for contemporaneous observations with the Spectro-Polarimetric High-contrast Exoplanet REsearch (SPHERE) instrument mounted on the Very Large Telescope (VLT) and the Multi-AperTure mid-Infrared SpectroScopic Experiment (MATISSE) mounted on the Very Large Telescope Interferometer (VLTI) *(31, 32)*. Fourteen sources are AGB stars and three are red supergiants.

Summarized in Table S1 are the (circum)stellar parameters of the fourteen AGB sources in the ATOMIUM sample. Variability is a common feature of AGB stars and is mainly caused by pulsations. Mira variables have regular, large amplitude variations in the visible light, with $\delta V > 2.5$ mag; semiregular variables of type a (SRa) are similar to regular Mira variables, but have smaller $V$-band amplitude. Semiregular variables of type b (SRb) show poor regularity with a small amplitude *(3)*. A source is classified as a long-period variable (LPV) if no regular pulsation period could be deduced from observations. Stellar pulsations are an important trigger for the onset of the dust-driven AGB wind *(33)* by increasing the density scale height. It follows that, in general, stars with regular long-period large amplitude pulsations have larger the mass-loss rate. For Mira-type variable stars with a luminosity $L$ above $\sim$2 000 $L_\odot$ and pulsation period $P$ between $\sim$300 – 800 days, a linear relation exists between the pulsation period and the logarithm of the mass-loss rate ($\sim 10^{-7}\,M_\odot\,yr^{-1} < \dot{M} < 3\times10^{-5}\,M_\odot\,yr^{-1}$) *(34,35)*. Semiregular variables with a pulsation period less than 200 days cover essentially the same mass-loss rate regime as the Mira variables with pulsation period between 200 – 400 days, while a maximum mass-loss rate of a few $10^{-5}\,M_\odot\,yr^{-1}$ is reached for $P \gtrsim 800$ days *(34, 35)*. Between $\sim$60 and $\sim$300 days, an approximately constant mass-loss rate of $\sim 3.7\times10^{-7}\,M_\odot\,yr^{-1}$ is found, while for $P <$ 60 days the mass-loss rate is a factor of $\sim$10 smaller *(33, 36)*.

**Distance:** Obtaining accurate parallax observations for the ATOMIUM targets is challenging owing to the movements of the photocenter as the star pulsates, and the motions of the convective cells around the central star. Both effects produce correlated errors in the proper motion, however the excess noise in the astrometric solution can be used to estimate the magnitude of these effects which should be random and average out on long timescales. The Gaia Data Release 2 (DR2) and Hipparcos reductions *(37, 38)* are independent. If the two reductions agree to within the estimated uncertainties, it implies the parallax is correct. Excess noise of any kind will generally result in a larger measured parallax and will manifest itself as a closer apparent distance.

Table S1: **Summary of (circum)stellar parameters of the ATOMIUM sample.** The first column gives the target name, the second column the AGB variability type, the third column the pulsation period $P$, the fourth column the distance $D$, the fifth column the stellar angular diameter $\theta_D$, the sixth column the effective temperature $T_{\rm eff}$, the seventh column the stellar luminosity $L$, the eighth column the mass-loss rate $\dot{M}$, the ninth column the wind velocities as determined from the ALMA ATOMIUM $^{12}$CO$J = 2 \rightarrow 1$ observations, the tenth column the local standard of rest velocity $v_{\rm LSR}$ used as input for the ALMA observations, and the last column an estimate of the local standard of rest velocity using the ALMA ATOMIUM data. "-" indicates there are insufficient measurements available. Targets are ordered according to increasing mass-loss rate.

| Name | Variability type | Pulsation period (days) | Distance $D$ (pc) | Angular diameter (mas) | $T_{\rm eff}$ (K) | $L$ ($L_\odot$) | Mass-loss rate ($M_\odot\,yr^{-1}$) | $v_{\rm wind}$ ALMA CO (km s$^{-1}$) | $v_{\rm LSR}$ ALMA obs. (km s$^{-1}$) | $v_{\rm LSR}^{\rm new}$ (km s$^{-1}$) |
|---|---|---|---|---|---|---|---|---|---|---|
| S Pav | SRa | 381 *(39)* | 190 *(37)* | 11.61 $^\star$ | 3100 *(40)* | 4900 $^\dagger$ | $8 \times 10^{-8}$ *(39)* | 14 | $-20.0$ | $-18.2$ |
| T Mic | SRb | 347 *(39)* | 210 *(37, 38)* | 9.26 $^\star$ | 3300 *(40)* | 4700 $^\dagger$ | $8 \times 10^{-8}$ *(39)* | 14 | 25.3 | 25.5 |
| U Del | SRb | 119 & 1170 *(41)* | 279 *(37, 38)* | 7.90 *(42)* | 3000 $^\S$ | 4000 $^\dagger$ | $1.5 \times 10^{-7}$ *(39)* | 17 | $-6.4$ | $-6.8$ |
| RW Sco | Mira | 389 *(43)* | 514 *(37)* | 4.87 $^\star$ | 3300 *(40)* | 7700 $^\dagger$ | $2.1 \times 10^{-7}$ *(43)* | 19 | $-72.0$ | $-69.7$ |
| V PsA | SRb | 148 *(39)* | 278 *(37)* | 12.80 $^\star$ | 2400 *(39)* | 4100 $^\dagger$ | $3 \times 10^{-7}$ *(39)* | 20 | $-11.1$ | $-11.1$ |
| SV Aqr | LPV | - | 389 *(37)* | 4.36 $^\star$ | 3400 *(40)* | 4000 $^\ddagger$ | $3 \times 10^{-7}$ *(39)* | 16 | 8.5 | 6.7 |
| R Hya$^\|$ | Mira | 366 *(44)* | 165 *(45)* | 23.0 *(42)* | 2100 *(35)* | 7400 $^\dagger$ | $4 \times 10^{-7}$ *(35)* | 22 | $-11.0$ | $-10.1$ |
| U Her | Mira | 402 *(44)* | 572 *(46)* | 11.0 *(42)* | 2100 $^\S$ | 8000 $^\dagger$ | $5.9 \times 10^{-7}$ *(47)* | 20 | $-14.5$ | $-14.9$ |
| $\pi^1$ Gru$^{\|,\P}$ | SRb | 150 *(35)* | 197 *(37, 38)* | 21.0 *(48)* | 2300 *(35)* | 4700 $^\dagger$ | $7.7 \times 10^{-7}$ *(49)* | 64 | $-13$ | $-11.7$ |
| R Aql | Mira | 268 *(44)* | 230 *(37, 38)* | 12.0 *(42)* | 2800 $^\S$ | 4900 $^\dagger$ | $1.1 \times 10^{-6}$ *(47)* | 16 | 47.0 | 47.2 |
| W Aql$^{\|,\P}$ | Mira | 479 *(44)* | 321 *(37)* | 11.0 *(42)* | 2800 *(35)* | 9700 $^\dagger$ | $3 \times 10^{-6}$ *(50)* | 25 | $-25.0$ | $-21.6$ |
| GY Aql | Mira | 468 *(44)* | 152 *(37)* | 20.46 $^\star$ | 3100 *(40)* | 9600 $^\dagger$ | $4.1 \times 10^{-6}$ *(51)* | 18 | 34. | 34.0 |
| IRC $-$10529 | Mira | 680 *(35)* | 760 *(35)* | 6.47 $^\star$ | 2700 *(35)* | 14400 $^\dagger$ | $4.5 \times 10^{-6}$ *(35)* | 20 | $-18.0$ | $-16.3$ |
| IRC +10011 | Mira | 660 *(35)* | 740 *(52)* | 6.53 $^\star$ | 2700 *(35)* | 13900 $^\dagger$ | $1.9 \times 10^{-5}$ *(35)* | 23 | 10.0 | 10.1 |

$^\star$ calculated from $L$, $T_{\rm eff}$ and distance $D$; $^\dagger$ using $M_{\rm bol}(P, L)$-relation *(35)*; $^\ddagger$ assumed following Olofsson et al. *(39)*; $^\S$ from $L$ and $R_\star(\theta_d, D)$; $^\|$ known binary system (see Sect. S9); $^\P$ S-type star with carbon over oxygen ratio (C/O) slightly below 1.

We adopt the Hipparcos parallax if its fractional uncertainty is <0.4; we adopt the Gaia parallax if its fractional uncertainty is <0.25; and we adopt the weighted average of both parallaxes if both conditions are met (where the weight is the inverse of the parallax uncertainty). However, for U Her a maser parallax is known *(46)*, and is used because it is well known that maser parallaxes give a better solution than optical parallaxes for AGB and red supergiant stars. If the parallax is unavailable, or if neither Hipparcos or Gaia parallax meets these criteria, the distance is determined from the period-luminosity relationship *(53)*, where we used the $K$-band magnitude from the Two Micron All-Sky Survey (2MASS). The distance $D$ was obtained from an inversion of the adopted parallax.

**Angular diameter:** Listed in Table S1 is the stellar angular diameter, $\theta_d$, measured from interferometric observations, otherwise $\theta_d$ was calculated from the luminosity $L$, the effective temperature $T_{\rm eff}$, and distance $D$, where the luminosity is derived from the $M_{\rm bol}(P, L)$-relation *(35)*, with $M_{\rm bol}$ the absolute bolometric magnitude and $P$ the pulsation period.

**Mass-loss rate:** The wind of all ATOMIUM targets seems to be shaped by a companion, resulting in a non-spherical envelope. In these circumstances, the AGB wind mass-loss rate is obtained from low-excitation CO rotational lines observed with large (single-dish antenna) beams *(54, 55)* so that the presence of a geometrically compact equatorial density enhancement (EDE) or a (broken) spiral-like structure (in the case of high mass-loss rate objects; see Sect. S3) only translates into an uncertainty of the mass-loss rate smaller than $\lesssim 30\%$. This uncertainty also accounts for the mass-accretion efficiency by the companion, which is well below 30% in most simulations *(21)*. All mass-loss rates in Table S1 are retrieved from low-excitation CO lines, typical uncertainties are a factor of $\sim$3 *(56)*.

**Local standard of rest velocity:** The local standard of rest velocity $v_{\rm LSR}$ in Table S1 is used as input for the ATOMIUM observations (Sect. S2.1). The ATOMIUM observations yield a revised estimate of the $v_{\rm LSR}$ by comparing the difference between the astronomical frequencies and the laboratory measured rest frequencies of identified molecular lines (see below). An estimate of $v_{\rm LSR}$ obtained from the low and medium resolution ATOMIUM measurements is given in Table S1, the uncertainty is $\sim 1.5\,{\rm km\,s^{-1}}$ .

**Wind velocity:** The wind velocity derived from the low and medium resolution data of the rotational line $^{12}$CO $J = 2 \rightarrow 1$ in the ground vibrational state — with $J$ specifying the rotational quantum number — is listed in Table S1 (see also Sect. S2). For each target, the integrated flux is extracted for a range of apertures. The extraction aperture yielding the largest integrated flux of the CO line is used to measure the extent of the blue and red wing (in $\mathrm{km\,s^{-1}}$) with respect to the $\upsilon_{\rm LSR}$ (for intensities greater than 3 times the $\sigma_{\rm rms}$ noise). Listed in Table S1 is the maximum velocity obtained from the blue and red wings, and for the observations at low and medium spatial resolution. The uncertainty in the velocity measure is taken from the spectral resolution of the data, being $\sim$1.25 $\mathrm{km\,s^{-1}}$.

## S2 ALMA observations and data reduction

### S2.1 ALMA observations

ALMA data have been secured in band 6 between 213.83 and 269.71 GHz. Fig. S1 shows the frequency coverage. Each set of four spectral windows (spw) in each array configuration comprises one Science Goal (SG), which together sample $\sim$30 GHz. The requested spectral resolution was $\sim$1.5 $\mathrm{km\,s^{-1}}$. Typical line widths range between 5–60 $\mathrm{km\,s^{-1}}$, where the smaller line widths probe the wind acceleration region.

All targets have been observed with ALMA at a spatial resolution of $\sim$25–50 mas depending on the stellar angular diameter. This setup was offered for the antenna configuration C43-8/C43-9 (with minimum and maximum baseline between 110 m – 13 900 m) with maximum recoverable scale (MRS) $\sim$0.38″–0.62″. To get the full line strength of the transitions, we complement these observations with the antenna configuration C43-5/C43-6 (minimum and maximum baseline between 15 m – 2 500 m, angular resolution of 0.24″/0.13″) to encompass a MRS of $\sim$2″. Even with C43-5/C43-6 the CO and SiO transitions in the ground-vibrational state might be resolved out. Hence, for 15 sources, C43-2 observations at an angular resolution of $\sim$1″ were requested (MRS ranging between 2″–10″, minimum and maximum baseline between 15 m – 314 m). The ALMA antenna configurations during Cycle 6 allowed for most targets to be observed at medium spatial resolution from 3 October – 23 November 2018, at low spatial resolution from 26 December 2018 – 23 January 2019 and from 3 March – 20 March 2019, and at high spatial resolution from 12 June – 13 July 2019.

The positions and proper motions were taken from the ALMA Observing Tool (OT) using its SIMBAD look-up facility, which at the time were taken from the Hipparcos catalogue *(57)*. Each target was observed multiple times during Cycle 6, and each observation was centered on the position and the assumed $\upsilon_{\rm LSR}$ (Table S1) which were adjusted at the start of the execution (Table S5). Three sources (RW Sco, R Aql and W Aql) lie close to the Galactic plane, with other sources in the constellations Aquila and Delphinus within 20 degrees. Some of these suffer from Galactic CO contamination, especially when they are observed at the lowest resolution, but this can be distinguished from circumstellar emission by velocity offsets and by its diffuse nature which is unrelated to the morphology of other lines from the circumstellar envelope. The highest proper motions are $\sim$70 mas $\mathrm{yr^{-1}}$, with a precision better than a few mas $\mathrm{yr^{-1}}$. During the data reduction phase, the positions were aligned to that of the earliest execution. The ALMA astrometric accuracy is better than 1/3 synthesized beam, and although this and errors in the assumed proper motion could cause a position offset of a few mas between executions, self-calibration (Sect. S2.2) prevents this from causing image artefacts.

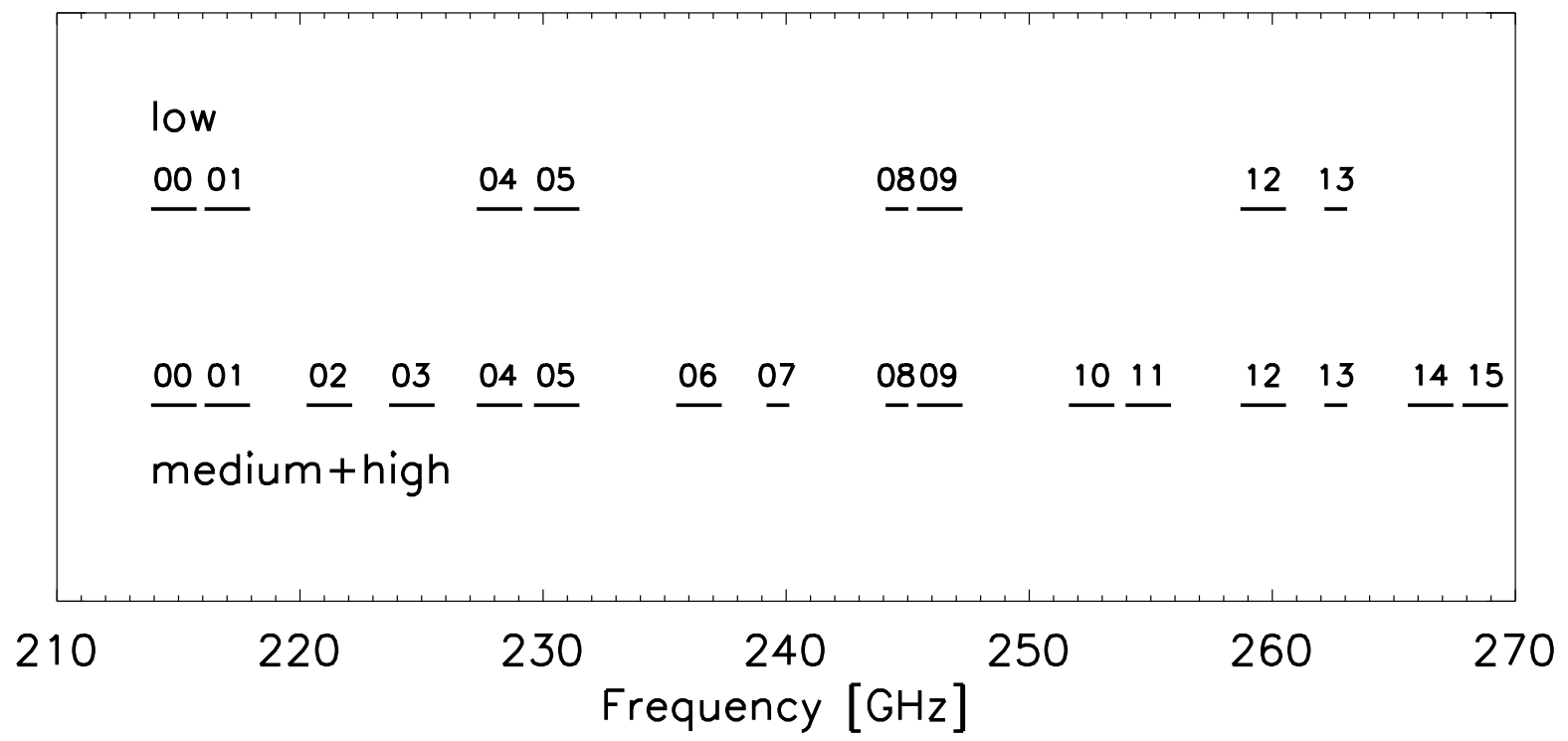

Fig. S1: **The frequency coverage for the ATOMIUM observations in each array configuration.** Each bar represents the frequency coverage of one spectral window, re-numbered in frequency order for convenience. Each individual Science Goal covered four spectral windows, grouped as follows: [0,1,4,5], [2,3,6,7], [8,9,12,13], [10,11,14,15]. The coverage in the high spatial resolution configuration is the same as in the medium-resolution configuration, while for the low-resolution observations only 8 spectral windows were requested. The exact frequency extent for each target depends on the correction for the assumed $v_{\rm LSR}$ on the dates of observation (Table S5).

### S2.2 ALMA data reduction

The medium resolution data are most suitable for retrieving the morphological characteristics of the AGB stars in the ATOMIUM sample, because these data sample the stellar wind at proper angular scales of $0\farcs2$, corresponding to $\sim$50 au for a target at 250 pc. Each fully observed Science Goal was first processed using the ALMA calibration and imaging pipelines *(58)*. The pipelines apply all calibration derived instrumentally (for example from water vapour radiometry) and from calibration and phase-reference sources (see Table S5). The line-free channels are identified from the visibility data and subtracted, and data cubes are made for each spectral window. We inspected the web logs; occasionally a few instances of over- or under-flagging were identified, but the former were negligible and the latter were remedied during our processing. This comprised the following steps:

1. Split out two copies of the calibrated target data, one at a 'continuum' spectral resolution of 15.625 MHz and one at a 'line' spectral resolution of 0.9765625 MHz, adjusted to constant $v_{\rm LSR}$ in the target frame. These were then concatenated to make continuum and line data sets containing the full spectral coverage for each star and array configuration.

2. Use LUMBERJACK *(59)* to identify line-free channels from the pipeline image cubes. The selection is adjusted to correspond to the channelisation of the continuum and line data sets, and checked interactively.

3. Image the continuum-only channels of the continuum data set and self-calibrate, starting with phase-only. This removes any small offsets due to differences in calibration or proper motion uncertainty, as well as improving image quality. If the signal-to-noise ratio is sufficient, an image using a first-order spectral index is made to provide a model for more cycles of self-calibration, including amplitude self-calibration. The dominant cause of small amplitude scaling offsets between executions and Science Goals is the ALMA flux density uncertainty of up to 5%, in practice only statistically significant above the noise in sources bright enough to remove any offset by self-calibration.

4. Apply the corrections to the line data set, check the selection of line-free channels and subtract the continuum using a linear fit.

5. Make a spectral image cube for each spectral window and configuration large enough to encompass all detectable emission. This was checked against the pipeline cubes which are made to the 20% primary beam response radius for the lowest-resolution observations, most sensitive to extended emission. This tends to be resolved-out at higher resolution, so smaller images (in angular size) can be made. In making the separate per-configuration cubes, the ALMA default weighting for optimum signal-to-noise was used, resulting in a restoring beam varying slightly depending on target elevation and exact antenna positions. The cubes used for analysis are primary-beam corrected, but (unless otherwise stated) the $\sigma_{\rm rms}$ is estimated near the center of the field avoiding any emission.

The image cube and continuum image properties can be found in Tables S6–S7 (Sect. S8).

# S3 ATOMIUM morphological classification

## S3.1 ATOMIUM wind morphologies

The lines of particular relevance for this study are the rotational transitions in the ground vibrational state of $^{12}$CO $J=2\rightarrow1$ (230.538 GHz), $^{28}$SiO $J=5\rightarrow4$ (217.105 GHz), and $^{28}$SiO $J=6\rightarrow5$ (260.518 GHz) observed at medium spatial resolution. The high fractional abundance of both molecules ([CO/$H_2$] $\sim$4$\times$10$^{-4}$, [SiO/$H_2$] $\sim$3$\times$10$^{-5}$ *(60)*) facilitate their detection. The low-excitation CO line is predominantly collisionally excited and is a tracer of the density distribution in the outer wind region because the excitation energy is low (upper state energy, $E_{\rm up}$ of 16.6 K), the dipole moment $\mu$ is only 0.11 Debye, and the Einstein $A_{u\rightarrow l}$ coefficient (with $u$ indicating the upper energy level, and $l$ the lower level) is very small ($A_{u\rightarrow l}$ = 6.91 $\times$ 10$^{-7}$ s$^{-1}$, *(61–63)*). The Einstein $A_{u\rightarrow l}$ coefficients of the two rotational lines of SiO ($E_{\rm up}$ = 31.26 K and 43.76 K for the $J=5\rightarrow4$ and $6\rightarrow5$ line) are three orders of magnitude higher ($A_{5\to4}$ = $5.1965\times10^{-4}$ s$^{-1}$, and $A_{6\to5} = 9.1161\times10^{-4}$ s$^{-1}$, with $\mu$ = 3.08 Debye; *(62,64)*). As a result the lines of SiO are sensitive to radiative (de-)excitation effects and are diagnostic of the dynamics in the inner wind region. SiO molecules can be depleted in the inner wind region if silicate grains are forming *(65)*. No conclusive evidence can yet be drawn on the depletion efficiency, but observation and modelling efforts indicate at least a fraction of SiO remains in gaseous form outside the main dust condensation region before photodissociation by the interstellar ultraviolet (UV) photons sets in *(66–69)*. The weaker isotopologue lines in the ground vibrational state ($^{13}$CO $J=2\to1$ at 220.399 GHz, $^{29}$SiO $J=5\to4$ at 214.386 GHz, and $^{30}$SiO $J=6\to5$ at 254.217 GHz) provide complementary information, because the lines of the dominant isotopic species can be prone to high optical depth effects. Owing to lost flux of the $^{12}$CO $J=2\to1$ line in the medium resolution observations of some sources (largest angular scale, LAS, of $\sim$ 1.5$''$), the low resolution data were included in the analysis to encompass a larger angular scale. The resolved-out flux density question affects physical parameters derived in a radiative transfer analysis. However, our analysis of the prevailing wind morphology is not affected because any asymmetry detected at a given angular resolution is a real feature imprinted in our data. The filtering of extended structures provides higher dynamic range for identifying asymmetric and clumpy structures. The derivation of the wind velocity from the CO and SiO lines (see Table S1 and Table S2) is also not affected because resolved-out flux is mainly seen as a depression of the line flux around the central velocities and not in the line wings.

Figure 1 was obtained by selecting three velocity frames of the channel map of the $^{12}$CO $J=2\to1$ line: one at rest velocity, one blue shifted, and one red shifted with respect to the local standard of rest velocity $v_{\rm LSR}$. For two targets (RW Sco and SV Aqr) the signal-to-noise of the data was too low to produce a three-colour map, although the individual channels show clear signs of asymmetry (Fig. S20, Fig. S28). Figure 1 shows a range of morphologies, with none of the stars in our sample displaying a spherically symmetric wind geometry. All these morphologies have a counterpart in the more evolved post-AGB stars and planetary nebulae (PNe); some prominent examples are given below.

- The 'rose-like' structure of R Aql resembles the inner structure of the Eskimo Nebula, a bipolar double-shell PN *(70)*;

- the 'eye'-like morphology of U Del bears a morphological resemblance with the outer regions of Helix Nebula *(71, 72)*, a bipolar planetary nebula;

- the biconical shape of R Hya is seen in various post-AGB stars and PNe including the post-AGB star IRAS 17150$-$3224 which also has a highly equatorially enhanced shell *(73)* and the Owl Nebula (NGC 3587), a planetary nebula which has a barrel-like structure in its inner shell caused by bipolar cavities *(74)*;

- regularly spaced arcs embedded in the bipolar outflow of $\pi^1$ Gru (see Fig. S42) are reminiscent of the Red Rectangle *(75)*, a pre-PN or to the outer halo structure of the Cat's Eye Nebula *(76)*, a planetary nebula. A regular spiral structure (as in IRC $-10529$ or IRC $+10011$) seen at high inclination angle would also manifest as regularly spaced arcs in the plane of the sky. The spiral-arm spacing of 1″ in the wind of the oxygen-rich AGB star OH 26.5+0.6 (see Table S3, Sect. S9; *(55)*) (for a distance of 1 370 pc) is similar to the regular arc spacing in the outer halo of the Cat's eye nebula, which is ∼1.3″ for a distance of 1 001 pc *(76)*;

- (stable) disks have been detected in numerous post-AGB stars and PNe, such as AR Puppis *(77)*;
- the kinematical behaviour of disks surrounding AGB stars and post-AGB star shows similarities. In particular, the Red Rectangle is an oxygen-rich post-AGB binary system which has a Keplerian (rotating) dosk and an outflow, the latter mainly being formed of gas leaving the disk *(78)*. The velocity vector field in the disk surrounding the Red Rectangle resembles that of the oxygen-rich AGB star R Dor (see Table S3, Sect. S9; *(79)*). Another example is the oxygen-rich post-AGB binary system AC Her which has a disk with a Keplerian velocity field *(80)* similar to the kinematical structure of the disk surrounding the oxygen-rich AGB star $L_2$ Pup (see Table S3, Sect. S9; *(81)*).

Even during the early stage of the AGB phase, characterized by a low mass-loss rate, the targets quite frequently have a pronounced aspherical morphology with an axisymmetric geometry. The resemblance between the morphologies of AGB stellar winds and post-AGB stars and PNe supports the idea that planetary nebulae are descendants of evolved low and intermediate mass AGB stars *(82)*. We conjecture that the same formation mechanism gives rise to the plethora of morphologies observed in AGB winds, post-AGB objects, and planetary nebulae.

### S3.2 Observational classification of ATOMIUM wind morphologies

The rotational emission lines of CO and SiO are the basis of our observational classification of the prevailing wind morphologies. All data used for the morphological classification are displayed in Figs. S8–S65. Our morphological classification scheme is intended to elucidate the predominant morphology, however in each source smaller scale structural features often enrich this picture. Our classification follows a stepwise approach and encompasses three steps (Fig. S2).

- Step 1: CO morphological information: The prevalence of arc-like structures is investigated in the medium and low-resolution CO channel maps (Table S2). Some sources display only one arc with an extent <180°, other sources are surrounded by a regularly spaced spiral structure, while an elliptical/circular structure surrounds the central target in a fraction of sources.
- Step 2: Inner versus outer wind kinematics: In the next step, we investigate the kinematics in the inner versus the outer wind region. Winds of AGB stars are radiation driven, so the absorption of the outwards-directed stellar radiation by newly formed dust grains produces a net force that can overcome gravity *(83)*. The gas is then accelerated beyond the escape velocity with the main acceleration occurring in the first few stellar radii. The velocity profile of AGB stellar winds is parametrised using the $\beta$-type law *(84)*

$$v(r) = v_0 + (v_\infty - v_0)\left(1 - \frac{r_{\mathrm{dust}}}{r}\right)^{\beta} , \tag{S1}$$

with $r$ the radial distance to the star, $v_0$ the velocity at the dust condensation radius $r_{\mathrm{dust}}$, $v_\infty$ the terminal wind velocity, and larger values of $\beta$ indicating a lower wind acceleration. If SiO is excited closer to the central star than CO, as confirmed by the measured angular extents, the width of the SiO lines should be smaller or equal to that of CO. However, in some ATOMIUM AGB sources, a different behavior is observed with $v_{\rm wind}$(CO) < $0.85 v_{\rm wind}$(SiO); see Table S2. The uncertainty in the derived fraction is ~0.11, owing to the velocity resolution of ~$1.25\,{\rm km\,s^{-1}}$.

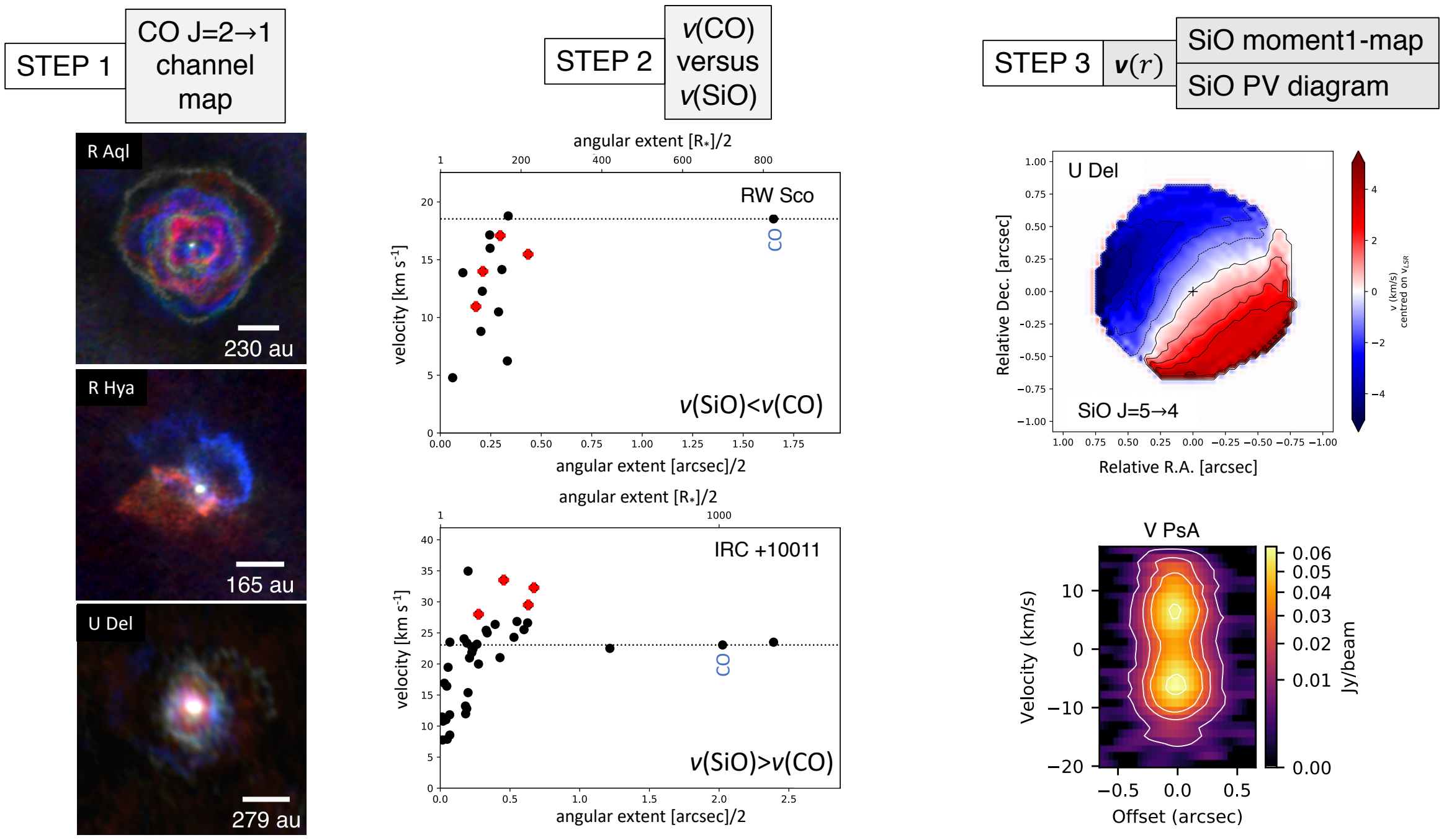

Fig. S2: **Decision tree used to classify the prevailing wind morphology for the ATOMIUM AGB stars.** Step 1 assesses the presence of arcs and spiral-like structures based on the $^{12}$CO $J = 2 \rightarrow 1$ channel maps. Step 2 investigates the difference between the inner and outer wind kinematics. Step 3 deduces systematic, although sometimes complex, inner wind dynamics. The images and plots at each step are illustrative of the data shown in Figs. S8–S65. Step 1 is illustrated using R Aql, R Hya, and U Del as an example for a complex spiral-like structure, a bipolar structure, and an 'eye'-like feature, respectively. Step 2 shows the velocity measurements in RW Sco and IRC +10011. All dots represent measurements of various molecular species in the ATOMIUM data. The measurement of the $^{12}$CO $J = 2 \rightarrow 1$ is indicated in blue, the SiO lines with red crosses. For RW Sco $v(\mathrm{SiO}) < v(\mathrm{CO})$, while for IRC +10011 $v(\mathrm{SiO}) > v(\mathrm{CO})$. Step 3 is illustrated using the SiO $J = 5 \rightarrow 4$ moment1-map of U Del, showing a clear sign of bipolarity or a rotating flow, and the SiO $J = 5 \rightarrow 4$ PV diagram of V PsA in which 2 bright blobs can be discerned.

- Step 3: SiO morpho-dynamical information: The SiO data were then analyzed using stereograms, moment1-maps, and position-velocity (PV) diagrams (see Sect. S7). Stereograms and moment1-maps are used to identify possible positional offsets between blue- and red-shifted emission, hence yielding information on the velocity vector field ($\mathbf{v}$) and rotation (Table S2). As examples, see the data for S Pav, T Mic, U Del, and $\pi^1$ Gru (Fig. S10, Fig. S14, Fig. S18, and Fig. S43). PV diagrams are constructed for two orthogonal slits so that the asymmetry between the two PV diagrams is maximized to unravel the morpho-kinematical behavior *(85)* (Table S2). A rotating Keplerian disk or an EDE with constant radial velocity appear in one (or both) of the PV diagram(s) as a 'butterfly'-like signature *(85)*. See, for example, the PV diagrams of $^{12}$CO in $\pi^1$ Gru shown in Fig. S44.

The outcomes of each of these three steps are described below and summarized in Table S2. This stepwise procedure allows the characterization and classification of 13 out of the 14 ATOMIUM AGB stars. U Her seems an outsider: the CO channel maps show pronounced asymmetric arcs although they seem not to be linked to a spiral-like structure.

- Step 1: Based on the CO channel maps, we discern a spiral-like structure in five targets (R Aql, W Aql, GY Aql, IRC-10529, and IRC+10011; Fig. S46, Fig. S50, Fig. S54, Fig. S58, and Fig. S62). The CO data of both R Hya and $\pi^1$ Gru show a clear 'bipolar' signature. For R Hya ($\pi^1$ Gru) the blue-shifted emission is east (north) with respect to the central star, while the red-shifted emission is west (south) (see the CO channel maps in Fig. S32 and Fig. S41, and the CO moment1-maps in Fig. S36 and Fig. S45). In both sources, an hourglass signature can be discerned. An 'eye'-like morphology is recognized in the channel maps of U Del, RW Sco, V PsA, and SV Aqr (Fig. S16, Fig. S20, Fig. S24, Fig. S28).

- Step 2: Five sources have an SiO line width considerably larger than the CO line width (Table S2): S Pav, T Mic, W Aql, IRC −10529, IRC +10011.

- Step 3: The analysis of the SiO stereograms and moment1-maps reveals two sources (U Del and $\pi^1$ Gru) with a definite offset between red-shifted and blue-shifted emission (Fig. S18, Fig. S43), due to either a rotating inner wind structure or a bipolar outflow. This is recognizable in the moment1-maps as an almost straight-line division between red and blue. For U Del and V PsA two bright blobs are discerned in the PV diagrams of SiO, independent of the slit position angle (PA) (Fig. S19, Fig. S27). This pattern cannot be explained by a detached SiO shell *(86)*, because that would produce lower emission for the central velocities, which is not seen. The SiO signatures of U Del, $\pi^1$ Gru, and V PsA can be understood as indication of 'bipolarity'. We use the term 'bipolarity' to indicate a directed bipolar flow or an EDE/disk-like structure, which may display Keplerian rotation, so that lower density biconical poles are created perpendicular to the EDE. In five sources (S Pav, T Mic, R Hya, W Aql, and GY Aql) a systematic, but complex, flow can be seen. The $v = 0$ signature in their moment1-maps is strongly skewed (Fig. S10, Fig. S14, Fig. S34, Fig. S52, and Fig. S56). The signal-to-noise ratio of the SiO lines of RW Sco and SV Aqr is too low to determine the inner wind dynamics (Fig. S23, Fig. S31).

A correlation emerges between the mass-loss rate (and hence pulsation properties of the AGB star) and the prevailing morphological appearance (Table S2) allowing us to define three Classes (Table 1). 'Class 1' sources are low mass-loss rate AGB stars ($\dot{M} \lesssim 1\times10^{-7}\,\mathrm{M_\odot yr^{-1}}$) characterized by a systematic, but complex, inner wind dynamics with signs of distorted rotation. Because rotation, or in general any tangential velocity component, implies a fraction of the material is not radially streaming, this characteristic often implies an EDE structure will be formed *(85)*. We use the term 'EDE' instead of 'disk' or 'torus', because the data does not allow us to determine whether the material is bound to the AGB star. 'Class 2' sources have medium mass-loss rates between $\sim 1\times10^{-7} - 1\times10^{-6}\,\mathrm{M_\odot yr^{-1}}$ and display a bipolar morphology including bipolar outflows and/or (compact) disk-like structures at smaller spatial scales. A spiral signature is only seen in the 'Class 3' sources with mass-loss rates greater than $\sim 1\times10^{-6}\,\mathrm{M_\odot yr^{-1}}$. These sources might also exhibit an EDE in their inner wind region, but the CO channel map (at sufficiently high angular resolution) is dominated by the spiral.

The correlation is quantified by a Kendall's rank $\tau_b$ correlation coefficient of 0.79, indicating a high correlation between the two observables (see Sect. S3.3). We also apply these three Classes to previous observations of other AGB sources in the literature (see Sect. S4.1 and Sect. S3.3). More than twenty nearby AGB stars are then recognized as pronouncedly non-spherical, and can be classified to first order on the basis of the mass-loss rate. This suggests there is a single mechanism which modifies the AGB wind morphology, and the mass-loss rate is an important parameter establishing the morphological outcome. Of the proposed models aiming to explain asphericity in AGB winds, (sub-)stellar binary interaction can explain the observed ATOMIUM morphologies (see Sect. S4.1) and the mass-loss rate relation (see Sect. S4.2). Single-star models based on rapid stellar rotation or strong magnetic fields *(11)* cannot explain these characteristics in a generalized way, without invoking target-specific concepts. Binary-induced characteristics include (see Sect. S4):

- The presence of a companion can perturb the AGB wind flow into a spiral structure either by: i) the direct gravitational pull of the companion which produces a bow shock around it and a tail in its wake, reminiscent of the Bondi-Hoyle-Lyttleton (BHL; *(87, 88)*) configuration; or ii) the orbital motion of the primary AGB star around the common center-of-mass *(19)*.
- Binary interaction can also result in the formation of a circumbinary disk, or an accretion disk around the companion. In each case, a systematic rotating flow will arise, axially shaping the disk and producing a velocity structure that is far more complex than a (simple) Keplerian flow *(20)*.
- For five ATOMIUM sources, the low and medium-resolution ALMA observations indicate that the kinematical structure departs from the classical $\beta$ velocity law and that $v_{\mathrm{wind}}$(SiO)>

Table S2: **Description of the prevailing morphological and dynamical structure of the ATOMIUM AGB sample.** First column gives the name of the target, second column the mass-loss rate (see Table S1), third and fourth columns the information deduced from the CO $J = 2 \rightarrow 1$ channel maps, column 5 the ratio of the velocity deduced from the CO $J = 2 \rightarrow 1$ line (see Table S1) over the maximum velocity as deduced from one of the SiO lines in the survey, and columns 6–9 the information as deduced from the SiO stereograms, moment1-maps and position-velocity (PV) diagrams, in particular on the velocity field $\mathbf{v}$. Explanation of the notation:

$^{\star}$ (x) = faint arc, x = several clear arcs with extent $<$180° , c-xx = circular/elliptical arc centred around target, o-xx = arcs symmetrically offset from central target, a-xx = pronounced asymmetric arc, xxx = more than 1 arc with extent $>$270°, often at regular intervals.

$^{\dagger}$ r/b: minor red/blue shift, R/B: well-resolved red/blue shift, '-': no red/blue shift.

$^{\ddagger}$ c-rb: red/blue shift with complex signature, RB: clear red/blue shift indicating rotation or bipolarity, (o) shift with spatial offset w.r.t. central target, (rb) potential red/blue shift, '-': no stringent red/blue signatures.

$^{\S}$ (B): (weak) butterfly-like signature, BB: 2 bright blobs, (HG): weak hourglass signature, BW: blue wing absorption *(89)*, S: almost axi-symmetric PV for axis at 0″.

$^{\|}$ 'bipolar/rotating flow' indicates a directed bipolar flow or an EDE/disk-like structure, with may display Keplerian rotation; 'skewed rotating $\mathbf{v}$-field denotes that systematic, but complex, signs of rotation are detected with the $v = 0$ signature in the moment1-map being skewed; 'complex dynamics' refers to a clear blue/red signature in the moment1-maps, but no obvious systematic rotation can be deduced; '-' denotes that no conclusion could be drawn, sometimes owing to too low a signal-to-noise ratio of the SiO data ('low S/N').

| Name | Mass-loss rate | Morphology CO $J = 2 \rightarrow 1$ channel map | | Kinematics | Morpho-dynamics SiO | | | | Figures |
|---|---|---|---|---|---|---|---|---|---|
| | ($M_{\odot}$ yr$^{-1}$) | Arcs$^{\star}$ | Description | $v$(CO)/$v^{\rm max}$ | Stereogram$^{\dagger}$ | Moment 1$^{\ddagger}$ | PV$^{\S}$ | Description$^{\|}$ | |
| S Pav | $8 \times 10^{-8}$ | (x) | red/blue shift central emission | 0.77 | r/b | c-rb | (B) | skewed rotating $\mathbf{v}$-field | S8–S11 |
| T Mic | $8 \times 10^{-8}$ | x | several regular arcs | 0.82 | r/b | c-rb | (B) | skewed rotating $\mathbf{v}$-field | S12–S15 |
| U Del | $1.5 \times 10^{-7}$ | c-xx | 'eye'-like feature+arcs | 1.00 | R/B | RB | BB + (B) | bipolar/rotating flow | S16–S19 |

*Continued on next page*

Table S2 – *Continued from previous page*

| Name | Mass-loss rate | Morphology CO $J=2\rightarrow1$ channel map | | Kinematics | Morpho-dynamics SiO | | | | Figures |
|---|---|---|---|---|---|---|---|---|---|
| | ($M_\odot yr^{-1}$) | Arcs$^\star$ | Description | $v$(CO)/$v^{\max}$ | Stereogram$^\dagger$ | Moment 1$^\ddagger$ | PV$^\S$ | Description$^\|$ | |
| RW Sco | $2.1\times10^{-7}$ | c-xx | weak 'eye' + arcs | 1.00 | - | - | - | - (low S/N) | S20–S23 |
| V PsA | $3\times10^{-7}$ | c-xx | 'eye'-like feature + arcs + brighter red/blue | 1.00 | - | - | BB + S | bipolar outflow | S24–S27 |
| SV Aqr | $3\times10^{-7}$ | c-xx | weak 'eye'+brighter red/blue | 0.88 | - | - | BW | - (low S/N) | S28–S31 |
| R Hya | $4\times10^{-7}$ | o-xx | bipolar+hourglass structure | 1.00 | R/B (inner 0.8″) | c-rb (inner 0.8″) | - | skewed rotating **v**-field (inner 0.8″) | S32–S36 |
| U Her | $5.9\times10^{-7}$ | a-xx | pronounced asymmetric arcs | 0.88 | r/b | - | (B) + BW | complex dynamics | S37–S40 |
| $\pi^1$ Gru | $7.7\times10^{-7}$ | o-xx | bipolar+regular arcs+weak hourglass | 1.00 | R/B | RB | (B) | bipolar/rotating flow | S41–S45 |
| R Aql | $1.1\times10^{-6}$ | xxx | rose-like spiral, brighter in red | 1.00 | - | - | S | - | S46–S49 |
| W Aql | $3\times10^{-6}$ | xxx | filamentary spiral | 0.75 | r/b | c-rb (o) | BW | complex dynamics | S50–S53 |
| GY Aql | $4.1\times10^{-6}$ | xxx | spiral | 0.89 | - | c-rb (o) | - | complex dynamics | S54–S57 |
| IRC −10529 | $4.5\times10^{-6}$ | xxx | weak, but regular spiral | 0.74 | r/b | RB | - | bipolar/rotating flow | S58–S61 |
| IRC +10011 | $1.9\times10^{-5}$ | xxx | weak, but regular spiral | 0.72 | r/b | (rb) | - | complex dynamics | S62–S65 |

$v_{\rm wind}$(CO). This can be caused, for example, by a Keplerian disk-like structure for which the tangential velocity component is inversely proportional to the radial distance, i.e. $v^{\perp}_{\rm Kep.}(r) = \sqrt{G\,M_\star/r}$, with $G$ the gravitational constant, $M_\star$ the mass of the AGB star to which the disk is gravitationally bound, and $r$ the radial distance *(17)*. Another cause for this behavior might be the presence of a stellar companion so the companion's gravity locally enhances the velocity amplitude by lowering the effective gravity felt by a particle leaving the primary AGB star.

Other physical phenomena, such as stellar pulsations and a magnetic field might induce additional complexities on the binary-induced morphologies. For example, pulsations can cause an additional ripple-like structure, visible in the lower density bipolar lobes *(20)*.

### S3.3 Statistical correlation coefficient and probability

We quantify the correlation in Table 1 using the Kendall's rank correlation coefficient $\tau_b$, a statistical measure of the ordinal association between two quantities based on the ranks of the data. The rank correlation quantifies the strength of association based on the relative occurrence of concordant and discordant pairs. The $\tau_b$ statistic makes adjustments for ties *(90)*. The $\tau_b$ of 0.79 between the logarithm of the mass-loss rate and our 3 Classes is high, supporting the ordinal association between the mass-loss rate and the morphological Class. Adding the eleven other AGB sources whose geometry was previously deduced from observations with ALMA (see Sect. S4.1) yields a similar correlation coefficient of $\tau_b = 0.77$. The value of $\tau_b = 0.77$ comes with a two-sided $p$-value of $p = 2.6 \cdot 10^{-6}$ under the null hypothesis of $\tau_b = 0$.

The probability to belong to a given morphology class (a categorical variable) given the mass-loss rate $\dot{M}$ (a continuous variable) can be modelled using a multinomial logistic regression. We assume a linear model

$$\ln\left(\frac{\pi_n}{\pi_1}\right) = \theta_{0,n} + \theta_{1,n} x\,, \tag{S2}$$

where $\pi_n$ is the probability of belonging to class $n \in \{2, 3\}$, and the explanatory variable $x \equiv \log_{10} \dot{M}$. The resulting probability curves are shown in Fig. S3.

Class 3 is distinguished from Classes 1/2 using the mass-loss rate, but there is considerable overlap between Class 1 and Class 2. The logistic model fitting yields a pseudo-$R^2$ *(91)* of $R^2 = 0.71$, indicating the mass-loss rate is a relevant explanatory variable. This value of $R^2$ increases to 0.84 if we merge Classes 1 and 2 then fit a 2-class model.

This outcome is a consequence of the classification strategy and the underlying physics governing binary-induced phenomena:

– The most distinct outcome of Step 1 is the presence of a spiral-structure in a fraction of the ATOMIUM sources. Step 1 is based on the $^{12}$CO channel maps, and CO is a tracer of the density, and hence of the mass-loss rate, of AGB stars.

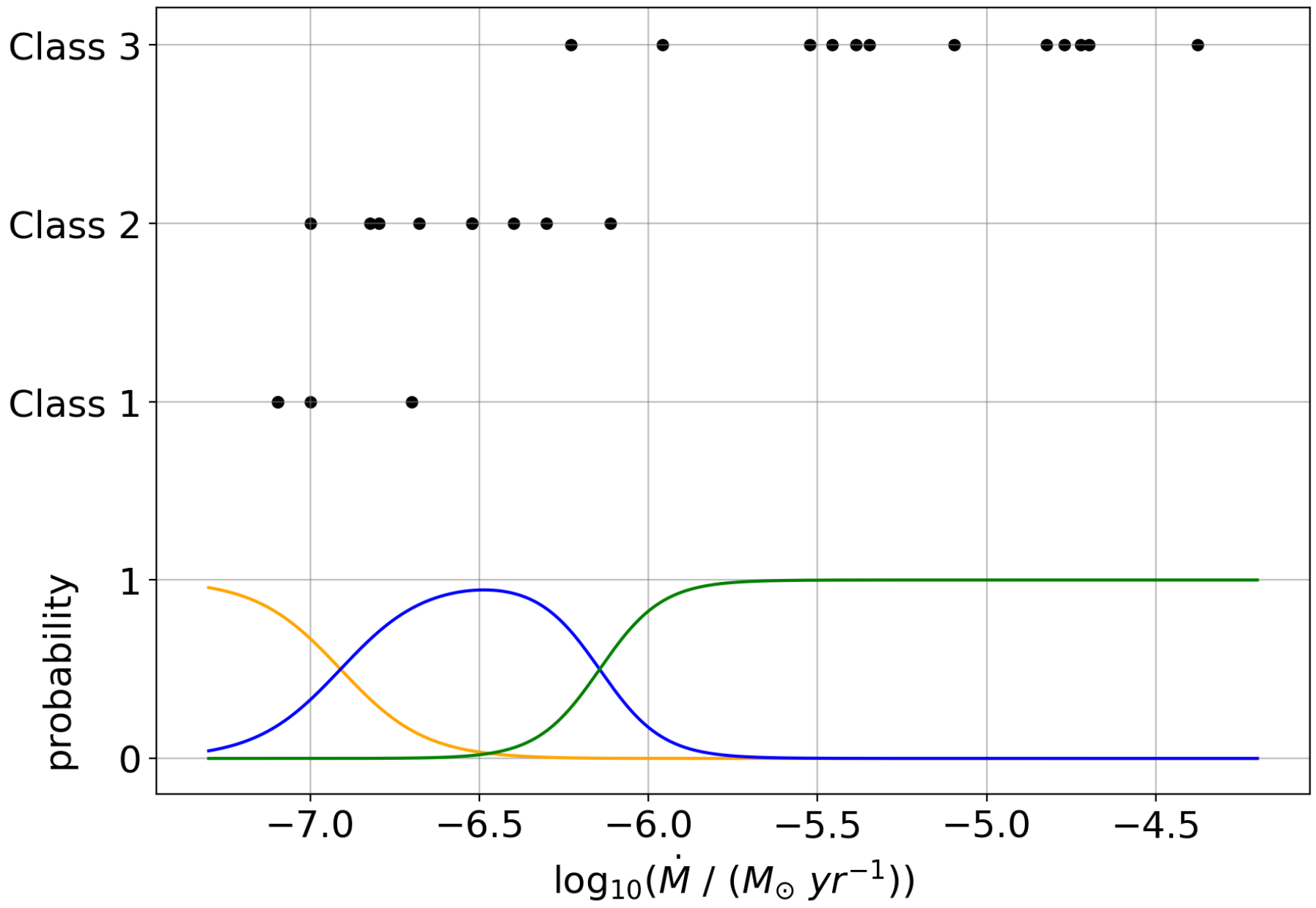

Fig. S3: **Probability curves for the ATOMIUM Classes.** Probability curves obtained through a linear multinomial logistic regression using the mass-loss rates and the morphology classification of 25 AGB sources (the 14 ATOMIUM AGB sources and the 11 AGB sources described in Sect. S4.1). The black dots are the observations, the orange, blue, and green curves are the probability curves for Classes 1, 2 and 3 respectively.

– Step 3 differentiates between Class 1 and Class 2 by using SiO to diagnose the morpho-kinematical behaviour of the inner wind region. The mass-loss rate is a relevant parameter establishing the morphological outcome, but the wind acceleration — and in particular the ratio of the wind speed over the orbital speed — is also involved (see Sect. S4.1 and Sect. S4.2). This implies that including the wind velocity pattern as a variable would change the multinomial logistic regression. However, the small sample size prevents us from adding another variable. The differentiation between Class 1 and Class 2 is visible in Fig. S3 because a lower mass-loss rate correlates with a lower wind acceleration (and terminal wind velocity via mass conservation), and hence the higher potential of forming an EDE with complex binary-induced interaction patterns (see Sect. S4.1). The differentiation between Class 1 and Class 2 is therefore not as strict as the differentiation between Classes 1/2 and Class 3.

The high probability of high mass-loss rate targets to be Class 3 is not caused by an observational bias. The CO intensity scales with $\sim\dot{M}/D^2$ *(35)*. The Kendall's rank correlation coefficient $\tau_b$ between $\log(\dot{M})/D^2$ and the 3 Classes is 0.26, implying that both variables are not correlated (at least not linearly), see Fig. S4.

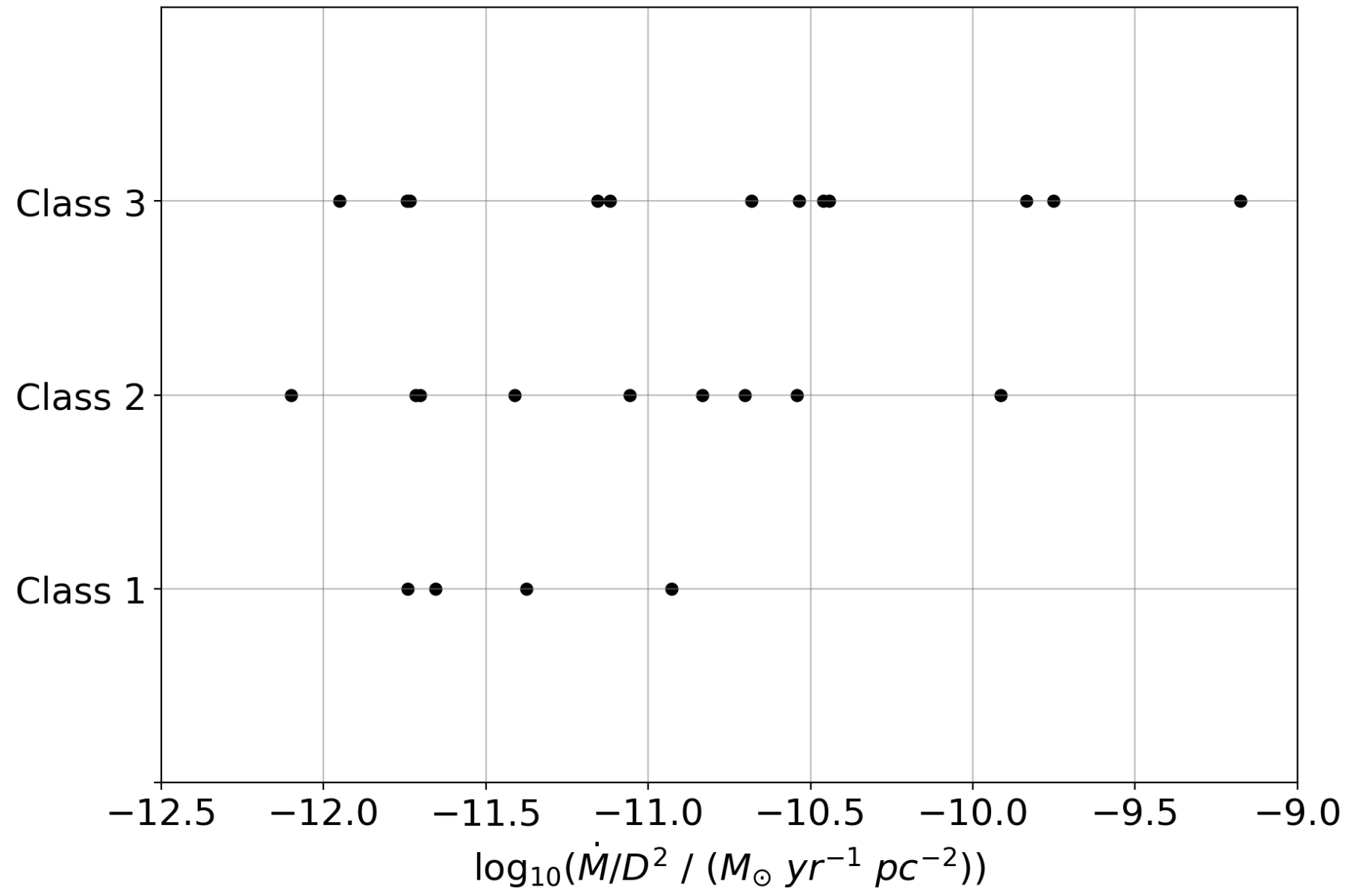

Fig. S4: **Testing of an observational bias.** Morphology classes of 25 AGB sources (same as in Fig. S3) plotted as a function of the logarithm of $\dot{M}/D^2$ (Table S1). The Kendall's rank correlation coefficient $\tau_b$ between both variables is 0.19, implying no linear correlation.

# S4 The ATOMIUM morphological classification in the context of binary-induced wind morphologies

## S4.1 Binary-induced wind morphologies

In this section, we first summarize the modelling outcomes for binary systems with a mass-losing AGB star as primary star. The presence of a binary companion can give rise to a plethora of specific morphological imprints in the AGB stellar winds. Some of these morphological imprints have already been detected in observations. A synopsis of the most recent ALMA results, prior to the ATOMIUM survey, for which the morphology of other AGB winds has been deduced, is given in Table S3. In a last part, we focus on the fact that all ATOMIUM sources are oxygen-rich AGB stars for which the wind acceleration might be much lower than in the case of carbon-rich AGB stars. This difference in wind acceleration profile can have a profound impact on the binary-induced wind morphologies, which here will be described.

__Morphologies in numerical simulations:__ Hydrodynamical simulations have demonstrated the influence of a companion on a wide orbit ($a \geq 3$ au) around a mass-losing AGB star *(92)*. For wide binary systems (with $a \geq 2$ au) the mass transfer does not occur through Roche-lobe overflow (RLOF), in which the primary star transfers material to its companion once it fills its Roche lobe *(93)*. Rather for these wide systems, mass transfer occurs via the stellar wind (wind Roche-lobe overflow, WRLOF) *(94)*. In the WRLOF situation, wind material from the mass-losing AGB star fills the giant's Roche lobe and is transferred to the companion through a narrow and compressed channel which generally does not pass through the inner Lagrangian point. WRLOF occurs when the Roche lobe surface is detached, while RLOF happens when the Roche lobe surface is connected. Simulations covering the binary parameter space have shown the WRLOF scenario can lead to wind capture rates by the companion, which differ by up to an order of magnitude compared to the BHL wind accretion scenario *(95, 96)*. Complex inner wind morphologies can arise *(20, 94, 97)* with a strongly non-radial velocity vector field near the EDE (see discussion in Sect. S8). Depending on the mass ratio, binary separation, mass-loss rate, wind velocity, eccentricity etc. a variety of morphologies arises including spiral structures (which can be bifurcated), accretion and circumbinary disks, bipolar outflows, EDEs with a regular (Keplerian) or complex velocity vector field, 'spider' or 'rose'-like structures etc.

**Previously observed morphologies:** The carbon-rich AGB star IRC +10216 ($\dot{M}\sim 1.5\times10^{-5}$ $M_\odot yr^{-1}$) has multiple, incomplete, concentric shells in its envelope *(98)*. These shells were attributed to mass-loss modulations with a time scale of ~200–800 year. A spiral structure in the form of a one-armed Archimedes spiral has been observed in the wind of the carbon-rich star AFGL 3068 ($\dot{M}\sim 4.2\times10^{-5}$ $M_\odot yr^{-1}$) *(16)*. The spiral was interpreted as being caused by binary interaction in which the mass losing AGB star undergoes reflex motion which shapes its wind with a spiral pattern *(92)*. Twelve AGB stellar winds have previously been observed at sufficiently high spatial resolution (see Table S3 and Sect. S9). The observations enumerated in Table S3 are consistent with our classification scheme: the effect of EDE/disk-like structures and bipolar outflows are more readily recognized in lower mass-loss rate targets, while high mass-loss rate AGB stars have spiral-like arcs.

Observations of carbon-rich AGB stars only show spiral-like structures; each of them falling into the category of high mass-loss rate stars ($\dot{M}>1\times10^{-6}$ $M_\odot yr^{-1}$; see Table S3). An EDE has been detected in a few carbon-rich sources, including IRC +10216 *(111)* and V Hya *(112)*. Nevertheless, we regard our classification scheme as primarily applicable to oxygen-rich AGB stars.

**The case of low wind acceleration:** Our sample is composed of oxygen-rich AGB stars. Observational studies have shown that in oxygen-rich AGB stars the wind acceleration might be much lower than assumed, with terminal wind velocities being reached at 50 – 200 stellar radii *(89, 113–115)*. Using Eq. (S1), this is quantified by $\beta$ greater than 1, while for carbon-rich stars $\beta$ is around 0.5 *(106)*, due to the opaque carbon dust grains facilitating photon momentum transfer. For some oxygen-rich AGB stars, $\beta \geq 5$ has been deduced *(89, 114)*, but the relationship between $\beta$ and the mass-loss rate is unknown.

This low wind acceleration, $\mathbf{v}/dt = \mathbf{a}$, can affect the wind morphology, because the wind velocity will be lower than or similar to the orbital velocity ($v_{\rm orb}(r)$) in a geometrically extended region (see Fig. S5, compare red line with the two star-symbols at a distance $r = a$), thereby strengthening the interaction between the wind and the companion star *(116)*. If the ratio of the wind over the orbital velocity at the orbital distance of the companion is <1, any binary-induced morphological feature will be more complex compared to the case of a high wind acceleration with $v_{\rm wind} \gg v_{\rm orb}$. For example, if an Archimedean spiral forms for a high wind acceleration scenario, the same binary setup applied to a low wind acceleration scenario can yield a 'broken/complex' spiral structure that will potentially only attain a self-similar regular structure far out in the wind. The 'rose-like' morphology of R Aql (Figure 1) appears similar to this [ *(116)*, their figure C1]. In addition, a lower wind velocity increases the capture radius (Eq. S3). This induces a difference between oxygen-rich and carbon-rich AGB stars and implies that (compact) EDE in the orbital plane of the binary system can be more pronounced for oxygen-rich stars, provided the filling factor of the dust condensation radius by the Roche lobe radius is large enough. The velocity vector fields in these EDEs strongly depart from a radial pattern. Moreover, the more progressive wind acceleration around oxygen-rich stars (i.e. with higher $\beta$-values) will also smear out the (spiral) shock compared to the wind around a carbon-rich star. Because the features are more dilute around an oxygen-rich star, we expect they will be more difficult to identify [e.g. *(116)*, their figure 5] .

Table S3: **Overview of the wind characteristics of previously-observed AGB sources.** Columns 1–5 contain the source name, spectral type ('M' indicating an O-rich AGB star, and 'C' a carbon-rich AGB star), pulsation variability type, mass-loss rate, and wind velocity. In Column 6, a description of the wind morphology as published in the literature is given. Column 7 indicates the ATOMIUM classification and references are given in Column 8.

| Name | Spectral type | Variability type | Mass loss ($M_\odot yr^{-1}$) | $v_{\rm wind}$ km s$^{-1}$ | Description | ATOMIUM classification | References |
|---|---|---|---|---|---|---|---|
| Mira | M | Mira | $1 \times 10^{-7}$ | 5 | complex +flat broken spiral in orbital plane + accretion disk | (Class 1)† | *(99, 100)* |
| R Dor | M | SRb | $1 \times 10^{-7}$ | 5.5 (+23)⋆ | differentially rotating disk | Class 2 | *(79)* |
| EP Aqr | M | SRb | $1.6 \times 10^{-7}$ | 11 | biconical + narrow spiral in differentially rotation disk | Class 2 | *(101, 102)* |
| R Aqr | M | Mira | $2 \times 10^{-7}$ | 14 | EDE+two plumes perpendicular to EDE | Class 1 | *(103, 104)* |
| $L_2$ Pup | M | SRb | $5 \times 10^{-7}$ | 3 (+17)⋆ | (sub-)Keplerian disk + bipolar | Class 2 | *(17)* |
| OH 26.5+0.6 | M | Mira | $3.5 \times 10^{-6}$ | 16 | spiral+EDE | Class 3 | *(55)* |
| CIT 6 | C | SRa | $8 \times 10^{-6}$ | 21 | spiral | Class 3 | *(105)* |
| IRC +10216 | C | LPV | $1.5 \times 10^{-5}$ | 14.5 | spiral | Class 3 | *(106, 107)* |
| OH 30.1−0.7 | M | Mira | $1.7 \times 10^{-5}$ | 23 | spiral+EDE | Class 3 | *(55)* |
| R Scl | C | SRb | $2 \times 10^{-5}$ | 14.3 | spiral | Class 3 | *(108, 109)* |
| AFGL 3068 | C | Mira | $4.2 \times 10^{-5}$ | 14 | (bifurcated) spiral | Class 3 | *(16, 110)* |

⋆ if two values are given, this last value between parenthesis reflects the Keplerian velocity at the inner boundary;
† ALMA observations indicate a Class 1 object, but the CO channel map exhibits a complex morphology due to WRLOF and a low wind velocity (see Sect. S9). We therefore exclude Mira ($o$ Cet) in the analysis performed in Sect. S3.3.

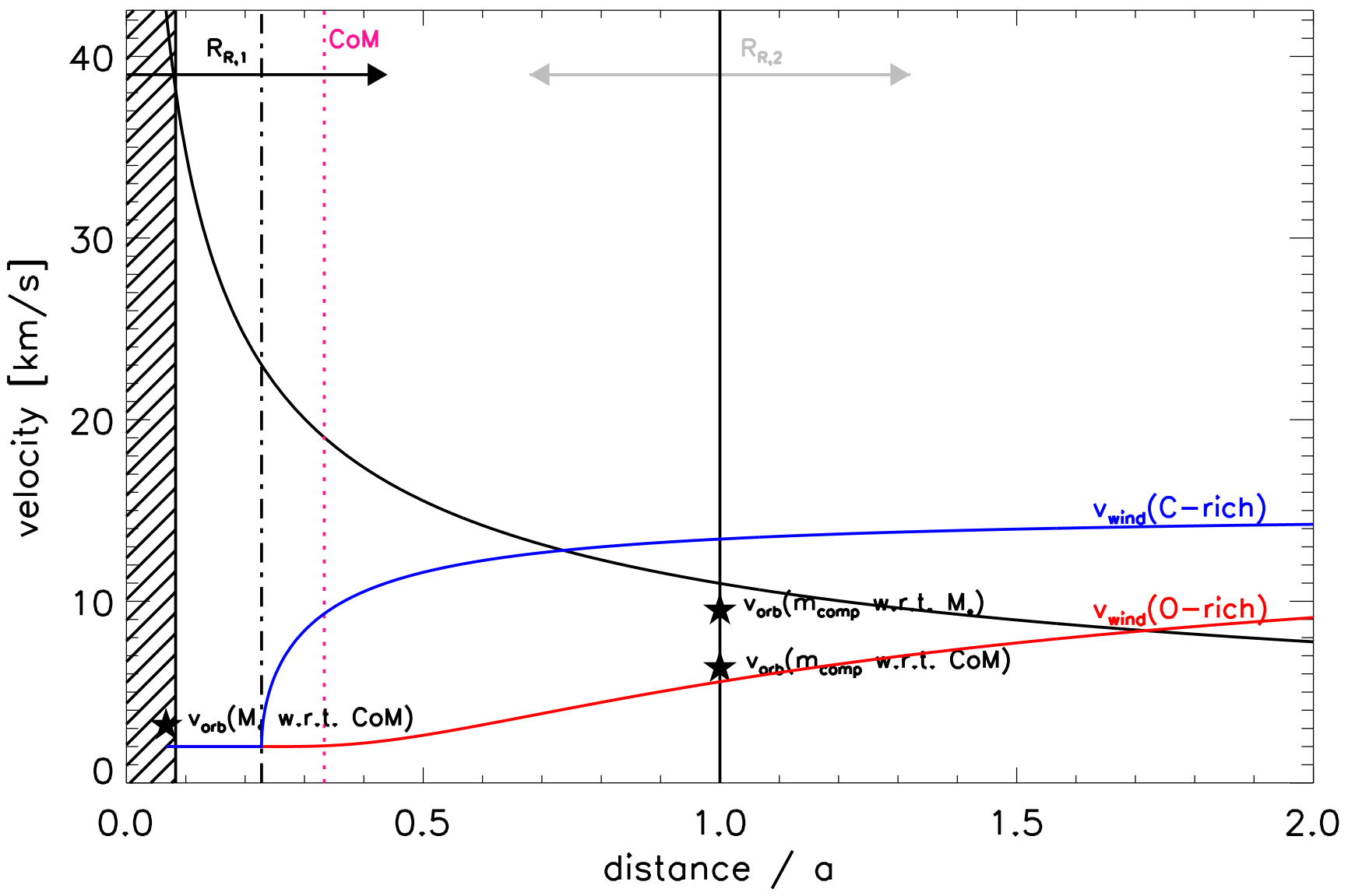

Fig. S5: **Velocity fields of importance for the binary interaction**. Given are the escape velocity (in black), the wind velocity of an oxygen-rich star assuming $\beta = 5$ (red), and of a carbon-rich star assuming $\beta = 0.5$ (blue) for a wind acceleration starting at 2.75 au. The central target is assumed to have a mass of 1.02 $M_\odot$, its Roche-lobe $R_{R,1}$ is indicated with the black arrow and its stellar size of 267 $R_\odot$ with the shaded region. For a companion mass of 0.51 $M_\odot$ at an orbital separation, $a$, of 15 au, the common center-of-mass (CoM) is at 5 au from the primary star (indicated with the dotted magenta line); its Roche-lobe $R_{R,2}$ is indicated with the gray arrow. The star-symbols indicate the orbital speeds of both stars with respect to the CoM, and of the companion with respect to the primary. The $x$-axis is in units of the binary separation $a$. In this example, the two Roche lobes are detached, but WRLOF can occur. In the particular case of a low velocity wind (red curve), the material has not yet reached the escape velocity at a distance $r = a$ and a strong interaction with the companion's gravity field can occur, resulting in complex wind morphologies *(116)*.

Hence, carbon-rich stars and oxygen-rich higher mass-loss rate AGBs with higher wind acceleration will have a lower $v_{\rm orb}(r)/v_{\rm wind}(r)$ ratio, resulting in a smaller geometrical region in which the dominant wind streaming lines are non-radial. For those systems, we expect the winds will more readily exhibit (Archimedean) spiral structures that expand radially in a self-similar way. An EDE can also form (depending on the system's properties), but the spiral will dominate the wind's appearance. An oxygen-rich AGB star with a low wind acceleration will be more prone to complex wind morphologies, such as 'rose'-like morphologies. These systems have a higher chance of forming a pronounced EDE with a large equator-to-pole density contrast, and hence the formation of bipolar cavities. Spiral structures could form as well, but a low wind velocity implies the spiral features be more broken/complex in structure and more dilute and hence more difficult to detect.

### S4.2 Analytical estimates of binary-induced AGB wind morphology

We consider a simplified binary set-up to produce an analytical relation between binary parameters and the departure from radial motion in the AGB stellar wind. We begin by defining the region where the gravitational field of the companion deflects the flow from its trajectory. The radius of the sphere of gravitational influence of the companion is denoted as $R_H$.

**Wind accretion scenario:** When the wind speed at the orbital distance $a$, $v_{\rm wind}(a)$, is large compared to the orbital velocity, $v_{\rm orb}$ ($v_{\rm wind}(a) \gg v_{\rm orb}$), $R_H$ can be approximated by the capture radius (sometimes referred to as gravitational or accretion radius):

$$R_H \approx R_{\rm capt} = \frac{2Gm_{\rm comp}}{v_{\rm wind}^2}\,, \tag{S3}$$

with $m_{\rm comp}$ the mass of the companion. This describes the critical impact parameter below which a planar flow coming from infinity with a certain kinetic energy per unit mass (and hence certain wind speed) is likely to be captured *(117)*. It relies on the classical BHL formalism *(87, 88)*. To derive the capture radius, we compare the kinetic energy per unit mass ($v_{\rm wind}^2/2$) to the gravitational potential of the companion ($G\,m_{\rm comp}/r$) and find the value of $r$ where both are equal. For $r < R_{\rm capt}$ the mechanical energy is negative, while in the opposite situation it is positive. For decreasing wind speeds and larger companion masses, the effective cross section set by the accretion radius increases, thereby enhancing the impact of the companion on deflecting the flow from the primary AGB star.

**Roche lobe overflow scenario:** The wind-accretion scenario is not valid for low wind speeds compared to the orbital speed ($v_{\rm wind}(a) \ll v_{\rm orb}$). In that case, the accretion radius can be larger than the Roche Lobe radius of the companion, $R_{R,2}$. We approximate the region of gravitational influence of the companion as its Roche lobe *(96)* following Eggleton's formula *(118)*:

$$R_H \approx R_{R,2} = \varepsilon(q)\, a = \frac{0.49q^{2/3}}{0.6q^{2/3} + \ln(1+q^{1/3})}\, a\,, \tag{S4}$$

where $q$ is the ratio of the mass of the companion star to the mass of the AGB star, $m_{\rm comp}/M_\star$. The function $\varepsilon$ only depends on the mass ratio. Accounting for the eccentricity of the orbit on the assumption of the mass of the companion being much smaller than the primary mass, the radius can be approximated by the Hill radius of the companion *(119)*

$$R_H \approx R_{\rm HL} = a(1-e)\left(\frac{m_{\rm comp}}{3\,M_\star}\right)^{1/3}\,, \tag{S5}$$

where $e$ is the orbital eccentricity and $a\,(1-e)$ is the pericenter distance. This approach is valid when the amount of kinetic energy per unit mass given to the flow by mechanisms other than gravity (for example radiative pressure on dust grains or resonant line absorption of UV photons) is negligible compared with the amplitude of the Roche potential ($\sim 2G(M_\star+m_{\rm comp})/a$).

The RLOF and wind accretion scenario are the two extreme situations. In between, WRLOF occurs *(94, 96)*, where both the mass ratio (as in the RLOF case) and the wind speed (as in the wind accretion case) are involved (see Sect. S4.1).

We introduce a dimensionless parameter $Q^p$ to predict the morphology of the outflow due to companion motion, defined as

$$Q^p \approx \frac{p_{\rm comp}}{p_{\rm wind}} = \frac{m_{\rm comp}\, v_{\rm orb}}{m_{\rm wind}\, v_{\rm wind}}\,, \tag{S6}$$

in which $p_{\rm wind}$ is the radial momentum of wind material the companion encounters in one orbit, $p_{\rm comp}$ the tangential momentum of the companion, $m_{\rm wind}$ the wind mass the companion encounters in one orbit, $v_{\rm wind}$ the local wind speed of the material lost by the AGB star at the location of the companion ($v_{\rm wind}(r=a)$), $m_{\rm comp}$ the mass of the companion, and $v_{\rm orb}$ the orbital speed of the companion at an orbital distance $a$ from the primary given by $\sqrt{G(M_\star + m_{\rm comp})/a}$. Large values of $Q^p$ reflect a strong departure from radial motion, such as that caused by the formation of a dense, potentially rotating, EDE or a (Keplerian) disk-like structure. Intermediate values of $Q^p$ reflect a situation in which an EDE is formed with not too high a density contrast between the equator and poles. Little departure from the radial flow will result in small values of $Q^p$, in the case of a quasi unperturbed spherical outflow or when a spiral structure expands self-similarly. The parameter $Q^p$ is defined for $r \approx a$. The formation of a circumbinary or accretion disk — reflected by high values of $Q^p$ — does not imply that no spiral shock forms, but the expanding spiral will only dominate at larger distances from the mass-losing AGB star (see Sect. S4.1); vice versa if the prevailing wind geometry is dominated by a spiral this does not imply that no compact EDE can be formed (see also Sect. S4.1, *(55)*).

In one orbit, the companion encounters a wind mass

$$m_{\rm wind} \simeq f_w\, 2\,\pi R_H^2\, \pi\, a\, \rho_{\rm wind}(r=a) + \pi\, R_{\rm H}^2\, \rho_{\rm wind}(r=a)\, v_{\rm wind}(r=a) \frac{2\pi a}{v_{\rm orb}}\,, \tag{S7}$$

with $\rho_{\rm wind}$ the density of the wind defined by the equation of mass conservation

$$\rho_{\rm wind}(r=a) = \frac{\dot{M}}{4\pi a^2 v_{\rm wind}(r=a)}\,. \tag{S8}$$

The first term on the right in Eq. (S7) describes the mass initially present at $r = a$ from the ongoing AGB stellar wind with $f_w$ a fraction $< 1$, while the second reflects the additional mass which can be captured by the companion. For the wind accretion scenario ($v_{\rm wind} \gg v_{\rm orb}$), Eq. (S7) reduces to

$$m_{\rm wind} \simeq 2\,\pi^2\, a\, R^2_{\rm capt}\, \rho_{\rm wind}(r=a)\, \frac{v_{\rm wind}(r=a)}{v_{\rm orb}}\,, \tag{S9}$$

while in the case that $v_{\rm wind} \ll v_{\rm orb}$, we can write that

$$m_{\rm wind} \simeq f_w\, 2\,\pi^2\, a\, R^2_{R,2}\, \rho_{\rm wind}(r=a)\,. \tag{S10}$$

For a low wind velocity ($v_{\rm wind} \ll v_{\rm orb}$) and using Eq. (S5), we can write $Q^p$ as

$$Q^p = 8.32\times 10^6 \frac{1}{(1-e)^2}\,\frac{1}{f_w}\left(\frac{m_{\rm comp}}{\rm M_\odot}\right)^{1/3}\left(\frac{M_\star}{\rm M_\odot}\right)^{7/6}\left(\frac{a}{1\,{\rm au}}\right)^{-3/2}\left(\frac{\dot{M}}{10^{-6}\,{\rm M_\odot/yr}}\right)^{-1}. \tag{S11}$$

Eq. (S11) shows that high values for $Q^p$ will be obtained for a high mass of the companion $m_{\rm comp}$, a small orbital separation $a$, and a low wind mass-loss rate $\dot{M}$. The companion mass only enters as $m^{1/3}_{\rm comp}$, so $Q^p$ is not strongly dependent on the companion mass. This implies that if low mass stellar companions are effective, massive planets might also be effective. Using the capture radius (Eq. (S3)) to define $R_H$, we obtain

$$Q^p = 0.33\times10^6\left(\frac{m_{\rm comp}}{\rm M_\odot}\right)\left(1+\frac{1}{q}\right)^2\left(\frac{a}{1\,{\rm au}}\right)^{-1}\left(\frac{\dot{M}}{10^{-6}\,{\rm M_\odot/yr}}\right)^{-1}\left(\frac{v_{\rm wind}}{10\,{\rm km\,s^{-1}}}\right)\left(\frac{v_{\rm wind}}{v_{\rm orb}}\right)^2. \tag{S12}$$

Eq. (S12) shows similar dependence on binary parameters to Eq. (S11), but there is an explicit dependence on the orbital and wind speed. However, a lower wind velocity implies a smaller value for $Q^p$, which would indicate a smaller departure from radial motion. The smaller value of $Q^p$ arises because lower wind velocities imply a larger capture radius therefore $m_{\rm wind} \propto v^{-4}_{\rm wind}$. Accounting for this difference in total intercepted mass, we estimate that more particles per unit mass will be deflected by the gravitational field of the companion in the case of a lower wind speed (see also Sect. S.4.4).

For convenience, we set

$$Q^1 = 10^{-6}\, Q^p \tag{S13}$$

to give a quantity close to unity. For the case of Jupiter orbiting around the Sun, and on the assumption the wind mass-loss rate of the Sun in its AGB stage will be around $5\times10^{-7}\,{\rm M_\odot yr^{-1}}$

— i.e., similar to what has been deduced for $L_2$ Pup, which has a near solar main-sequence mass *(17)* — $Q^1$ is around 0.35 using Eq. (S11) for $f_w = 1$. Accounting for the fact that the Sun will have lost part of its mass by the time it enters the regularly pulsating Mira AGB stage, $Q^1$ will be even lower (around 0.2) suggesting that the presence of Jupiter will leave the solar wind morphology nearly unperturbed and only a very weak BHL type spiral will form.

In summary, these simplified analytical estimates can explain the main morphological classification we have identified in the ATOMIUM sample (Sect. S3.2): a decrease in mass-loss rate increases $Q^p$ and reflects the conditions in which a (compact) EDE can be formed with pronounced contrast in density between the poles and the equator. The EDE can harbor a circumbinary disk or accretion disk around the companion (see Sect. S4.1). Higher mass-loss rates yield low values of $Q^p$ with minimal departure from radial motion. A spiral structure that is expanding in a self-similar way would fall into this latter category. The observed correlation between the mass loss and the morphology is caused by the ability of orbiting companions to inject angular momentum into the (mostly) spherical mass loss from the star. If the mass loss is too strong or fast, or if the companion is too far away (and hence has a low orbital velocity), the material in the circumstellar envelope is 'overwhelmed' by the radially streaming AGB wind and cannot shape the material into an EDE.

## S5 (sub)-Stellar binary population statistics

Here we summarize the occurrence of (sub-)stellar companions around stars with an initial mass between 0.8–8 $\mathrm{M}_\odot$. We differentiate between stars with initial mass less than $\lesssim$1.5 $\mathrm{M}_\odot$, i.e progenitor spectral types F, G, and K; and those with higher initial mass that represent A and B spectral types on the main sequence.

**Orbital period:** We first assess the minimum and maximum orbital radius $a$ and orbital period $P_{\mathrm{orb}}$ where companions can influence the shaping of the wind. A conservative upper limit is $a < 500\,\mathrm{au}$ *(120)*, or $\log P_{\mathrm{orb}}(\mathrm{days}) < 6.1 - 6.5$ ($\sim$4–10 kyr for masses $\sim$1–8 $\mathrm{M}_\odot$, applying Kepler's third law). This is similar to the maximum recoverable scale of our medium resolution ALMA data ($\sim 2''$) of $\sim$600 au at 300 pc; and is also the maximum orbital period where an orbiting body is likely to create arc-like structures. A conservative lower boundary would be an orbit on the radius of the AGB star ($\gtrsim 1.5 - 3\,\mathrm{au}$), and hence $\log P_{\mathrm{orb}}(\mathrm{days}) \gtrsim 2.7$ (or $\sim$2 yr) — ignoring the fact that orbits close to the star will have a tendency to inspiral during the upper RGB and AGB phases *(21, 121, 122)*. Because this will affect stars more than planets (see Sect. S6), we assume $\log P_{\mathrm{orb}}(\mathrm{days}) > 3$ for stars and $\log P_{\mathrm{orb}}(\mathrm{days}) > 2.7$ for planets.

**Initial mass:** Next, we investigate the initial masses of the ATOMIUM AGB stars, because i) the binary fraction is strongly mass-dependent *(23)* and ii) we need to determine how representative the ATOMIUM sample is of AGB stars in general. A first indication can be retrieved from the luminosity distribution of the ATOMIUM stars with $3.6 \lesssim \log L(\mathrm{L}_\odot) \lesssim 4.1$. Using relations between the pulsation period and the luminosity, the implied progenitor stars are at least $\sim$1.1 $M_\odot$ for a star to reach $\log L = 3.6$ during the interpulse period *(22, 123)*. However, stars whose mass is about a solar mass, will only seldom have a luminosity above that threshold, implying stars should have an initial mass $\gtrsim$1.5 $M_\odot$ so that $\log L \geq 3.6$ for an extended period.

For six ATOMIUM stars we can determine the initial mass by using the $^{17}$O/$^{18}$O isotopic ratio which stellar evolution models predict is a sensitive function of initial mass *(124, 125)*. The derived initial mass is $\sim$1.7 $M_\odot$ for R Aql ($^{17}$O/$^{18}$O = 1.77 *(125)*), $\sim$1.35 $M_\odot$ for R Hya ($^{17}$O/$^{18}$O = 0.54 *(125)*), $\sim$1.95 $M_\odot$ for U Her ($^{17}$O/$^{18}$O = 2.53 *(125)*), $\sim$1.57 $M_\odot$ for W Aql ($^{17}$O/$^{18}$O = 1.17, *(124)*), $\sim$1.06 $M_\odot$ for IRC +10011 ($^{17}$O/$^{18}$O = 0.26, *(124)*), and $\sim$2 $M_\odot$ for IRC $-$10529 ($^{17}$O/$^{18}$O = 3.1 (ALMA proposal ADS/JAO.ALMA2019.1.00187.S, PI T. Danilovich.). Most ATOMIUM stars are fundamental mode pulsators, however U Del is a first-overtone pulsator and long secondary period pulsator *(41)*, suggesting it is a higher-mass star with $M \gtrsim 4\,M_\odot$ *(123)*. Therefore five out of the seven stars with mass measurements have an estimated initial mass greater than 1.5 $M_\odot$.

There is a selection bias towards stars with high initial mass. One of our selection criteria was the mass-loss rate, which we required be greater than $\sim 1\times10^{-7}\,M_\odot yr^{-1}$ to improve the signal-to-noise ratio. However, a star with an initial mass of $\sim$1 $M_\odot$ will only have a very short period in its AGB phase during which the mass-loss rate is greater than $10^{-7}\,M_\odot yr^{-1}$ before it transforms into a post-AGB star and PN. More massive stars have a longer period during which the luminosity is above $\log L = 3.6$ and the mass-loss rate is $\geq 10^{-7}\,M_\odot yr^{-1}$. For an initial mass of $2\,M_\odot$, the luminosity exceeds $\log L = 3.6$ for 1–2 Myr and the mass-loss rates exceeds $10^{-7}\,M_\odot yr^{-1}$ for just over 1 Myr, while the C/O ratio only exceeds unity for $\sim$300 000 yr [ *(126)*, their figure 5]. Using evolutionary tracks *(22)*, the criteria of $\log L > 3.6$, $T_{\rm eff} < 4000$ K, and $\dot{M} > 1 \times 10^{-7}\,M_\odot yr^{-1}$ yield a 68% confidence interval that stars with mass $1.51 - 2.81\,M_\odot$ were selected, and a 95% confidence interval of $1.28 - 4.67\,M_\odot$. Consequently, we expect the majority of the ATOMIUM AGB stars to be more massive than $\sim$1.5 $M_\odot$.

This bias in the ATOMIUM AGB sample selection indicates our targets have a higher initial mass than typical AGB stars ($\sim$1.5 $M_\odot$ for bulge Miras of intermediate age (1–3 Gyr) *(127)*).

**Main-sequence stellar multiplicity factor:** We determined the main-sequence multiplicity fraction in the $2.7(3.0) \lesssim \log P_{\rm orb}({\rm days}) \lesssim 6.5$ period range (see Table S4). The stellar binary fraction of F/G/K stars (0.8$\lesssim$M$\lesssim$1.4 $M_\odot$) is well studied *(23)*, but the lack of spectral features at optical wavelengths makes the search for radial velocity features from companions to A stars ($\sim$1.4–3 $M_\odot$) difficult. The fraction of F/G/K stars that are part of binaries with mass ratio $q = m_{\rm comp}/M_\star > 0.1$ is $\sim$50%, while the fraction increases to $\sim$85% for A/late-B stars *(23)*. We multiply both fractions by 1.1, because about 10% have white dwarfs companions [ *(23)*, their figure 29]. Stellar companion binaries with $\log P_{\rm orb}({\rm days}) \leq 3$ will tighten their orbit, evolve to a common envelope situation and merge; and binaries with $\log P_{\rm orb}({\rm days}) > 6.1 - 6.5$ will not influence the wind shaping, so the multiplicity of F/G/K stars reduces to $\sim$30–37% and A/late-B stars reduces to $\sim$52–60% [ *(23)*, their figure 37] In addition, about 10% of solar-type main-sequence primaries are in triple/quadruple systems, while the triple and quadruple star

Table S4: **Main-sequence (sub-)stellar multiplicity fraction**. The first and second columns give the main-sequence initial mass and related spectral type on the main-sequence. Columns 3–6 list the (sub)-stellar multiplicity fraction for companions with $\log P_{\rm orb}(\rm days) < 6.5$ and $\log P_{\rm orb}(\rm days) > 2.7$ for planets (or $> 3$ for stars).

| | | Stellar companions | Brown dwarfs | Planets $M > 5\,M_{\rm Jup}$ $a = 10 - 100\,$au | Planets $M > M_{\rm Jup}$ $a = 2 - 10\,$au |
|---|---|---|---|---|---|
| $M_{\rm ini} < 1.5\,{\rm M}_\odot$ | FGK | ∼30–37% | ∼0.8% | ∼9% | ∼7% |
| $M_{\rm ini} > 1.5\,{\rm M}_\odot$ | AB | ∼52–60% | ∼0.8% | ∼40% | ∼7% |

fraction for systems with primary mass of about 8 $\rm M_\odot$ is around 37% *(23)*.

**Main sequence sub-stellar multiplicity factor:** For planets and brown dwarfs statistics, we used the results of the Gemini Planet Imager survey (GPIES) *(128)* and California-Kepler survey *(24, 129)*. First order estimates *(130)* indicate a conservative minimum mass (at a given radius) for which planets can influence the shaping of the wind is a few Jupiter masses ($M_{\rm Jup}$), a number confirmed by hydrodynamical simulations. The GPIES results show that ∼0.8% of stars host a brown dwarf ($13 < M < 80\,M_{\rm Jup}$) between 10 – 100 au, and ∼9% host a planet (5–13 $M_{\rm Jup}$) in the same orbital distance range. However, these results are strongly mass-dependent with ∼40% of stars with mass above ∼1.5 $\rm M_\odot$ having a planet *(129)*. Similarly, the quoted percentage for very low-mass stellar companions around the higher mass A/B stars is probably a conservative lower limit. The frequency of 0.1–19 $M_{\rm Jup}$ planets in the range of 2–10 au is ∼15%, with about half of these less than a Jupiter mass *(129)*.

**AGB (sub-)stellar multiplicity factor:** We next assess how this (sub-)stellar binary fraction evolves from the main-sequence to the AGB evolutionary phase. Tidal inspiral and stellar mass loss will reduce the binary fraction by dragging nearest companions inwards and spiraling out the farthest ones. On the assumption our target stars are a part of their way through the AGB evolution (because they are strongly mass-losing) their masses are probably half their birth masses, hence the orbits of the majority of companions to the ATOMIUM stars will have expanded by a factor of ∼2 (see Sect. S6.1). Taking a range between $3 < a < 5\,00$ au (2.22 dex), losing ∼0.3 dex in orbital separation results in a ∼14% reduction in parameter space and ∼12% reduction in multiplicity fraction for the stellar and brown dwarf companions listed in Table S4. Orbital expansion induces Kozai-Lidov perturbations in multiple planet systems, which will tighten the orbit and cause inspiral or ejection of many companions. This process also occurs for some systems on the main sequence, and (rare) chance stellar encounters can also cause ejections. Observations of free-floating planets *(131)* show this is likely to cause a few percent reduction of the planet fraction in Table S4. Depending on the amount of photo-dissociating UV irradiation from the host star, a fraction of the Jupiter mass gas giants can evaporate before stars reach the AGB phase. This depletion factor is not well known, but is thought to be the cause for the absence of Neptune-mass planets at very close orbital radii *(132)*. We conclude that as stars evolve toward and on the AGB, the main-sequence (sub-)stellar binary fraction in Table S4 will decrease by roughly 10–20%.

Accounting for all these effects, it is evident that the accumulated multiplicity fraction increases by a factor 2–3 over the range of 0.8–8 $M_\odot$ with a (sub-)stellar multiplicity fraction for higher mass AGB stars ($M \gtrsim 1.5\,M_\odot$) of $\gtrsim 1$. This dependence of the multiplicity fraction on initial mass is also evident in the Gaia DR2 data *(133)*. As a result, i) the ATOMIUM data complemented with the additional ALMA data presented in Sect. S4.1 (Table S3) and Sect. S9, ii) the outcome of theoretical models for binary systems summarized in Sect. S4.1 and Sect. S8, and iii) the population statistics here described all indicate that binary interaction is the dominant wind shaping mechanism for these AGB stars. Our conclusion applies to all AGB stars for which observations indicate a mass-loss rate greater than $\sim 1\times10^{-7}\,M_\odot yr^{-1}$ — i.e., those with a mass-loss rate exceeding the nuclear burning rate *(4)* so the mass loss determines the further stellar evolution. A high fraction of these targets will have an initial mass greater than $\sim 1.5\,M_\odot$ (see above) for which (sub-)stellar population statistics indicate the multiplicity fraction is $\geq 1$.

# S6 Predicting the evolution of wind morphology

## S6.1 Change of orbital separation throughout the AGB evolution

We predict how the prevailing wind morphology might change throughout the AGB evolution. For a star of size 267 $R_\odot$ (see also Fig. S5), we follow its evolution on the AGB track using the relation between the pulsation period and the luminosity, here expressed in terms of the absolute bolometric magnitude *(134)*

$$M_{\rm bol} = -3.00 \log P + 2.85 \tag{S14}$$

with the pulsation period $P$ in days. The pulsation period is derived using the period-mass-radius relation *(34)*

$$\log P({\rm days}) = -2.07 + 1.94 \log R_\star/{\rm R}_\odot - 0.9 \log M_\star/{\rm M}_\odot\,, \tag{S15}$$

and the mass-loss rate for pulsation periods $P > 300$ days is given by *(34)*

$$\log \dot{M}({\rm M}_\odot\,{\rm yr}^{-1}) = -11.4 + 0.0123\,P({\rm days})\,, \tag{S16}$$

but cannot exceed the single-scattering limit *(34, 55)*. The mass-loss rate is assumed to be $2 \times 10^{-8}\,M_\odot yr^{-1}$ for thermally pulsating AGB stars with lower pulsation periods *(4)*.

For an AGB star of initial mass 1.5 $M_\odot$, the initial luminosity is $\sim$5 400 $L_\odot$, $P = 300$ days, and $\dot{M} = 2 \times 10^{-8}\,M_\odot yr^{-1}$. For various companion masses (1.2, 0.6, 0.3, 0.01 $M_\odot$) we compute the change in orbital separation due to angular-momentum loss [ *(135)*, their Eqs. (6)–(13)]. These analytical relations were derived for binary population synthesis codes, but cannot capture all details of the complex wind mass-transfer physics. The rate of orbital change, $\dot{a}/a$, is dependent on the current mass-loss rate, mass of the primary, mass-ratio, capture rate by the companion, and ratio of the orbital speed to the wind speed. The calculation ends when the AGB mass is less than $0.6\,\mathrm{M}_{\odot}$ *(136)* — by which time the AGB star has reached a luminosity of $14\,500\,\mathrm{L}_{\odot}$, a pulsation period of 685 days, and a mass-loss rate of $2 \times 10^{-5}\,\mathrm{M}_{\odot}\mathrm{yr}^{-1}$ — or when both stars merge. Fig. S6 shows the result for four initial orbital separations ($a = 2$, 4, 10, 25 au). For initial orbital separations greater than 25 au, the results are similar to the one with $a = 25$ au. If the companion has a mass $\sim 10\,M_{\mathrm{Jup}}$, the orbit always widens. The more massive the companion and the smaller the initial separation, the greater the chance for the orbit to shrink. For very small initial separations ($\lesssim$1.5 au) this eventually can lead to a merging of both stars. If the initial orbital separation is $\gtrsim$25 au, the orbital separation always increases independent of the mass of the companion. However, the maximum factors of the widening (1.12) and shrinking (0.76) are quite modest.

A more pronounced change occurs for primary AGB stars with higher initial mass, and hence more extreme values of the mass ratio (see Fig. S7). Given Eqs. (S14)–(S16), the luminosity, pulsation period, and mass-loss rate at the start of the calculations are $L \sim 2\,500\,\mathrm{L}_{\odot}$, $P = 161$ days, and $\dot{M} = 2 \times 10^{-8}\,\mathrm{M}_{\odot}\mathrm{yr}^{-1}$, respectively. In general, the more massive the AGB star the more extreme the orbit's expansion (if occurring) that can be attained with maximum derived widening factors of $0.0179 + 0.733\,M_{\star}$. A more massive AGB star also implies a large initial orbital separation for which all companions (independent of mass ratio) increase their separation during the AGB evolution. Carbon-rich AGB stars also develop larger orbital separations compared with oxygen-rich AGB stars during the AGB evolution.

The frequency distribution for $q > 0.1$ is roughly flat in $\log P_{\mathrm{orb}}$-space for main-sequence stars with mass between 2–5 $\mathrm{M}_{\odot}$ and $\log P_{\mathrm{orb}} = 3$–6.5 [ *(23)*, their figure 37]. This implies the majority of companions reside at larger orbital separation ($a \gtrsim 20$ au). Not accounting for any change in orbital separation between the main-sequence and start of the AGB phase *(1)*, Figs. S6–S7 show that at an initial separation of $\gtrsim 20$ au the majority of the stellar and substellar companions will widen their orbit during the AGB phase as stellar mass reduces due to mass loss, eventually reaching the BHL wind-accretion regime (see Fig. 2). Our results imply that early-type oxygen-rich AGB stars with a low mass-loss rate are more likely to host planet-mass companions.

### S6.2 Morphological characteristics on and beyond the AGB phase

Compact EDE/disk-like structures have a higher tendency to be formed around early-type oxygen-rich AGB stars with a slow wind acceleration and spiral structures are readily formed in late-type AGB stars with high mass-loss rate (Figure 2). Our results show that the aspherical geometries arise during the AGB phase when mass-loss rates exceed $\gtrsim 1\times10^{-7}\,\mathrm{M}_{\odot}\mathrm{yr}^{-1}$. This does not preclude an earlier timing of the shaping event, but this cannot be determined with the current data. The prevailing AGB wind morphology and companion properties specify the initial aspherical conditions when transitioning into the post-AGB and PN phase. Our model has implications for the morphology during the AGB, post-AGB (or pre-PN) and PN phases.

Detached shells are detected around carbon-rich stars *(137)*. These shells are thought to be caused by a thermal pulse yielding a peak in luminosity and hence in mass-loss rate $\dot{M}(t)$ or density. The chance for this initially spherical shell to be deformed is smaller for carbon-rich binary stars than for oxygen-rich ones because the wind streaming lines are more radial (see Sect. S4.1). These detached shells might be the precursors of ring-like features detected in carbon-rich post-AGB stars, such as IRAS 19700+3457 *(138)*.

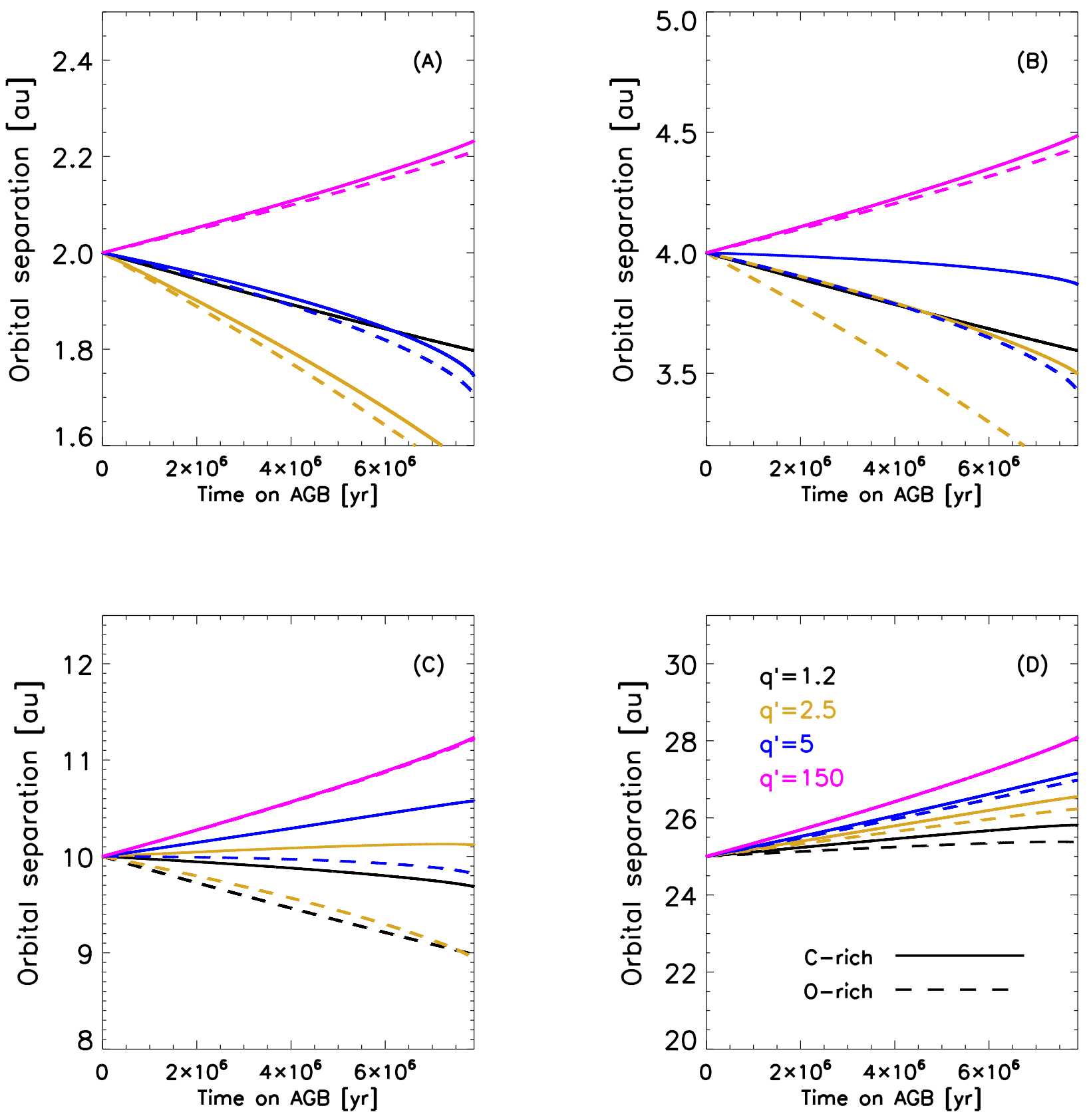

Fig. S6: **Evolution of the orbital separation in function of time for a primary AGB star of mass 1.5 $M_\odot$.** The $x$-axis represents the time on the AGB, with the zero point defined as when the luminosity rises above 5 400 $L_\odot$. The mass of the companion is 1.2 $M_\odot$ ($q' = 1/q = 1.25$ in black), 0.6 $M_\odot$ (gold), 0.3 $M_\odot$ (blue), or 0.01 $M_\odot$ (magenta). The initial orbital separation is Fig. S6A: 2 au, Fig. S6B: 4 au, Fig. S6C: 10 au, or Fig. S6D: 25 au. Simulations for a C-rich AGB star with a wind acceleration given by a $\beta$-profile with $\beta$=0.5 are shown with full line, while the dashed line shows an O-rich AGB star with $\beta$=5. For $q' = 1.25$, the evolution of the orbital separation for situations with a low and high $\beta$-value coincide.

Our proposed scenario explains the formation of a silicate-rich EDE around carbon-rich stars (so-called silicate carbon stars; *(27)*). During the early AGB phase, when the mass-loss rate was still modest and the C/O ratio lower than 1, a stable disk-like structure can be formed which favours the formation of crystalline silicate dust *(55)*. Later, as mass-loss rate and C/O ratio increase due to dredge-ups, the star will gradually become carbon-rich implying a velocity vector field which is predominantly radial. Moreover, our scenario is consistent with the occurrence of mixed chemistry in post-AGB stars and galactic bulge PNe *(28)*. Carbon dredge-up cannot explain the mixed chemistry because the bulge stars are of too low mass. Ultraviolet irradiation of the dense torus (EDE in our scenario) induces the formation of hydrocarbon chains *(28)*. In addition, our scenario predicts that disks are predominantly found around oxygen-rich post-AGB stars. This is compatible with disk detections around post-AGB stars *(139)*.

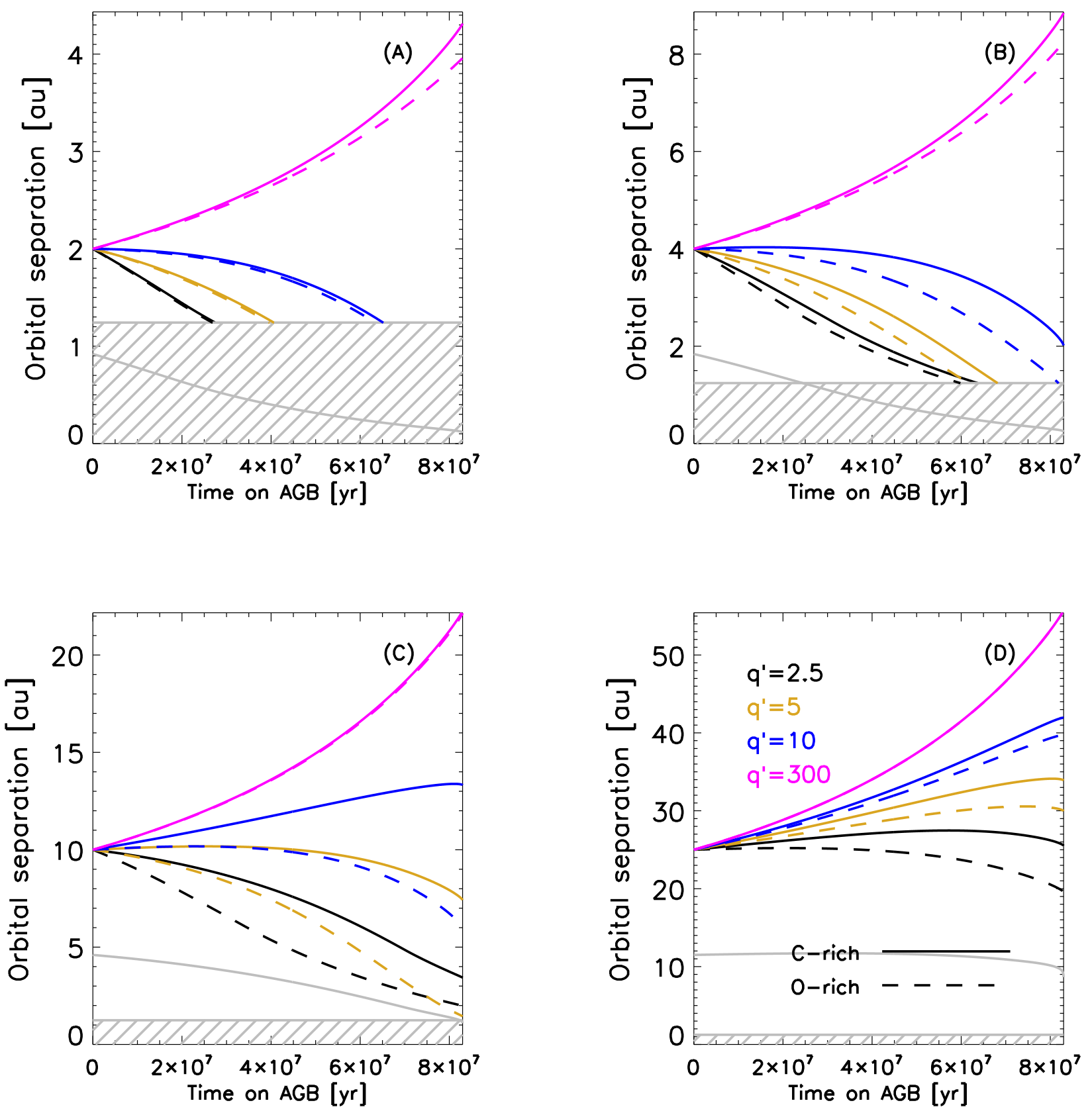

Fig. S7: **Same as Fig. S6, but for a primary AGB star of mass 3 $M_{\odot}$.** The zero point for time on the AGB is when the luminosity rises above 2 500 $L_{\odot}$. The hatched region indicates the stellar radius, the gray line the evolution of the Roche lobe radius of the AGB star for $m_{\rm comp} = 0.01\,M_{\odot}$.

The impact of the progenitor mass on the orbital decay or widening has implications for the morphological classification of post-AGB stars, which are divided in two classes: i) the 'sole' nebulae, for which the emission is star-dominated with faint extended nebulosity, and ii) the 'duplex' nebulae, with dust-dominated emission and a faint or completely obscured central star *(29)*. 'Duplex' sources tend to have a higher equator-to-pole density ratio, attributed to a more intense axisymmetric (equatorial) superwind mass loss at the end of the AGB, and hence formation of an EDE, and have been suggested to have higher-mass progenitors *(140)*. It follows from the calculations in Sect. S6.1 that the more massive the AGB progenitor star, the more likely the orbit of a stellar companion will tighten during the AGB evolution favouring the formation of a compact EDE, thereby producing the post-AGB morphological dichotomy. This progenitor-mass dependent evolution of the orbit also suggests that bipolar PNe are the descendants of high mass stars, so 'duplex' sources might be the precursors of bipolar PNe. This prediction is consistent with observations that most bipolar PNe have a central star of high initial mass *(73, 141, 142)*. 'Sole' post-AGB stars have been suggested to be precursors of elliptical PNe *(29)*. The lower tendency of carbon-rich AGB stars to form EDEs implies a higher fraction carbon-rich elliptical PNe, as observed *(141)*.

The orbits of binary post-AGB stars are often non-circular. Over 70% of the post-AGB binaries have nonzero eccentricities as high as 0.3, despite Roche-lobe radii that are smaller than the AGB stellar radius, and tidal circularization which should have been strong when the primary was on the AGB *(30)*. Tidal coupling to the circumbinary disk present during the post-AGB phase has been proposed as a mechanism to increase the eccentricities *(143)*. However, this scenario has difficulties explaining the eccentricities of post-AGB stars, and requires massive ($\gtrsim 10^{-2}\,M_\odot$) long-lived ($\gtrsim 10^5$ years) circumbinary disks which do not accrete *(144)* — a nontrivial combination, especially in the latter aspect. Our results show that circumbinary disk-like structures can be present in the early AGB phase, thereby lengthening considerably the timescale during which eccentricity pumping can occur as a result of gravitational interaction with the disk. Using parameters resembling the well-studied $L_2$ Puppis system *(17)*, results in $e = 0.29$ [using equation 5 of *(144)*]. Bifurcations in spiral-like patterns suggest there is a companion residing at an eccentric orbit around the carbon-rich AGB star CIT 6 *(105)* and so the eccentricity might already be (highly) nonzero at the beginning of the post-AGB phase.

The progenitor mass of PNe is greater than $\sim$1–2 $M_\odot$ *(6)*, a population for which the (sub-)stellar multiplicity rate approaches 100% (Sect. S5). We expect round PNe to be rare which is consistent with the small fraction of $\lesssim$20% observed *(9, 10)*. Aspherical PNe will not only arise due to the action of short-period binaries within a common-envelope evolution (orbital period $P_{\rm orb} \leq 10$ days; *(12, 145–149)*), but also the transitioning of wider AGB binaries into the post-AGB and PN phase can result in the formation of aspherical PNe. In particular the formation of an EDE will favour a polar ejection of material. This argument is supported by the growing number of aspherical PNe detected whose binary central stars have a long-period orbit (1 yr $\leq P_{\rm orb} \leq$10 yr) so that the common-envelope phase has been avoided *(150–153)*. Our results support the claim that binarity may be a prerequisite for the formation of observable aspherical PNe *(154)*. While magnetic fields and stellar rotation can play a role in shaping PNe (as well as AGB winds), the angular momentum from a binary is thought to be required to sustain the global magnetic field and rotation long enough to affect the geometry of the mass loss and the shaping of the PN *(13, 155)*.

Although none of the ATOMIUM sources displays a smooth, spherical wind, many nebulae surrounding post-AGB and PN stars exhibit even stronger asymmetries, notable examples include M1-92 *(156)*, the 'Butterfly' nebula M2$-$9 *(157)*, or HD 101584 *(158)*. Any asphericity present in the AGB wind will impact the post-AGB morphology. The development of an EDE during the AGB phase can lead to the creation of bipolar/multipolar shapes, due to the action of a collimated fast wind (CFW) during the late AGB or early post-AGB phase *(9)*. This CFW carves the dense AGB shell from the inside out. As the central star evolves further towards the PN phase, the existing aspherical post-AGB morphology will be further shaped due to the action of spherical, radiatively driven, fast wind from the hot PN central star *(9)*. Additional mechanisms such as jets launched from the accretion disk around the companion *(159)*, or the combination of a fast collimated wind with toroidal magnetic field *(160)*, could also carve the post-AGB and PNe morphology.

# S7 ALMA ATOMIUM observational results

In this section, we present all the data used in our morphological characterization of the ATOMIUM AGB stars. These include: channel maps of the $^{12}$CO $J=2\rightarrow1$ line at medium and low spatial resolution; stereograms of the $^{28}$SiO $J=5\rightarrow4$ line; and moment1-maps and position-velocity diagrams of the $^{12}$CO $J=2\rightarrow1$ and $^{28}$SiO $J=5\rightarrow4$ lines.

**Channel maps:** Fig. S8, Fig. S9, Fig. S12, Fig. S13, Fig. S16, Fig. S17, Fig. S20, Fig. S21, Fig. S24, Fig. S25, Fig. S28, Fig. S29, Fig. S32, Fig. S33, Fig. S37, Fig. S38, Fig. S41, Fig. S42, Fig. S46, Fig. S47, Fig. S50, Fig. S51, Fig. S54, Fig. S55, Fig. S58, Fig. S59, Fig. S62, Fig. S63: The minimum and maximum velocity offset with respect to the local standard-of-rest velocity are determined from the line profile, so that the integrated line intensities are greater than 3 times the $\sigma_{\rm rms}$ noise (see Table S6). We set the minimum intensity in the color scale is set to $(-1)\times\sigma_{\rm rms}$. The maximum of the color scale is chosen to better visualize the weaker features, and is set to ~0.5 times the maximum intensity of the data. The CO channel map of five targets display spiral arcs (R Aql, W Aql, GY Aql, IRC-10529, and IRC+10011; Fig. S46, Fig. S50, Fig. S54, Fig. S58, and Fig. S62). R Hya and $\pi^1$ Gru have a biconical / hourglass structure (Fig. S32 and Fig. S41).

**Stereograms:** Fig. S10A, Fig. S14A, Fig. S18A, Fig. S22A, Fig. S26A, Fig. S30A, Fig. S34A, Fig. S39A, Fig. S43A, Fig. S48A, Fig. S52A, Fig. S56A, Fig. S60A, Fig. S64A: Stereogram plots of the $^{28}$SiO $J=5\rightarrow4$ line show the offset between blue and red shifted emission. The blue and red contours are constructed from 49% of the blue and red-shifted emission of the molecular lines, respectively. The black dotted contours are constructed with the remaining 2% of the velocity channels which are centered on the local velocity of rest $v_{\rm LSR}$. The contours are evenly spaced between $2\times\sigma_{\rm rms}$ in the (red, blue, black) velocity channel maps and the maximum of the moment0-map of the respective (red, blue, black) channels, where the moment0-map is the emission map integrated over all frequency channels with measurable emission in the line. A distinct difference in the position of the red and blue shifted emission is observed in U Del (Fig. S18A) and $\pi^1$ Gru (Fig. S43A); and a small systematic difference is also noted for S Pav (Fig. S10A) and T Mic (Fig. S14A), and the inner 0.8″ region of R Hya (Fig. S34A).

**First moment maps:** Fig. S10B – Fig. S10C, Fig. S14B – Fig. S14C, Fig. S18B – Fig. S18C, Fig. S22B – Fig. S22C, Fig. S26B – Fig. S26C, Fig. S30B – Fig. S30C, Fig. S34B – Fig. S34C, Fig. S36, Fig. S39B – Fig. S39C, Fig. S43B – Fig. S43C, Fig. S48B – Fig. S48C, Fig. S52B – Fig. S52C, Fig. S56B – Fig. S56C, Fig. S60B – Fig. S60C, Fig. S64B – Fig. S64C: First moment (or moment1) maps of two SiO isotopologue lines are used as a tool for visualizing structures in the velocity fields in the inner wind region. The maps are obtained by taking

$$\frac{\sum_{\nu_{\rm blue}}^{\nu_{\rm red}} I_\nu\, v_\nu\, {\rm d}\nu}{\sum_{\nu_{\rm blue}}^{\nu_{\rm red}} I_\nu\, {\rm d}\nu}\,, \qquad \text{(S17)}$$

with $v_{\rm blue}$ and $v_{\rm red}$ indicating the minimum and maximum velocity offset with respect to the local standard-of-rest velocity as determined from the line profile (see above), and $I_\nu$ the intensity at frequency $\nu$ with corresponding velocity $v_\nu$ with respect to the local standard-of-rest velocity. For most targets, the SiO emission mainly traces the inner 1″ region of the wind. For some targets the line velocity map exhibits distinct red-shifted and blue-shifted components, which is the classical signature of rotation *(17)* or bipolar outflow: U Del (Fig. S18B – Fig. S18C) and $\pi^1$ Gru (Fig. S43B – Fig. S43C) are examples of this type of signature. Another example is the $^{12}$CO moment1-map of R Hya (Fig. S36). In the inner 0.5″ region of R Hya there is some evidence of potential rotation/bipolar outflow in agreement with the offset between the red and blue shifted emission in the stereogram. S Pav (Fig. S10B – Fig. S10C) and T Mic (Fig. S14B – Fig. S14C) have signs of complex rotation or bipolar dynamics, with their moment1-maps resembling that of the red supergiant Betelgeuse *(161)*.

**Position-velocity diagrams:** Fig. S11, Fig. S15, Fig. S19, Fig. S23, Fig. S27, Fig. S31, Fig. S35, Fig. S40, Fig. S44, Fig. S49, Fig. S53, Fig. S57, Fig. S61, Fig. S65: Position-velocity (PV) diagrams of the $^{12}$CO $J=2\rightarrow1$ and $^{28}$SiO $J=5\rightarrow4$ emission are shown for each target. A PV diagram is obtained by taking a slice through the 3D ALMA data at an arbitrary angular axis in the plane of the sky, and hence is a 2D plot of the emission along this chosen axis versus velocity. In principle any slit width can be chosen. If the slit width is larger than one singular data pixel, the emission is collapsed onto each other by summing up the emission with identical PV coordinates. We here have chosen the slit width to encompasses the full emission in the moment0-maps. A set of two orthogonal pairs of PV diagrams are displayed with the position angle (PA) chosen to produce the greatest difference between the two PV diagrams, maximizing the asymmetry in the data.

### S7.1 S Pav

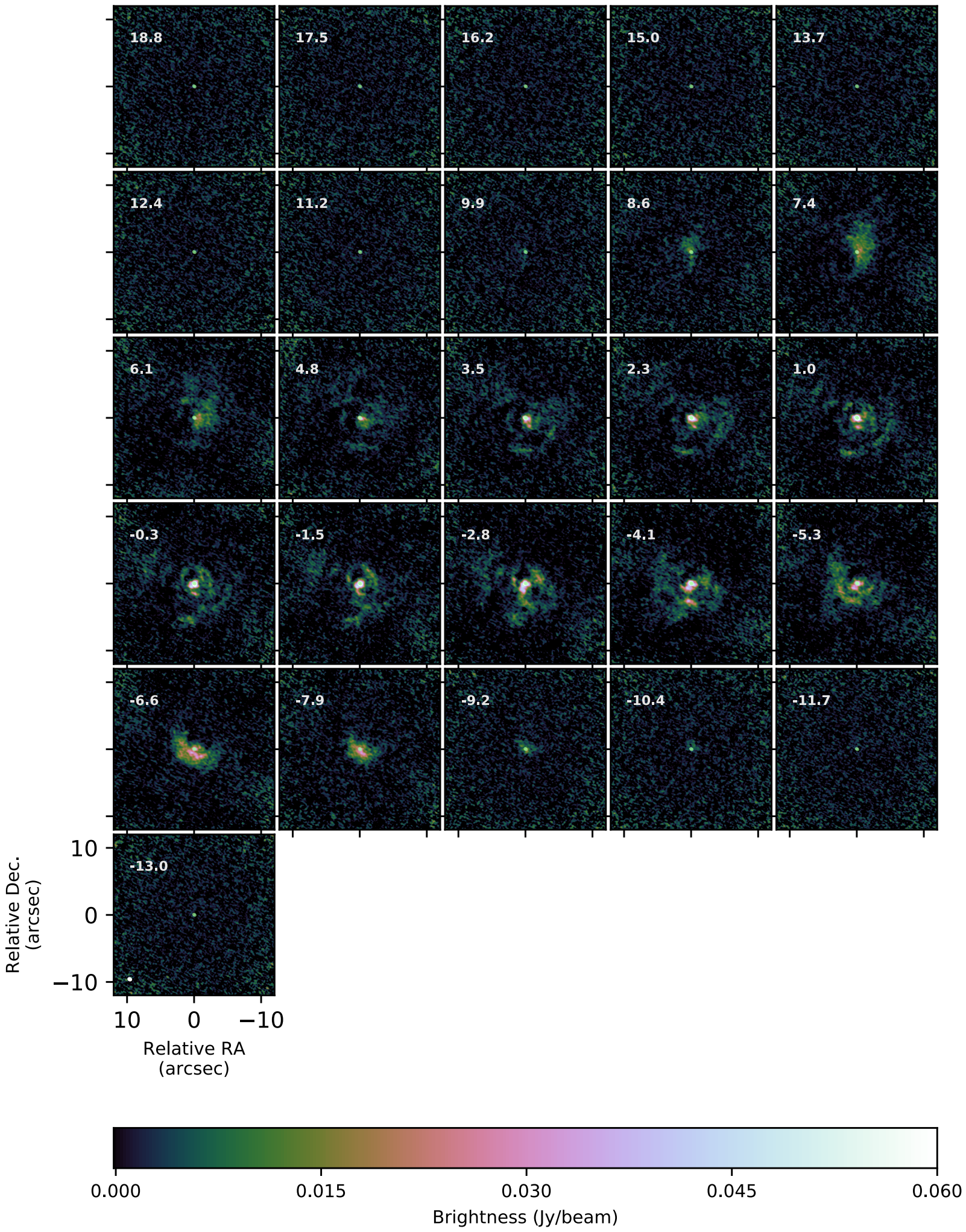

Fig. S8: **ALMA medium resolution $^{12}$CO J = 2 → 1 channel map of S Pav.** North is up, East is left. Contours show the continuum emission at [10, 30, 60, 90, and 99]% of the peak continuum emission. The ordinate and co-ordinate axis give the offset of the right ascension and declination, respectively, with respect to the peak of the continuum emission. The velocity with respect to the local standard of rest velocity ($\upsilon_{\rm LSR}$ as given in column 11 in Table S1) is given in the upper right corner of each panel (in units of km s$^{-1}$). The ALMA beam is shown as a white ellipse in the bottom left corner of the bottom left panel. The minimum and maximum velocity offset are determined from the line profile, accounting for all channels in which the integrated flux is greater than $3 \times \sigma_{\rm rms}$ (see Table S6). The minimum intensity in the color scale is set to $(-1) \times \sigma_{\rm rms}$, the maximum of the color scale is set to ~0.5 times the maximum intensity of the data.

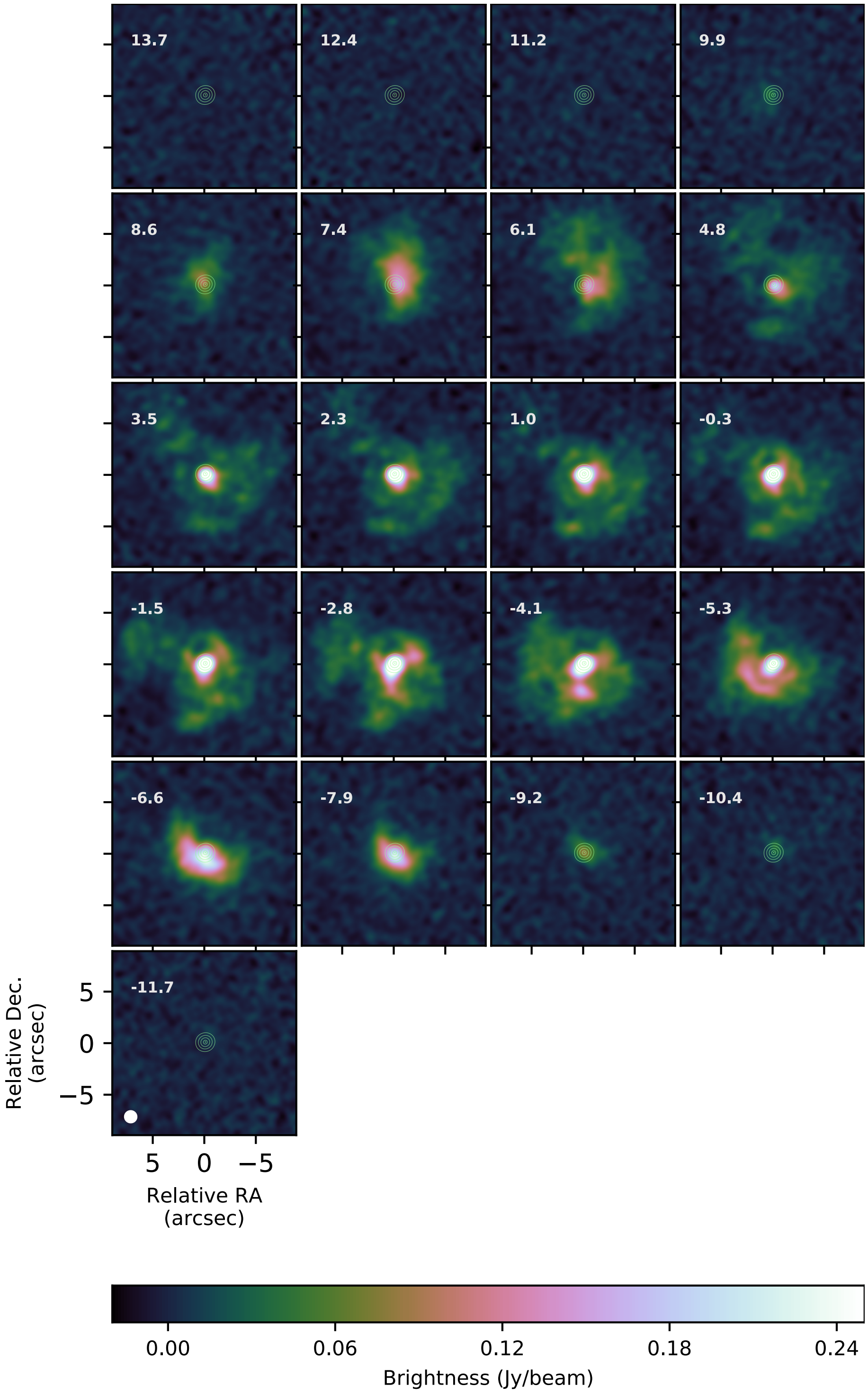

Fig. S9: **Low resolution $^{12}$CO J$=2\rightarrow1$ channel map of S Pav.** Same as Fig. S8 but for the low resolution observation.

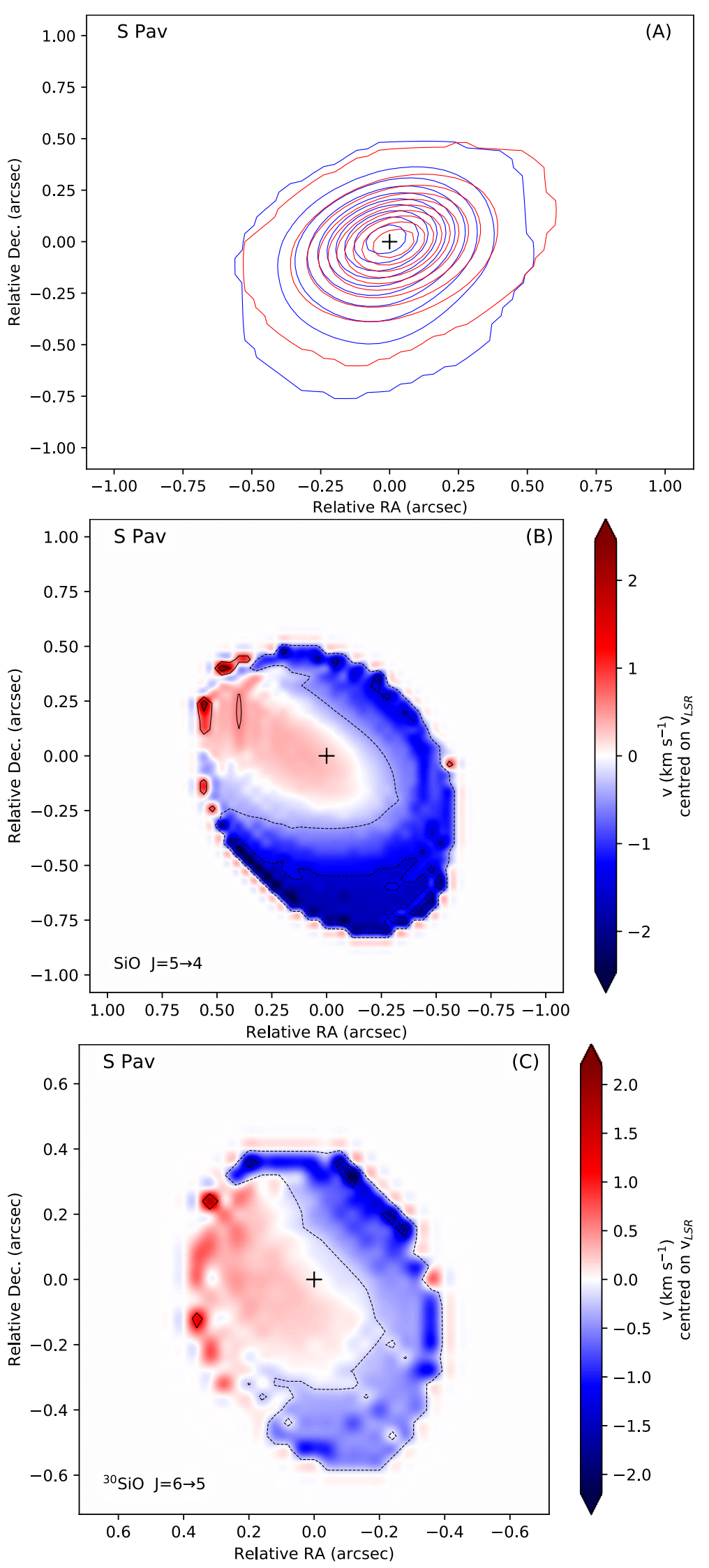

Fig. S10: **SiO stereogram and moment1-map of S Pav.** North is up and east is to the left. The cross indicates the position of the maximum continuum flux. The ordinate and co-ordinate axis give the offset of the right ascension and declination, respectively, with respect to the peak of the continuum emission. Fig. S10A: $^{28}$SiO $J=5\rightarrow4$ stereogram. The blue and red contours are constructed from 49% of the blue and red-shifted emission of the molecular lines, respectively. The black dotted contours are constructed with the remaining 2% of the velocity channels which are centered on the local velocity of rest $v_{\rm LSR}$. The contours are evenly spaced between $2\times\sigma_{\rm rms}$ in the (red, blue, black) velocity channel maps and the maximum of the moment0-map of the respective (red, blue, black) channels, where the moment0-map is the emission map integrated over all frequency channels with measurable emission in the line. Fig. S10B: $^{28}$SiO $J=5\rightarrow4$ moment1-map, Fig. S10C: $^{30}$SiO $J=6\rightarrow5$ moment1-map.

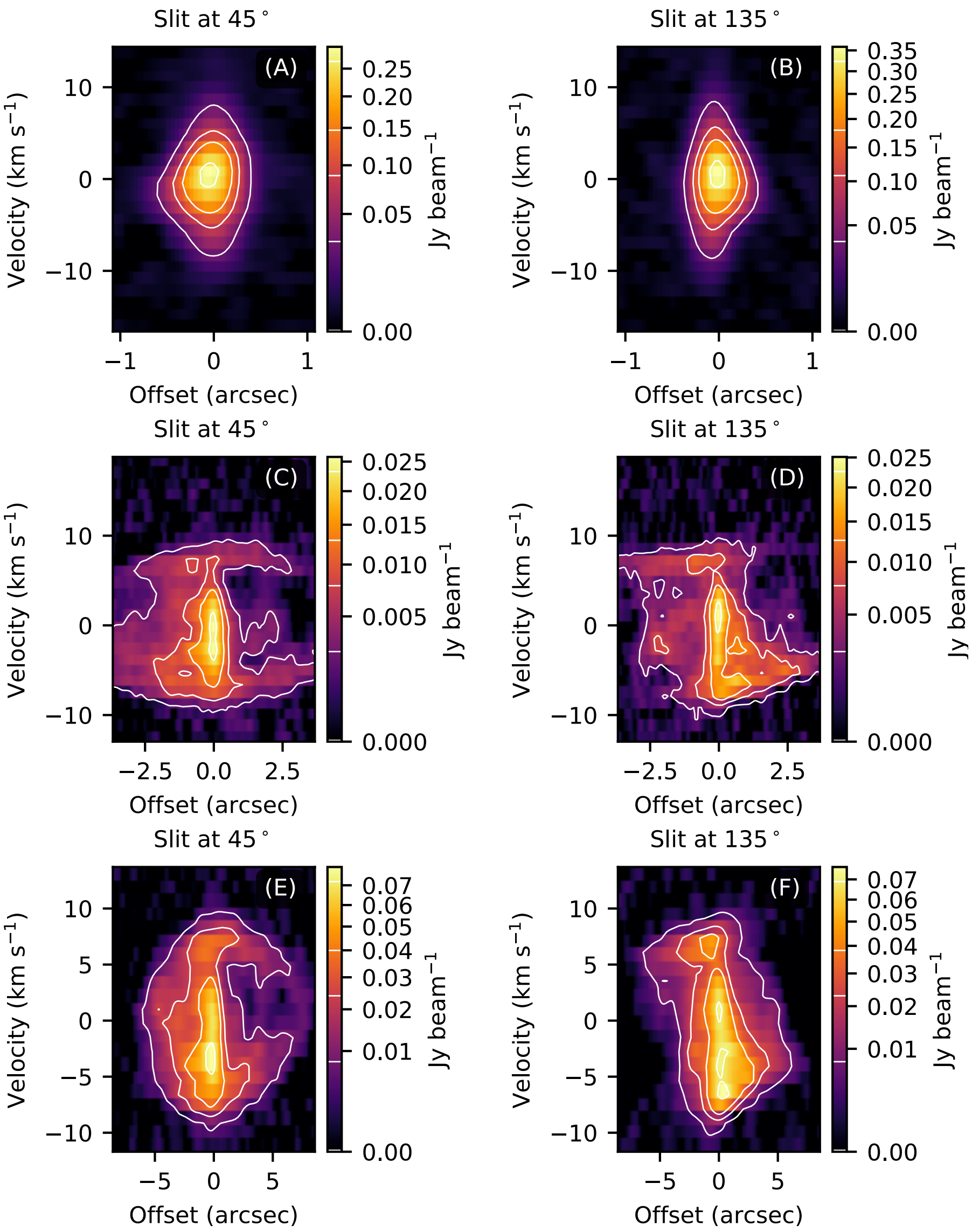

Fig. S11: **ALMA $^{12}$CO J$=2\rightarrow1$ and $^{28}$SiO J$=5\rightarrow4$ position-velocity (PV) diagram of S Pav.** The position angle (PA) is indicated at the top of each panel. Contours (in white) are plotted at [0.1, 0.3, 0.5, 0.9] times the maximum value. Fig. S11A – Fig. S11B: PV diagrams of the SiO $J=5\rightarrow4$ medium resolution data. Fig. S11C – Fig. S11D: PV diagrams of the CO $J=2\rightarrow1$ medium resolution data. Fig. S11E – Fig. S11F: PV diagrams of the CO $J=2\rightarrow1$ low resolution data.

### S7.2 T Mic

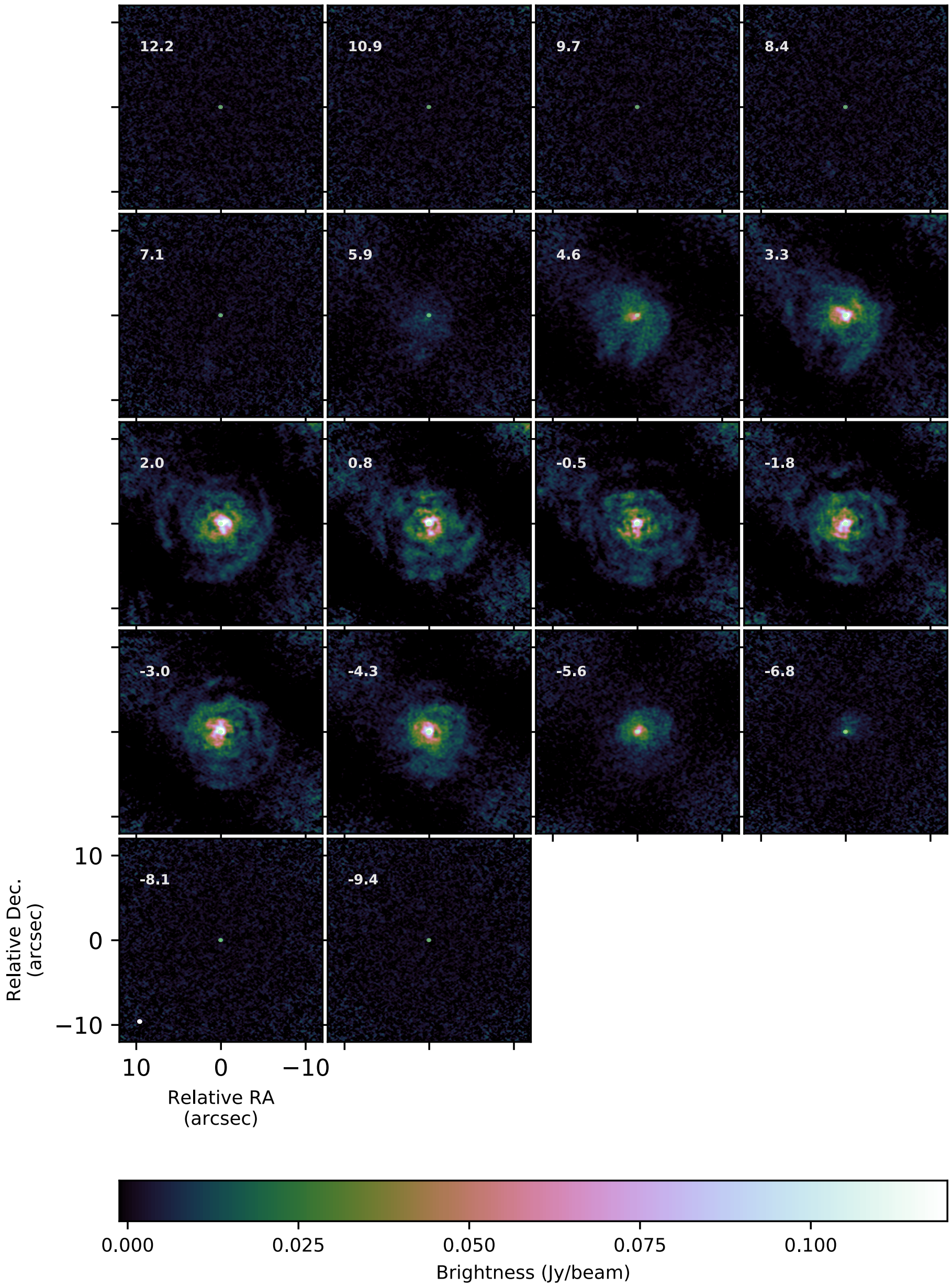

Fig. S12: **Medium resolution $^{12}$CO J = 2 → 1 channel map of T Mic.** Same as Fig. S8 but for T Mic.

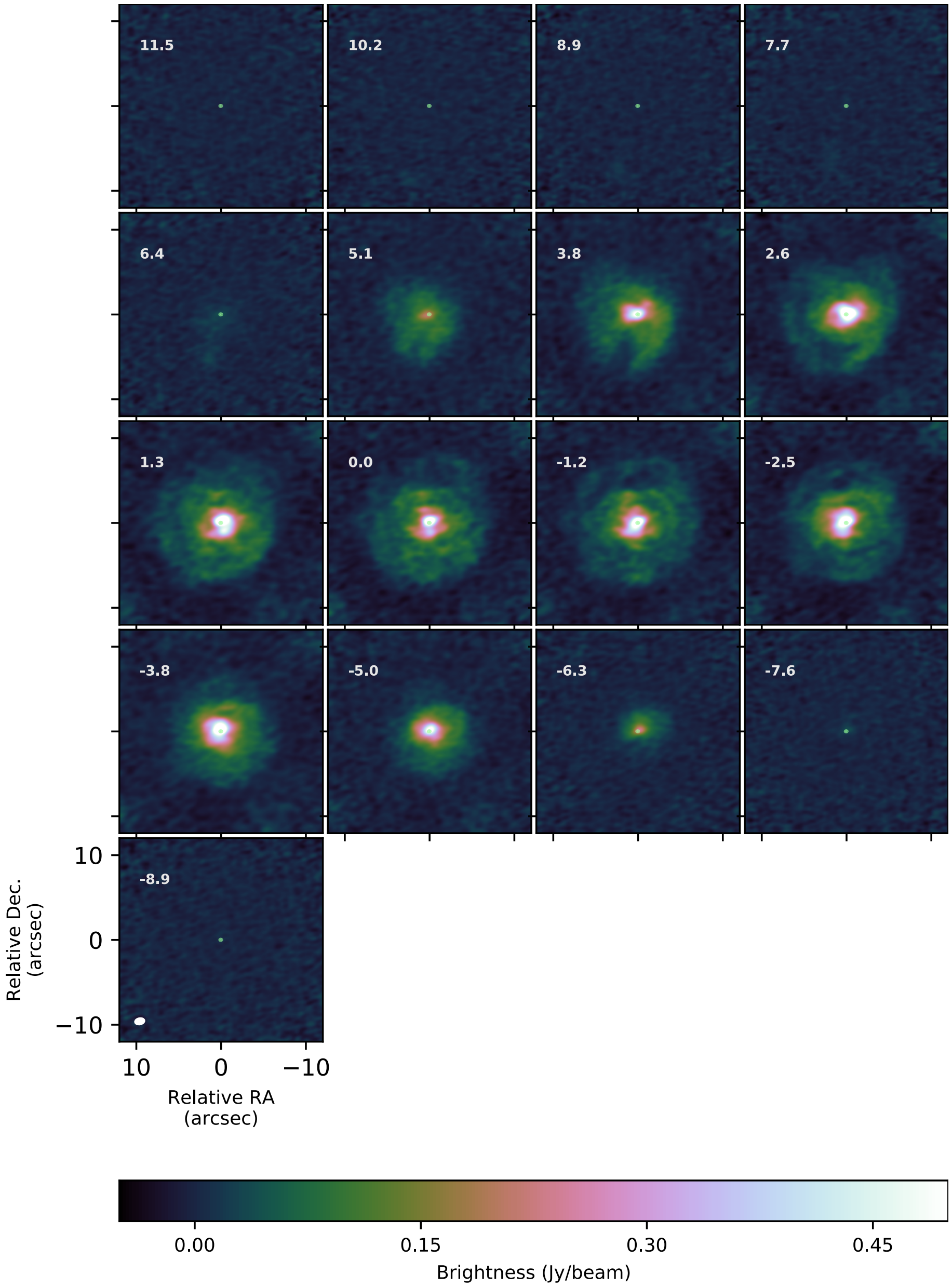

Fig. S13: **Low resolution $^{12}$CO J = 2 → 1 channel map of T Mic.** Same as Fig. S9, but for T Mic.

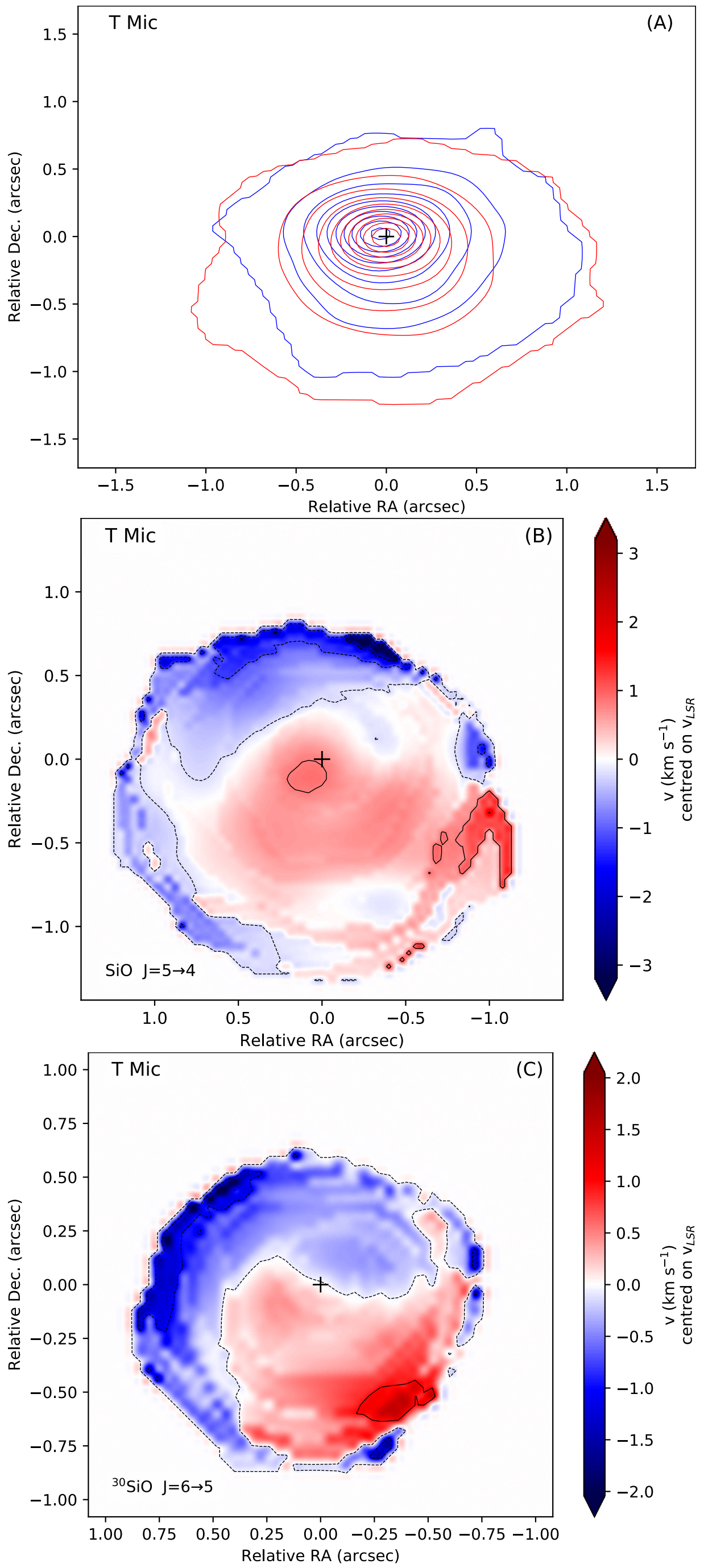

Fig. S14: **SiO stereogram and moment1-map of T Mic.** Same as Fig. S10, but for T Mic.

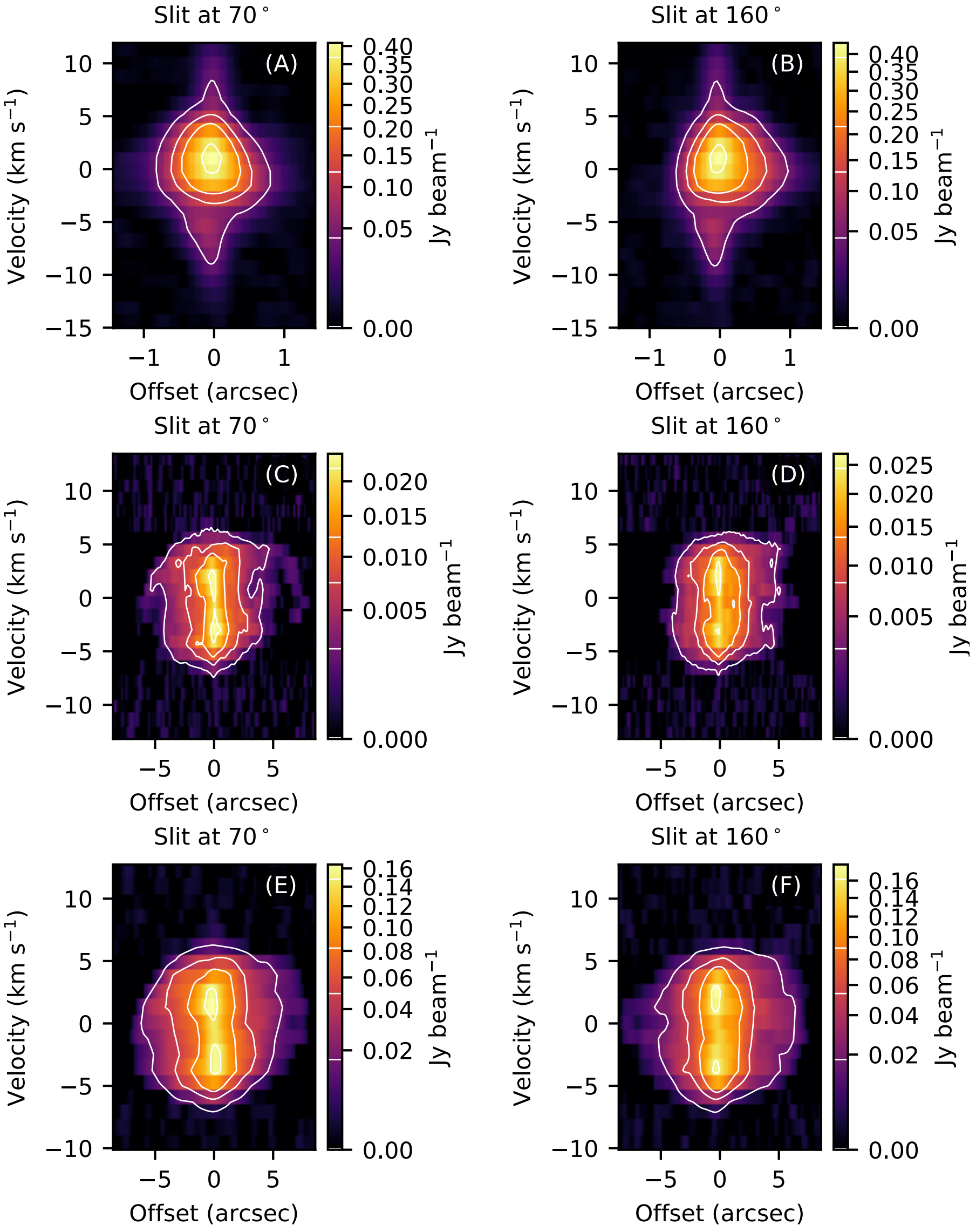

Fig. S15: **ALMA $^{12}$CO J = 2 → 1 and $^{28}$SiO J = 5 → 4 position-velocity (PV) diagram of T Mic.** Same as Fig. S11, but for T Mic.

### S7.3 U Del

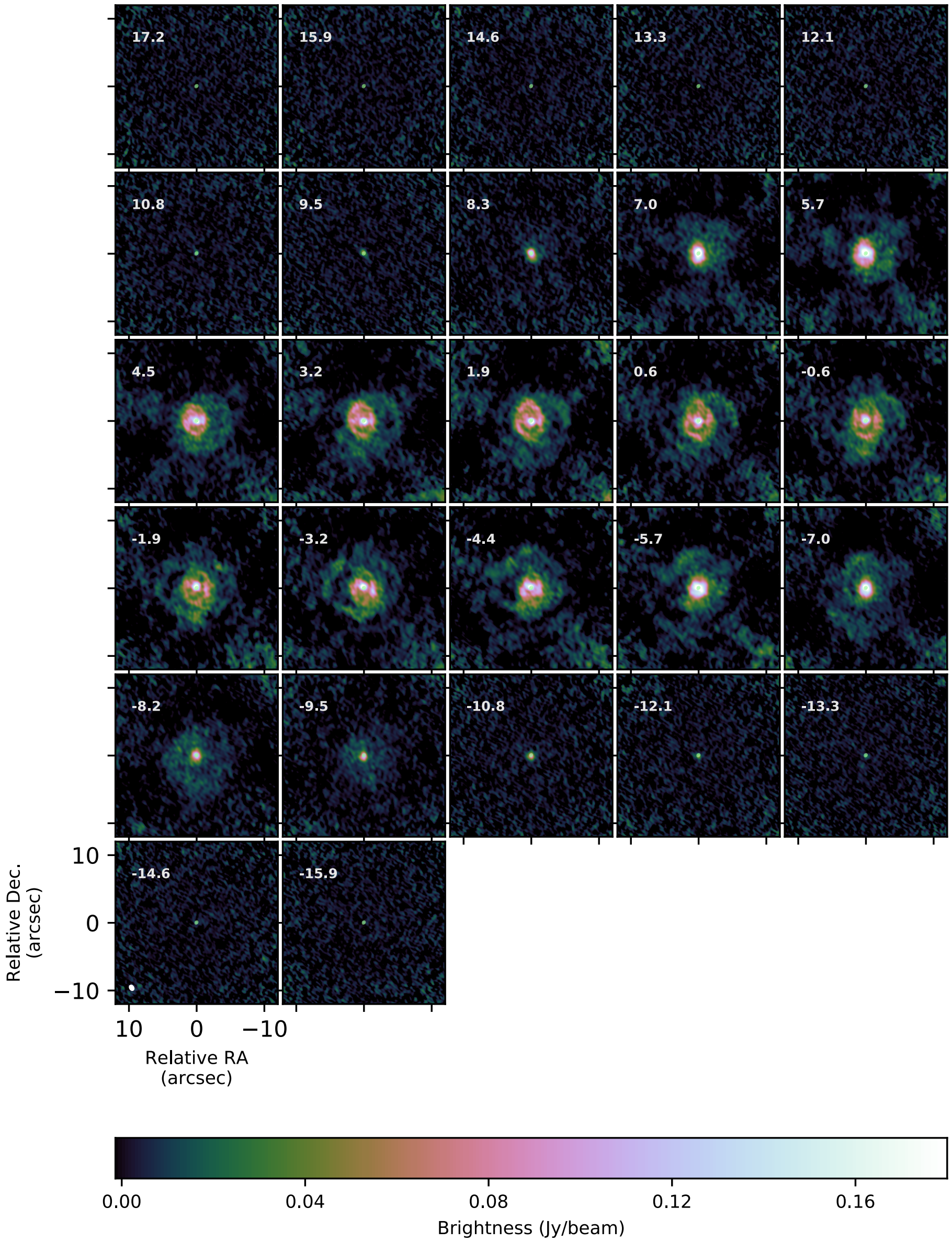

Fig. S16: **Medium resolution $^{12}$CO J = 2 → 1 channel map of U Del.** Same as Fig. S8, but for U Del.

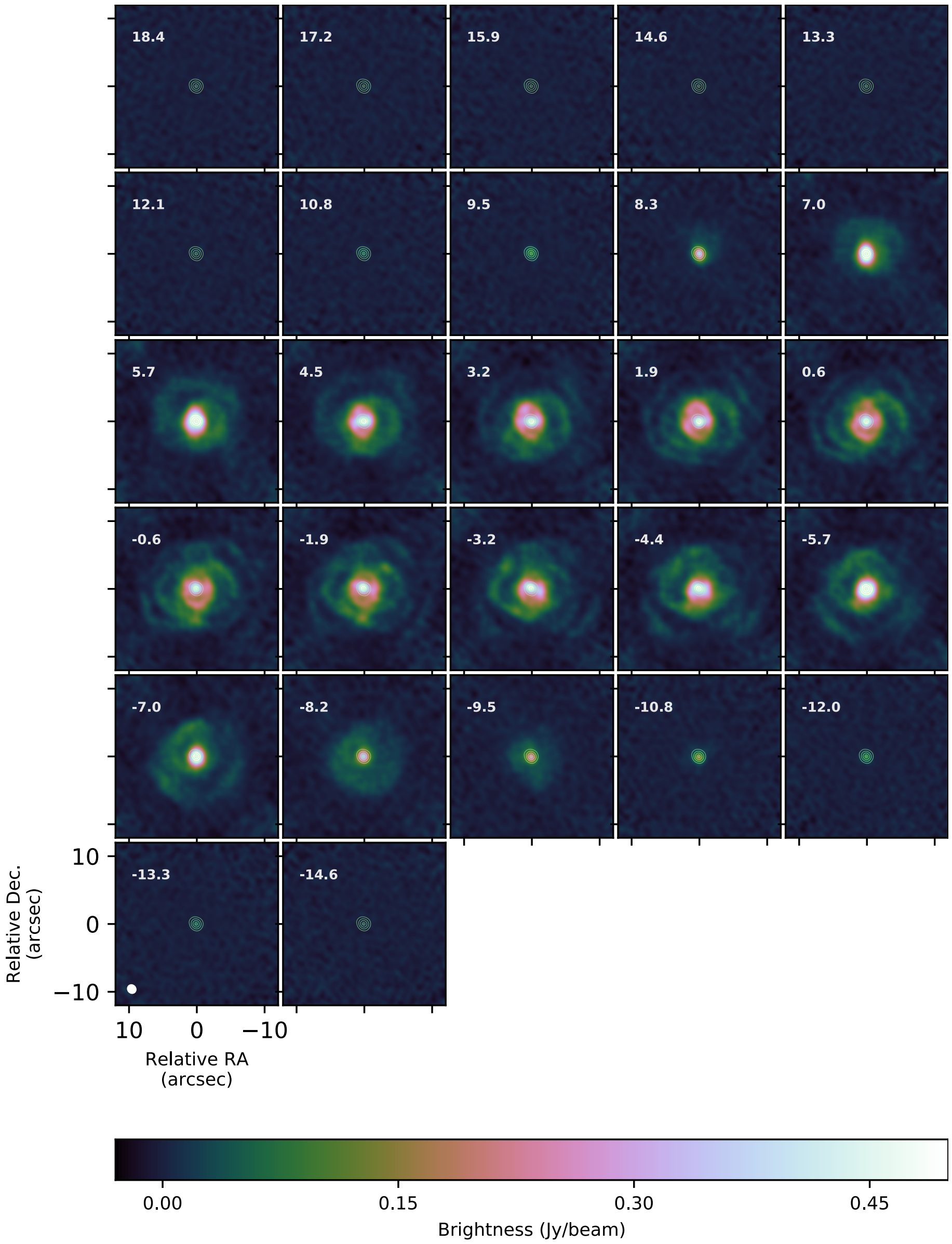

Fig. S17: **Low resolution $^{12}$CO J = 2 → 1 channel map of U Del.** Same as Fig. S9, but for U Del.

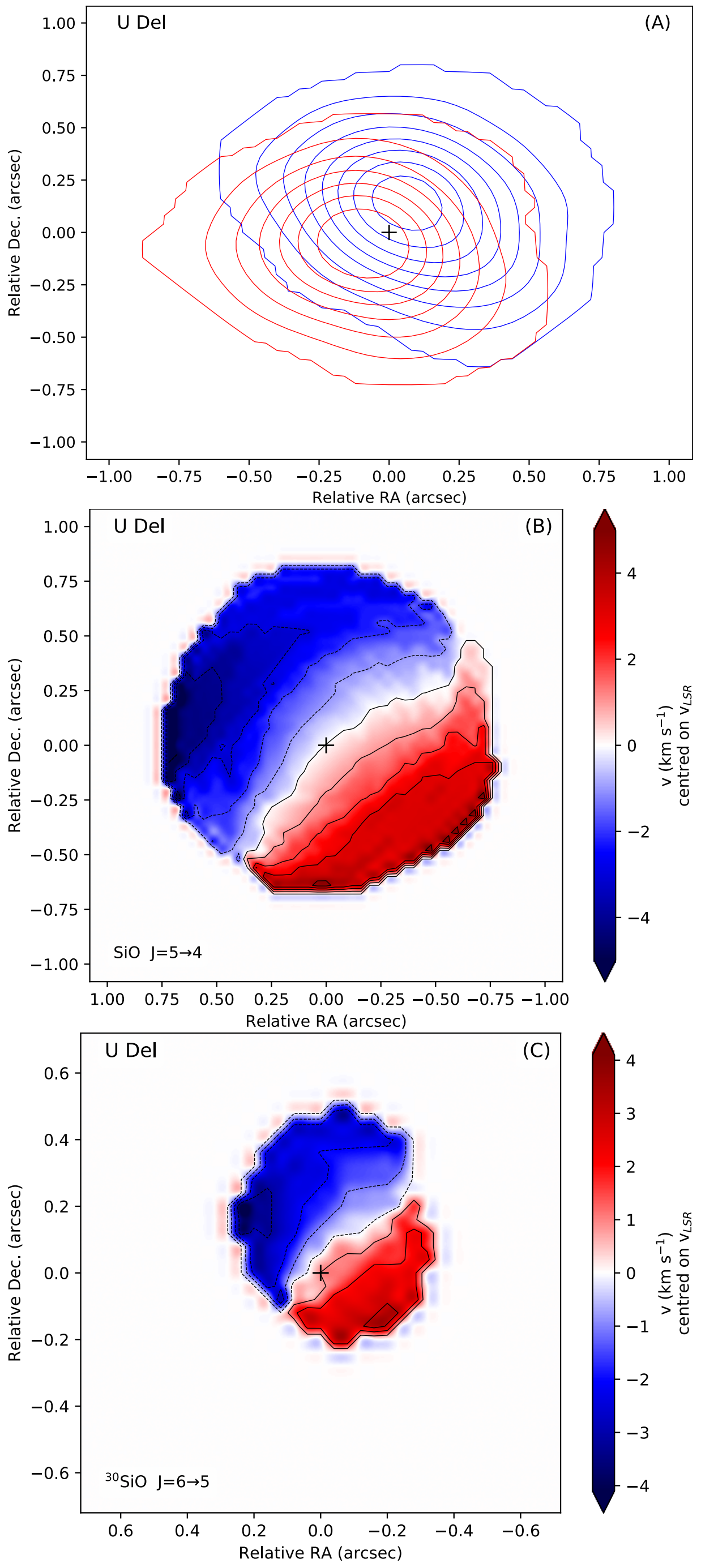

Fig. S18: **SiO stereogram and moment1-map of U Del.** Same as Fig. S10, but for U Del.

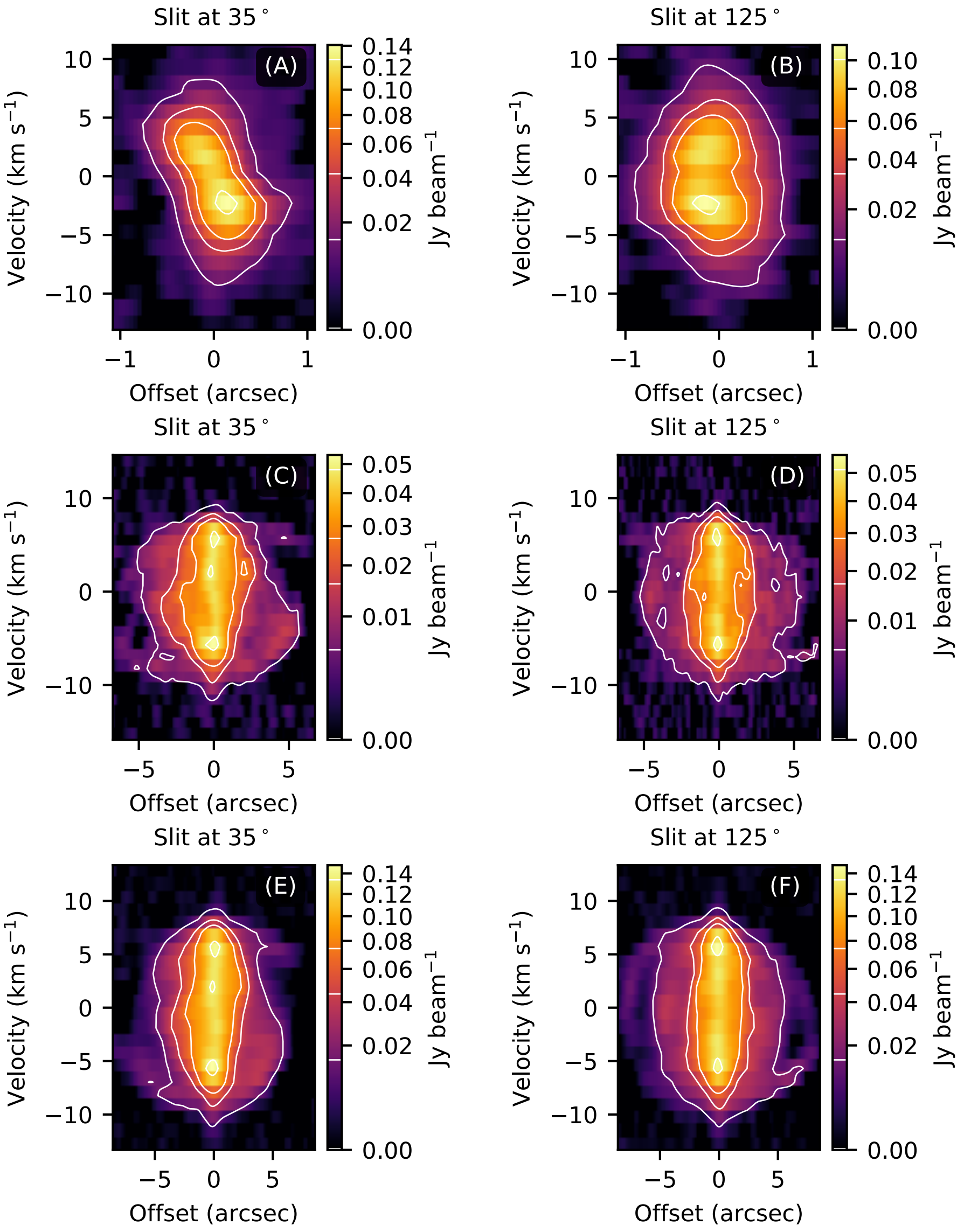

Fig. S19: **ALMA $^{12}$CO J = 2 → 1 and $^{28}$SiO J = 5 → 4 position-velocity (PV) diagram of U Del.** Same as Fig. S11, but for U Del.

### S7.4 RW Sco

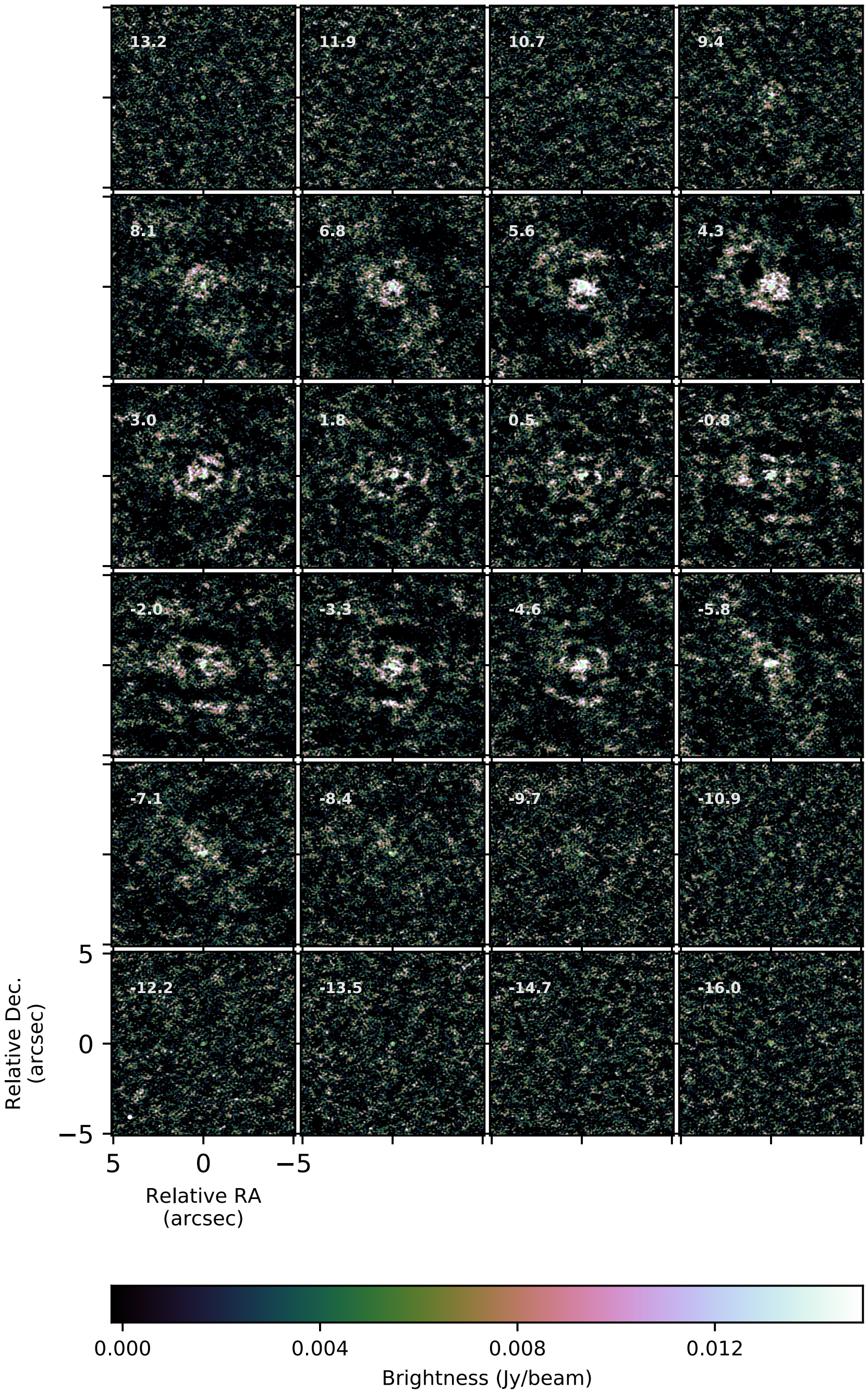

Fig. S20: **Medium resolution $^{12}$CO J = 2 → 1 channel map of RW Sco.** Same as Fig. S8, but for RW Sco.

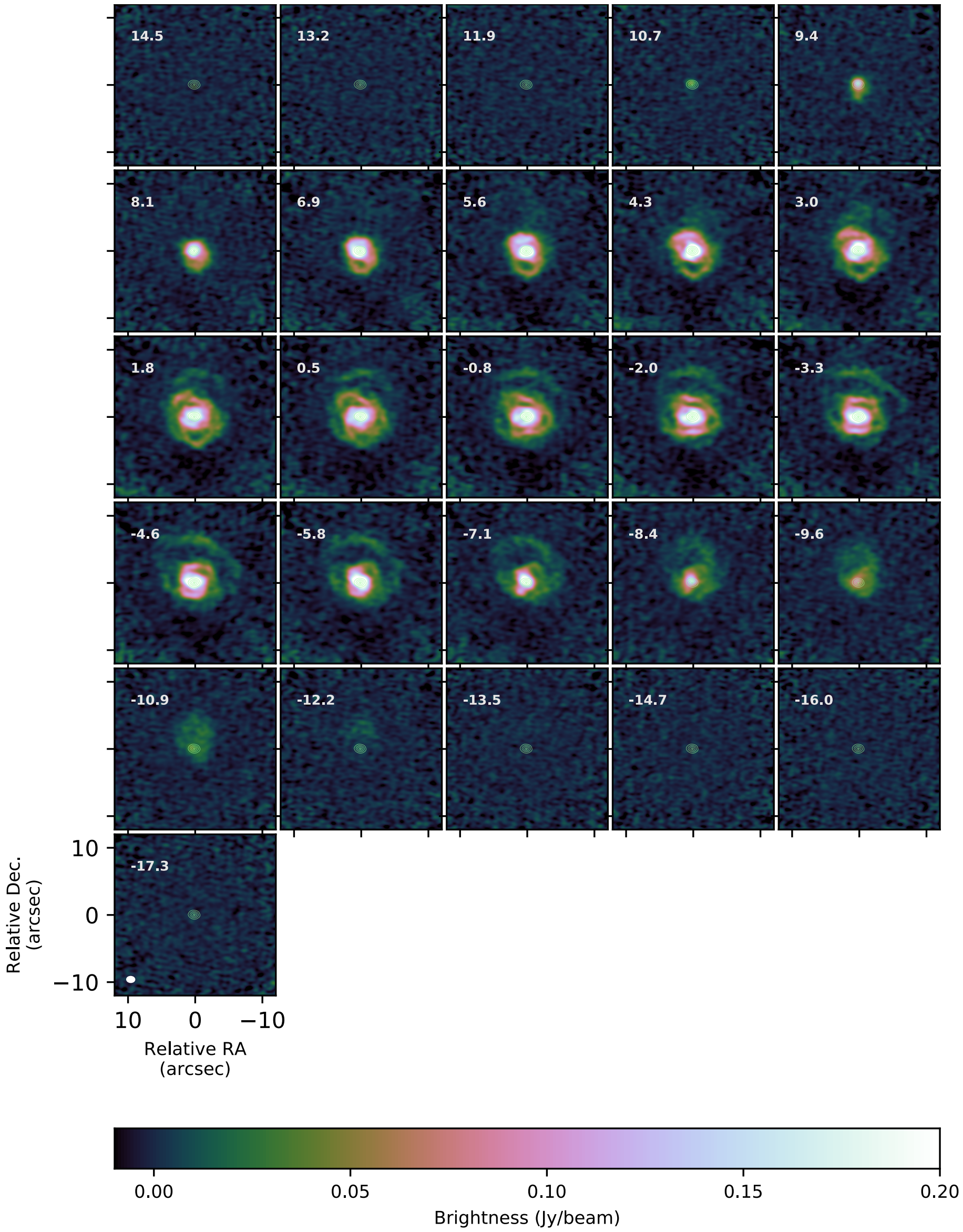

Fig. S21: **Low resolution $^{12}$CO J = 2 → 1 channel map of RW Sco.** Same as Fig. S9, but for RW Sco.

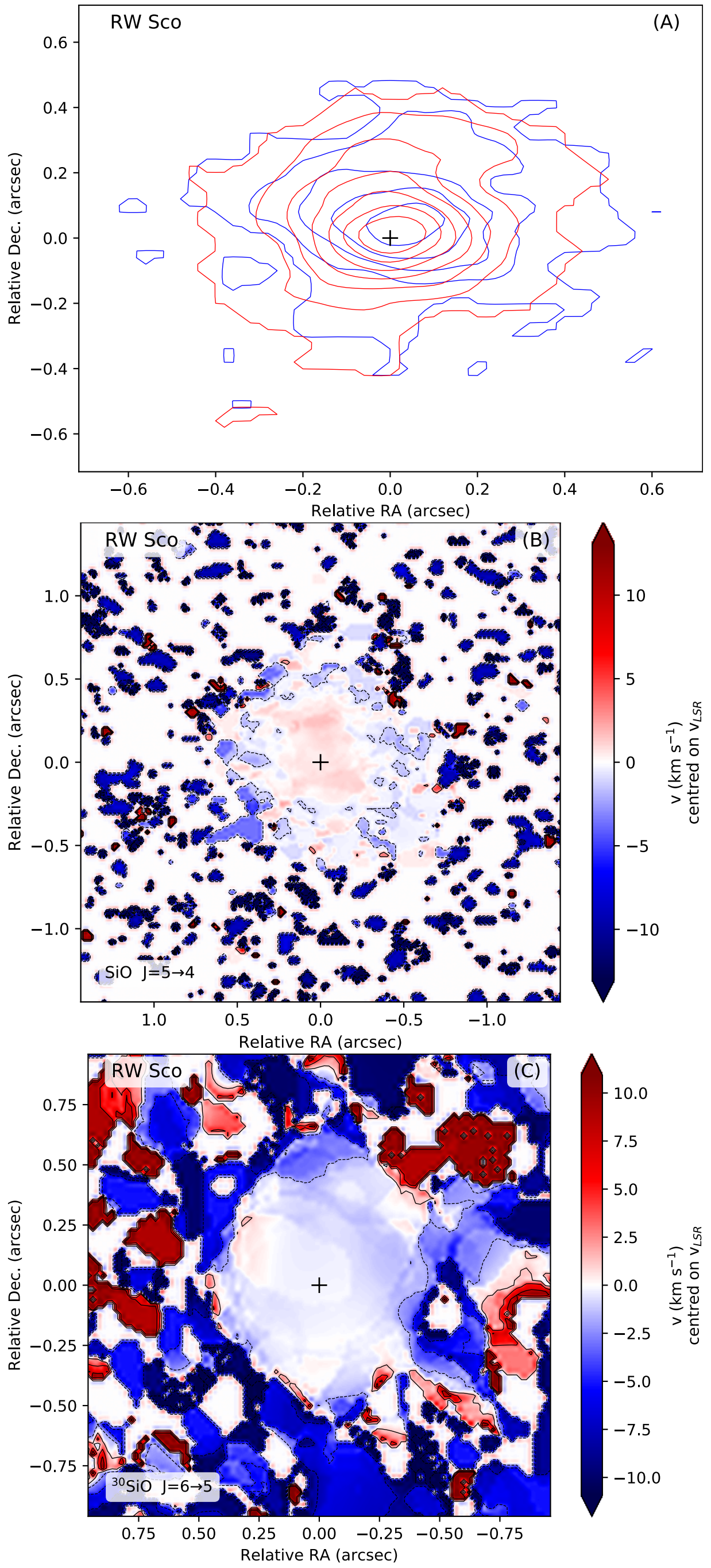

Fig. S22: **SiO stereogram and moment1-map of RW Sco.** Same as Fig. S10, but for RW Sco. Owing to the low signal-to-noise ratio of the SiO emission, these images should be interpreted with care (see also Sect. S3.2).

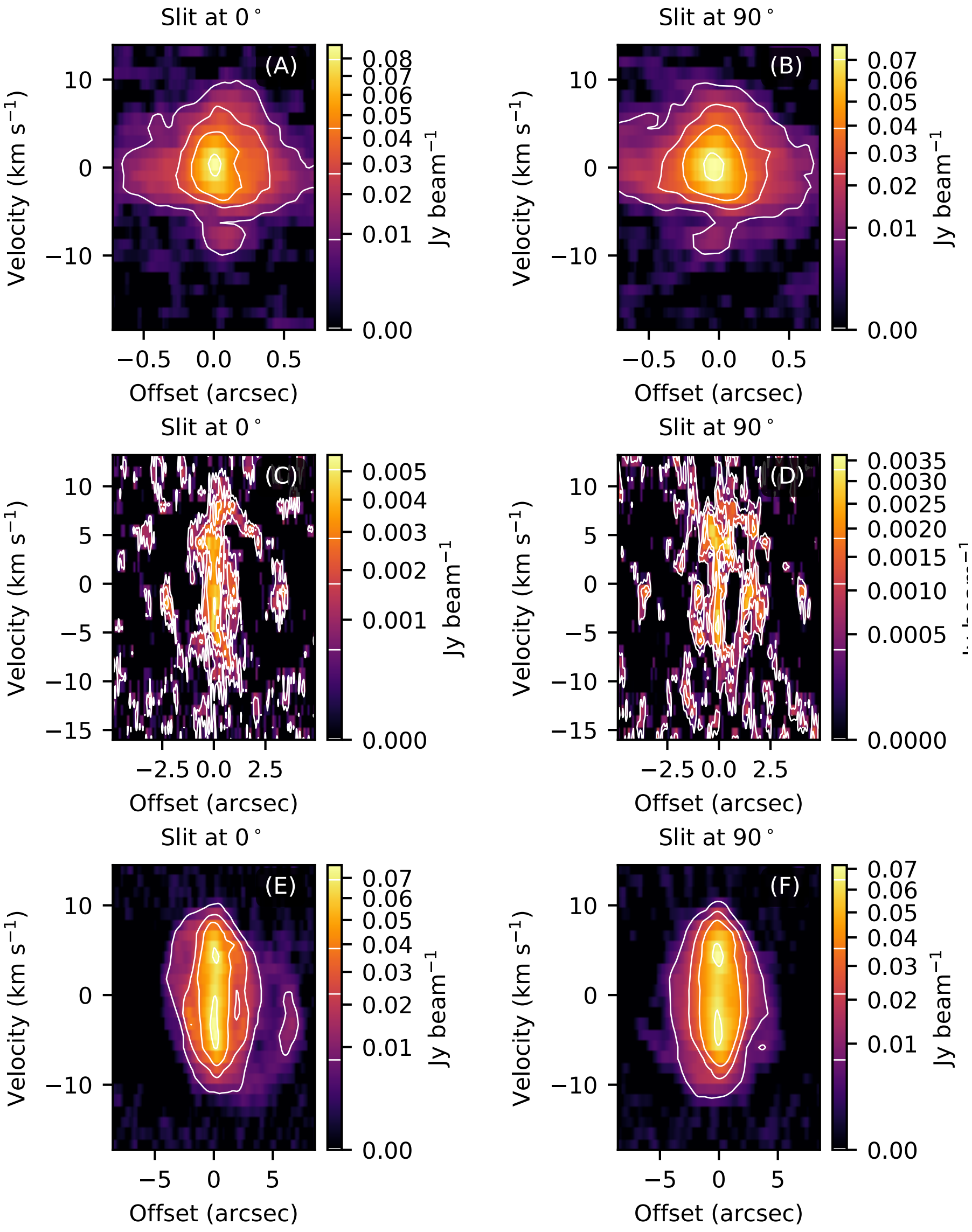

Fig. S23: **ALMA $^{12}$CO J = 2 → 1 and $^{28}$SiO J = 5 → 4 position-velocity (PV) diagram of RW Sco.** Same as Fig. S11, but for RW Sco.

### S7.5 V PsA

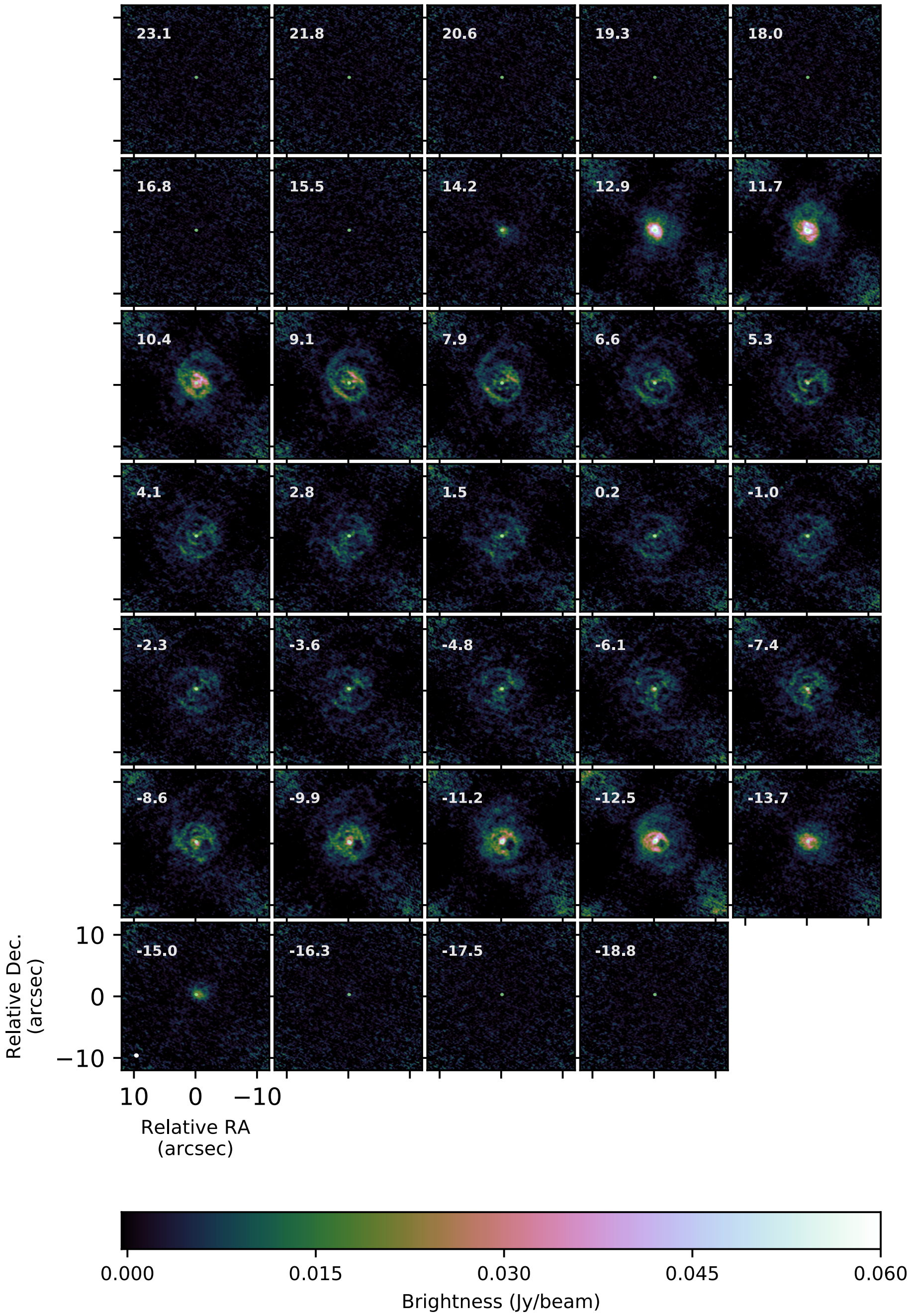

Fig. S24: **Medium resolution $^{12}$CO J = 2 → 1 channel map of V PsA.** Same as Fig. S8, but for V PsA.

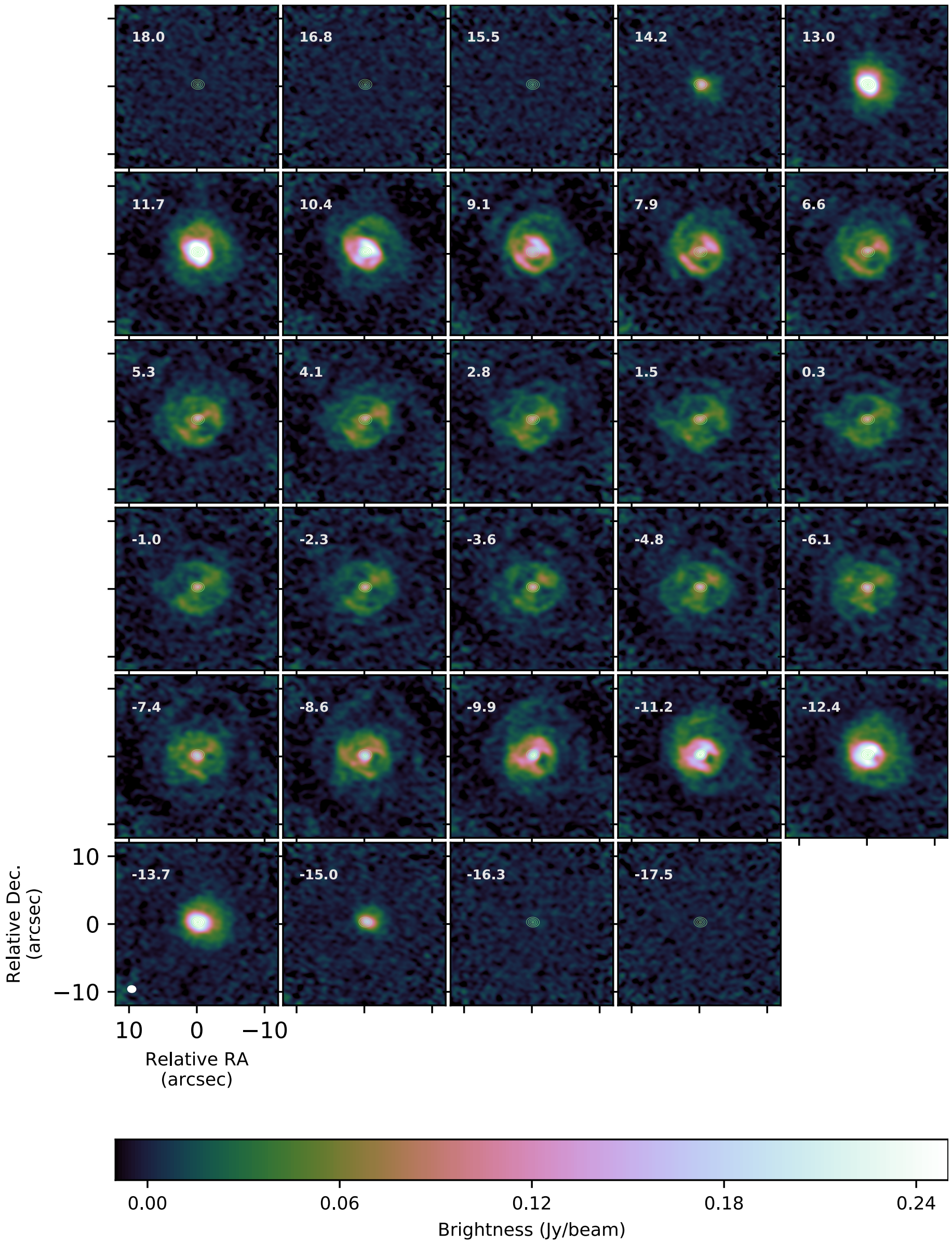

Fig. S25: **Low resolution $^{12}$CO J = 2 → 1 channel map of V PsA.** Same as Fig. S9, but for V PsA.

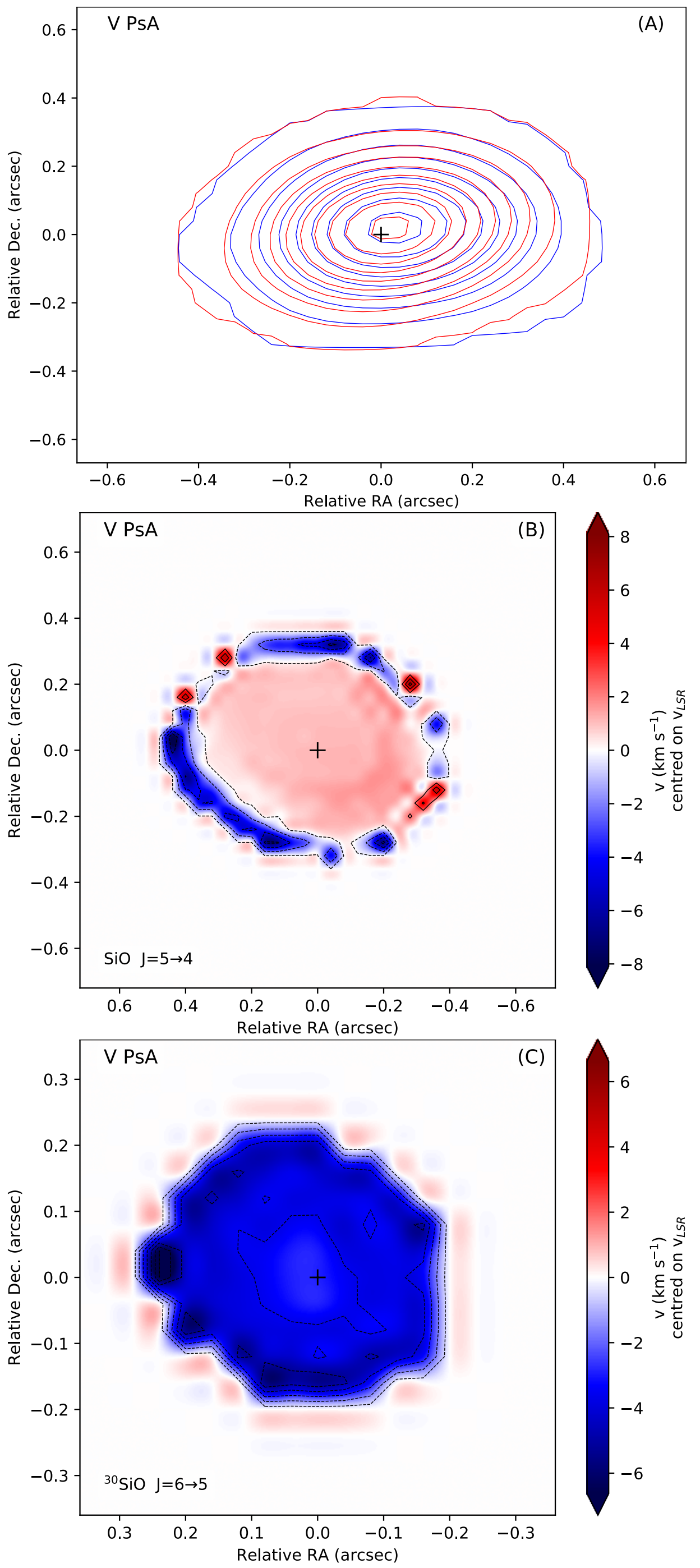

Fig. S26: **SiO stereogram and moment1-map of V PsA.** Same as Fig. S10, but for V PsA.

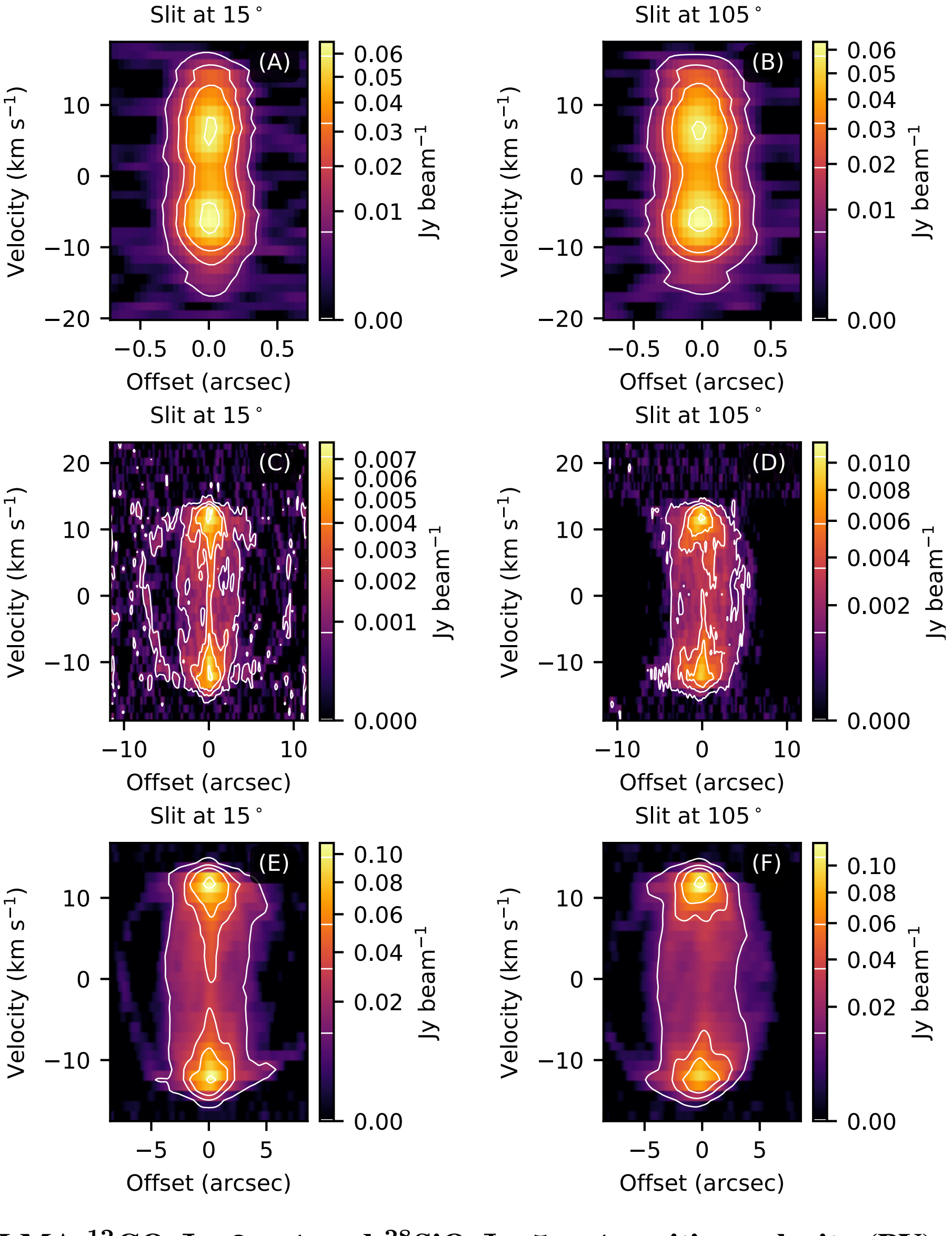

Fig. S27: **ALMA $^{12}$CO J = 2 → 1 and $^{28}$SiO J = 5 → 4 position-velocity (PV) diagram of V PsA.** Same as Fig. S11, but for V PsA.

## S7.6 SV Aqr

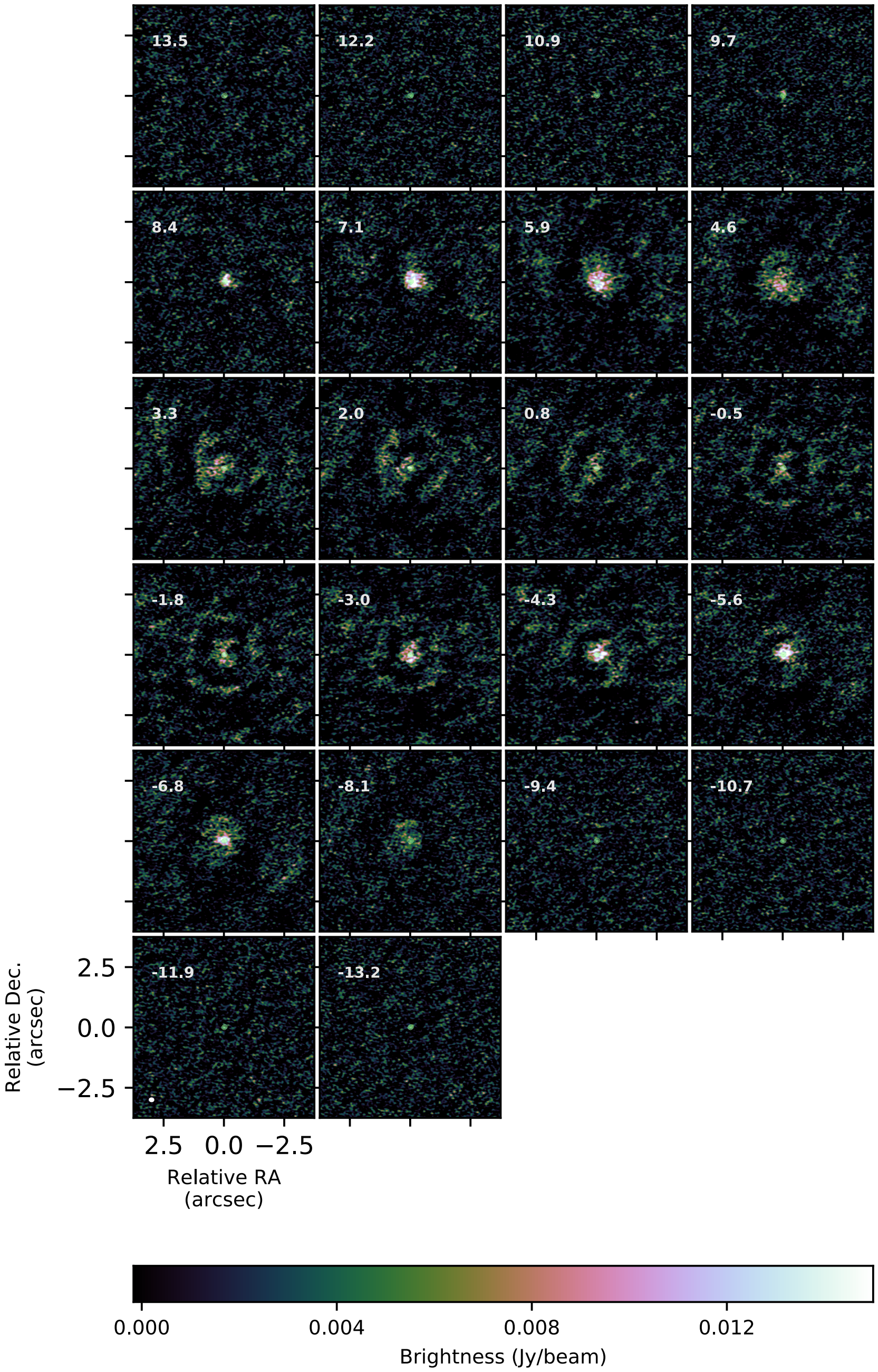

Fig. S28: **Medium resolution $^{12}$CO J = 2 → 1 channel map of SV Aqr.** Same as Fig. S8, but for SV Aqr.

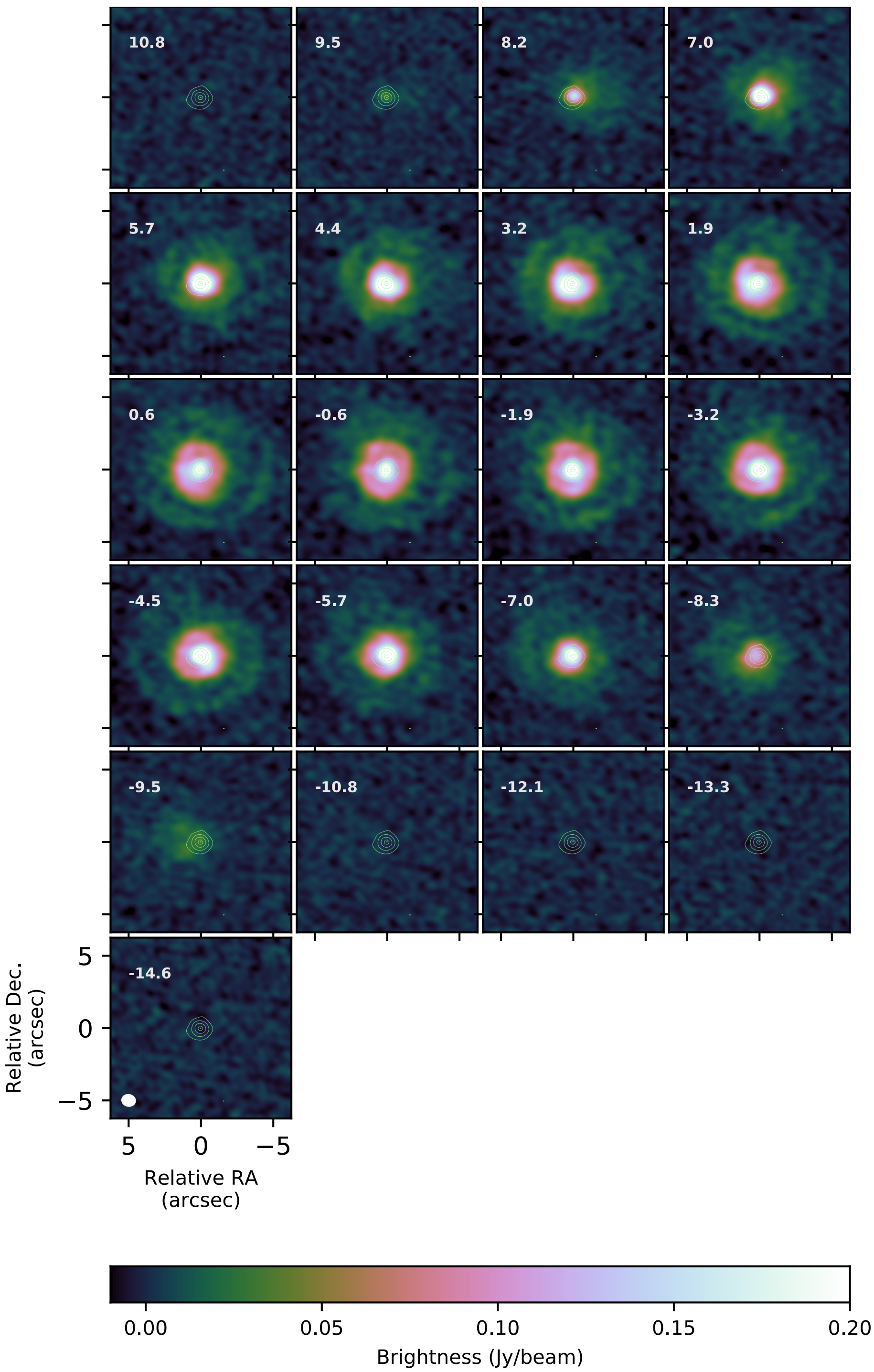

Fig. S29: **Low resolution $^{12}$CO J = 2 → 1 channel map of SV Aqr.** Same as Fig. S9, but for SV Aqr.

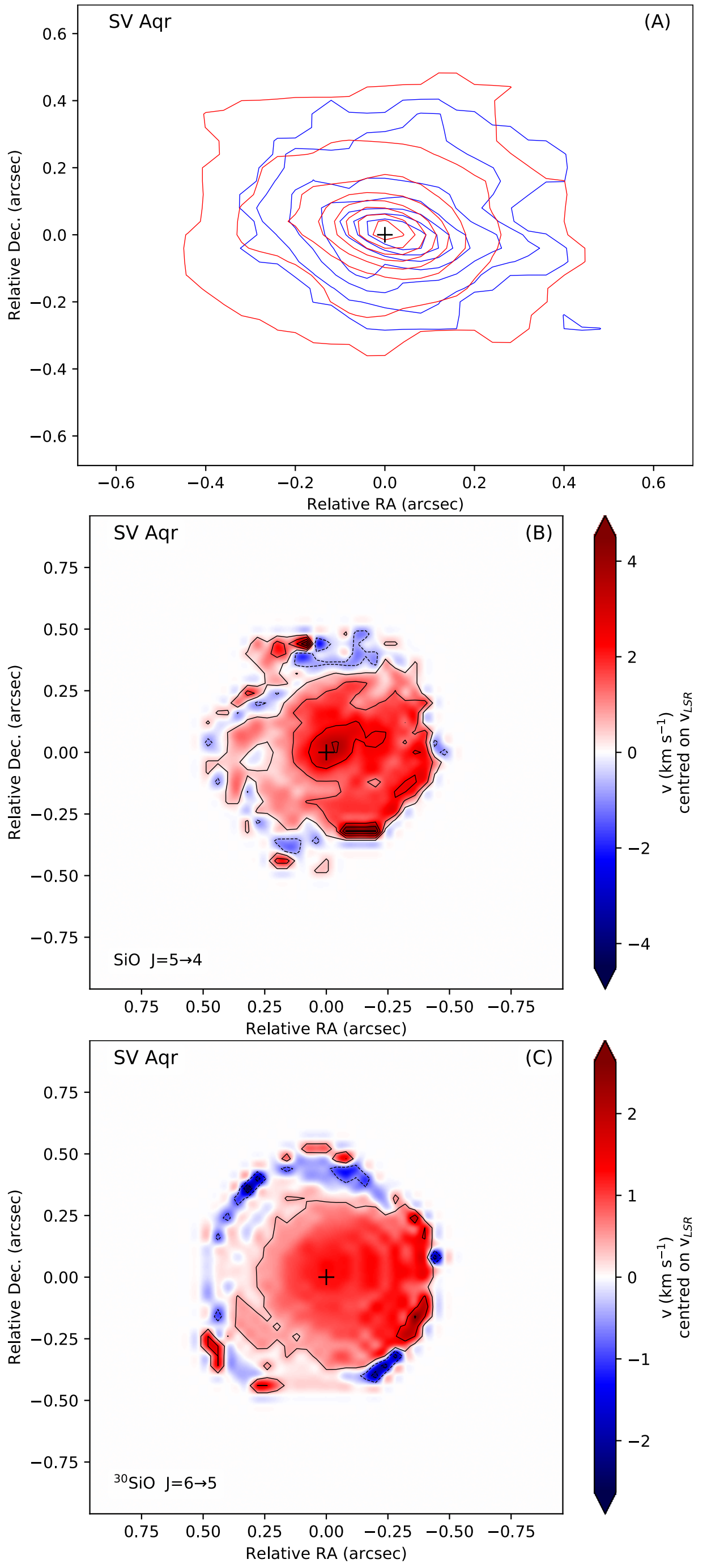

Fig. S30: **SiO stereogram and moment1-map of SV Aqr.** Same as Fig. S10, but for SV Aqr. Owing to the low signal-to-noise ratio of the SiO emission, these images should be interpreted with care (see also Sect. S3.2).

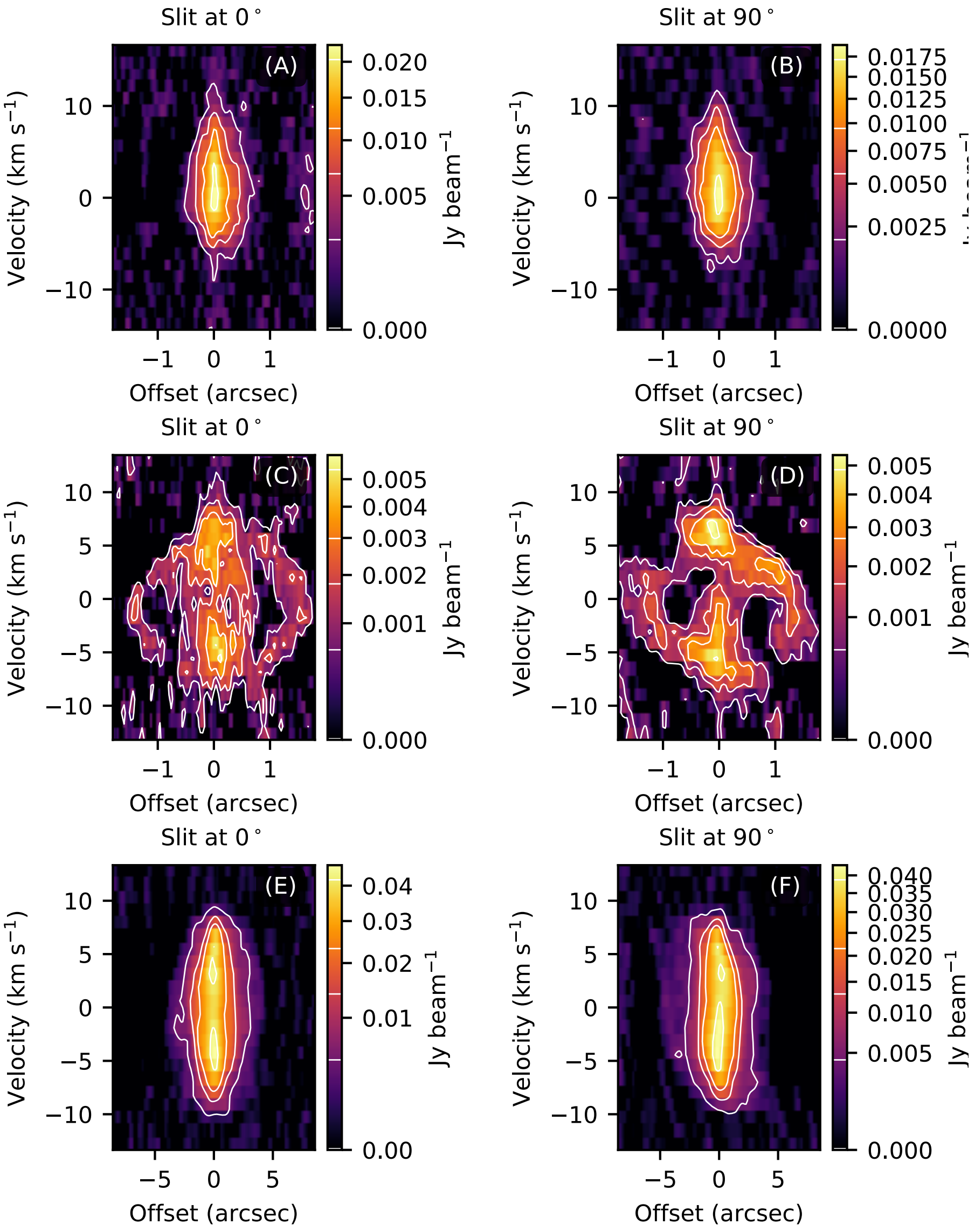

Fig. S31: **ALMA $^{12}$CO J = 2 → 1 and $^{28}$SiO J = 5 → 4 position-velocity (PV) diagram of SV Aqr.** Same as Fig. S11, but for SV Aqr.

## S7.7 R Hya

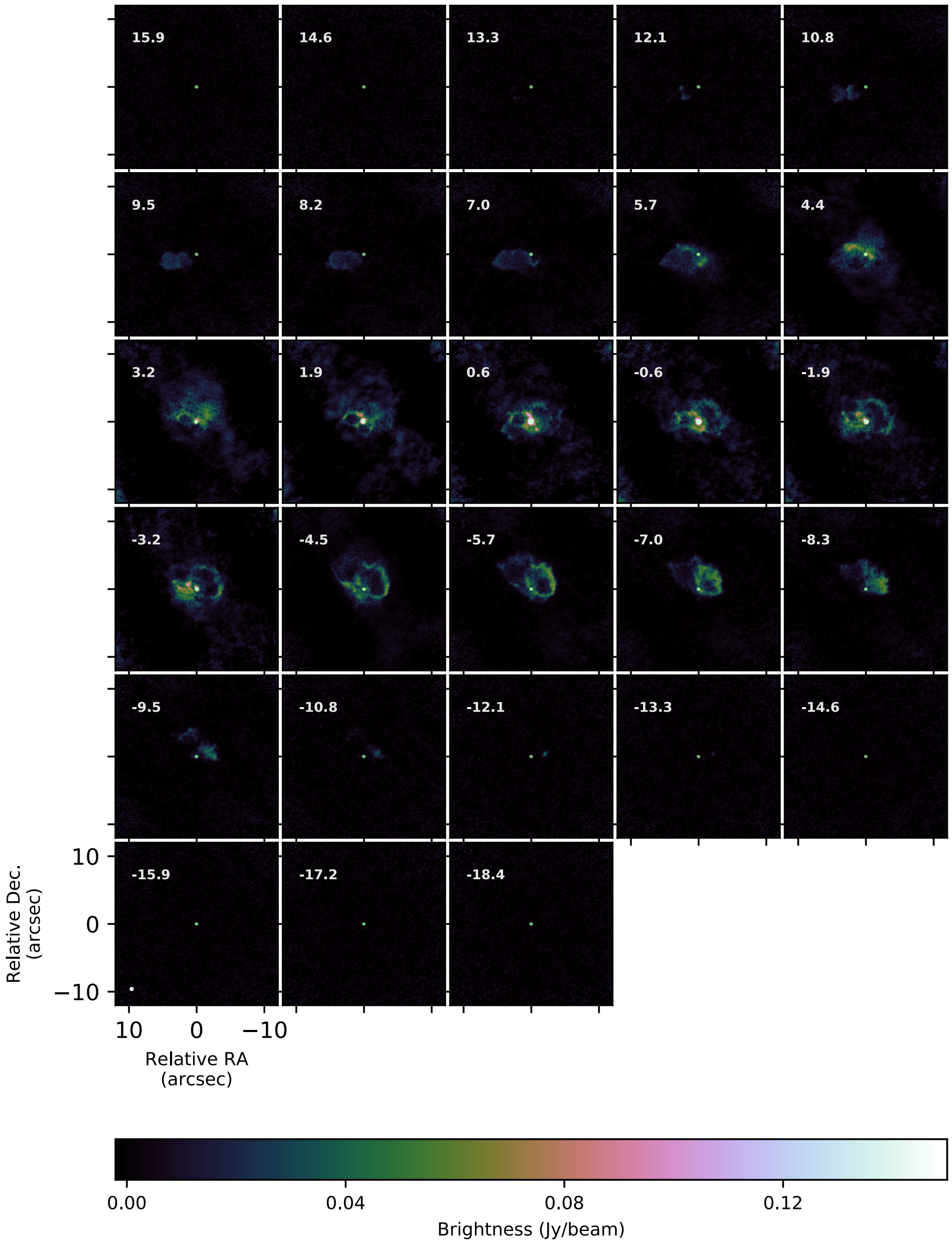

Fig. S32: **Medium resolution $^{12}$CO J = 2 → 1 channel map of R Hya.** Same as Fig. S8, but for R Hya.

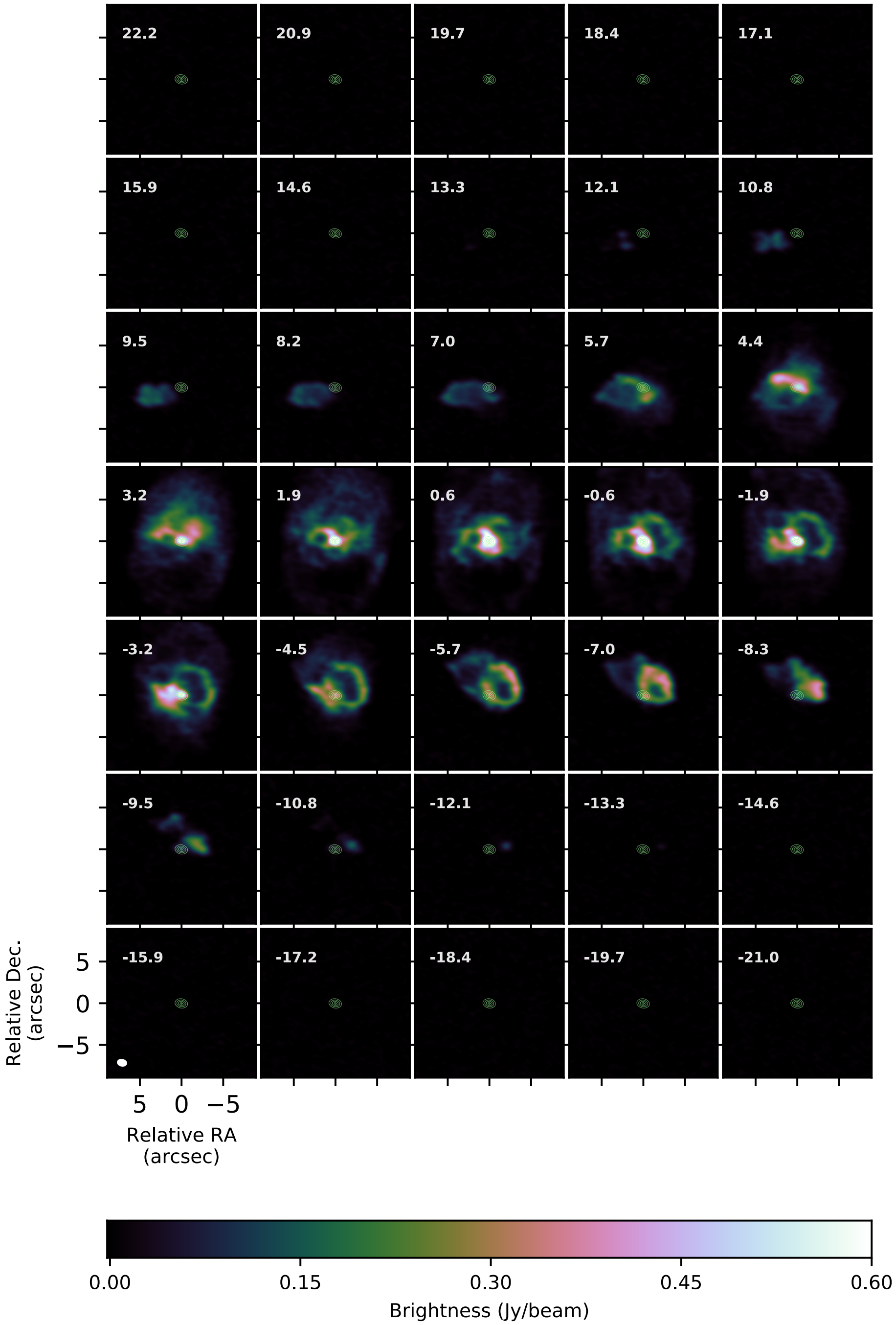

Fig. S33: **Low resolution $^{12}$CO J$=2\rightarrow1$ channel map of R Hya.** Same as Fig. S9, but for R Hya.

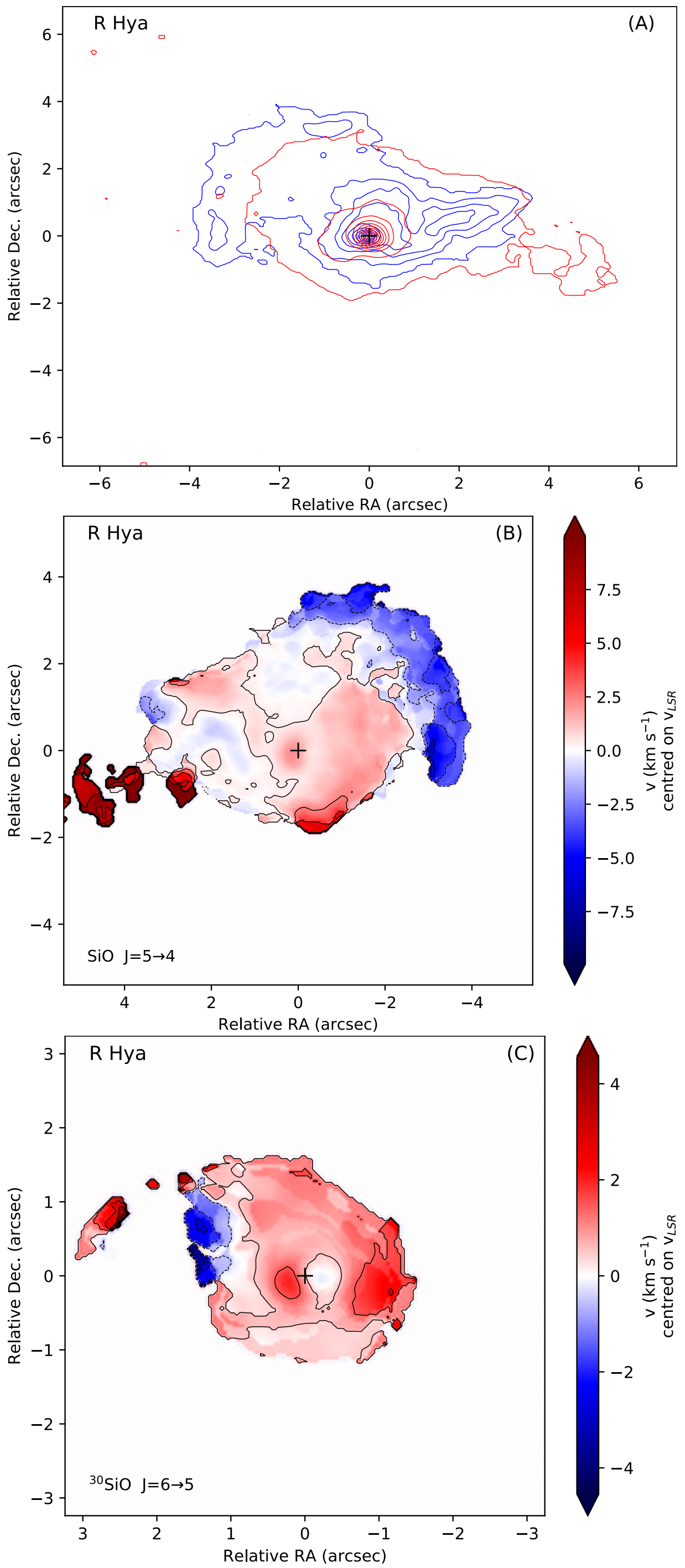

Fig. S34: **SiO stereogram and moment1-map of R Hya.** Same as Fig. S10, but for R Hya.

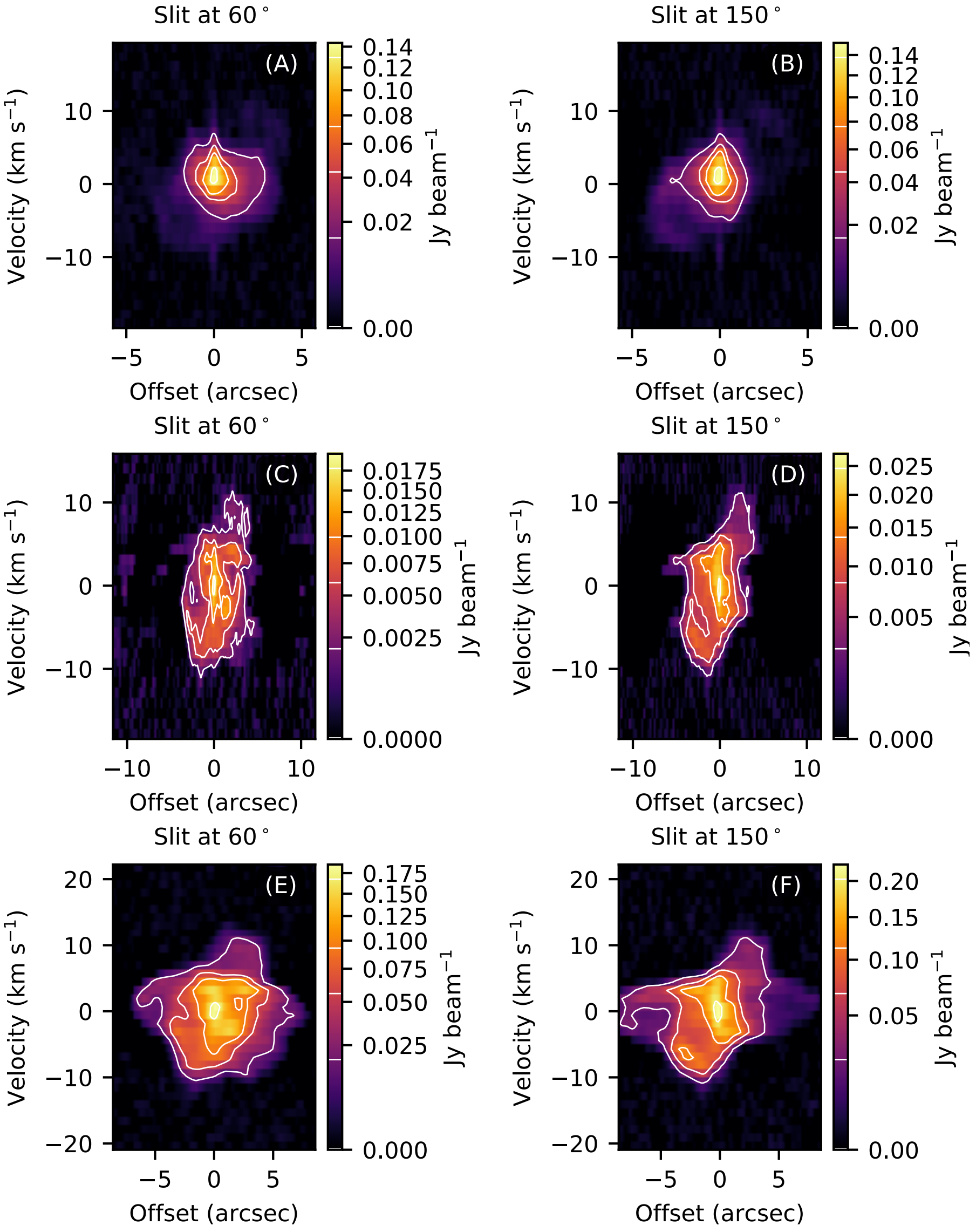

Fig. S35: **ALMA $^{12}$CO J = 2 → 1 and $^{28}$SiO J = 5 → 4 position-velocity (PV) diagram of R Hya.** Same as Fig. S11, but for R Hya.

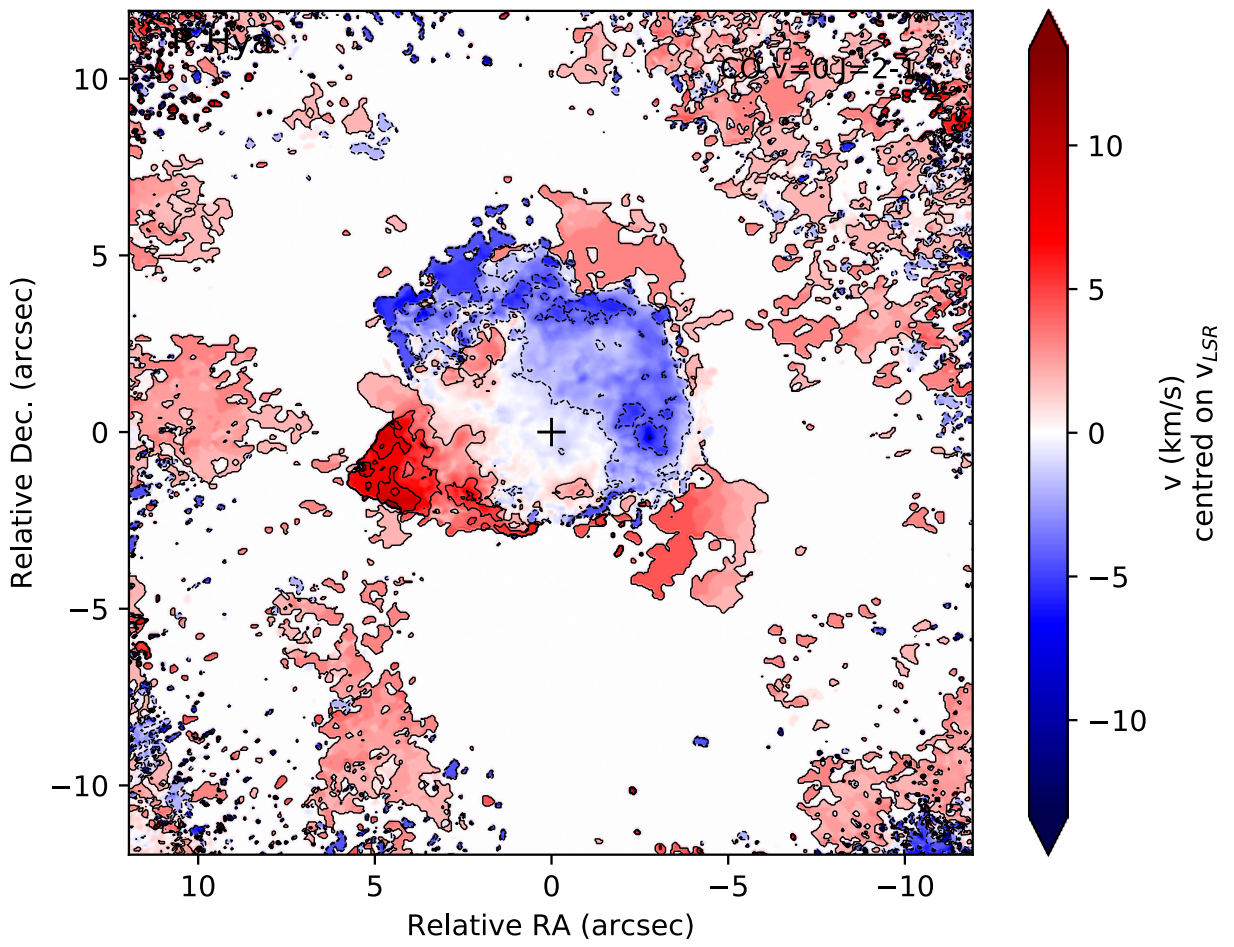

Fig. S36: **ALMA $^{12}$CO J = 2 → 1 moment1-maps of R Hya.**

### S7.8 U Her

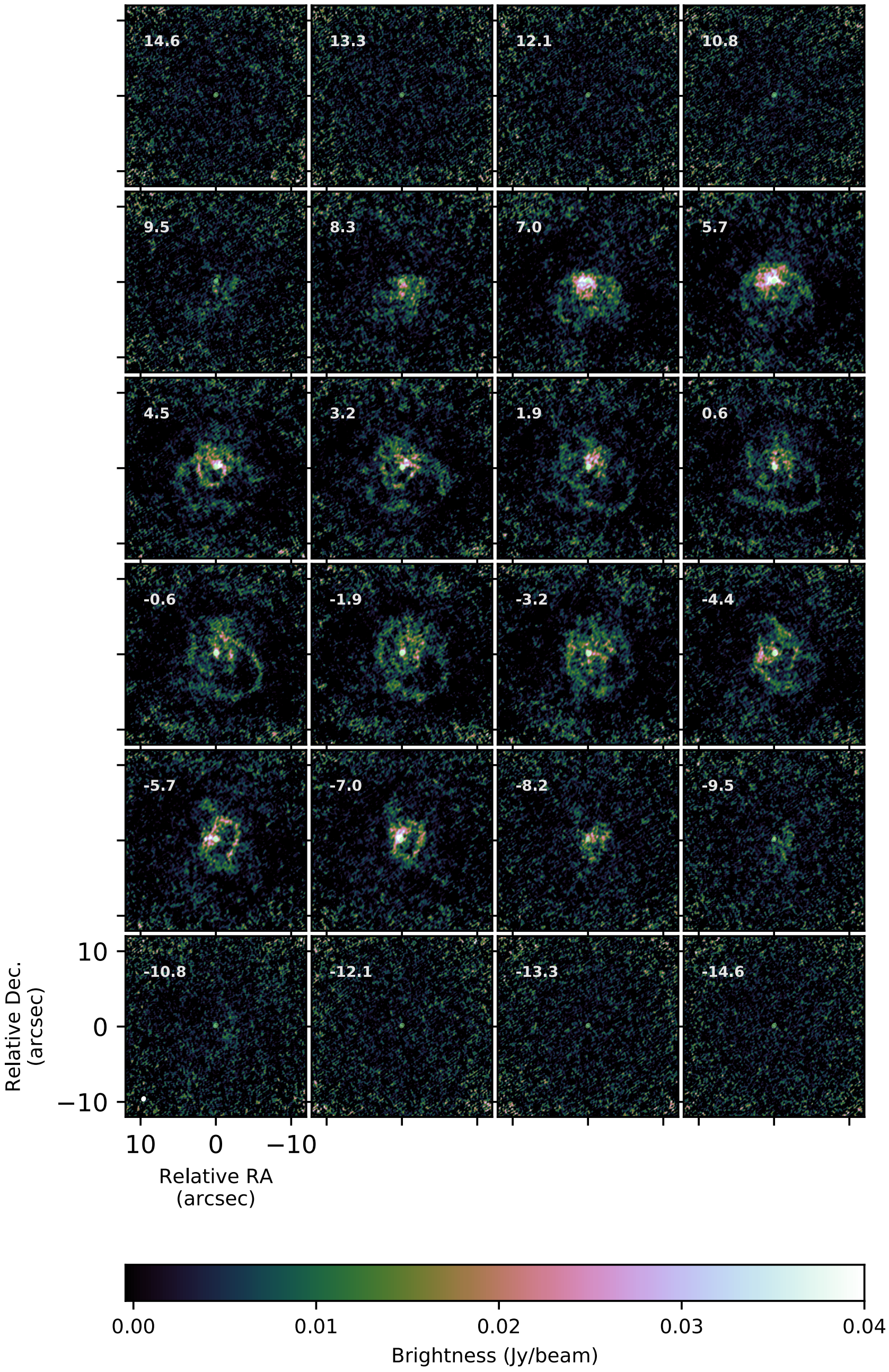

Fig. S37: **Medium resolution $^{12}$CO J = 2 → 1 channel map of U Her.** Same as Fig. S8, but for U Her.

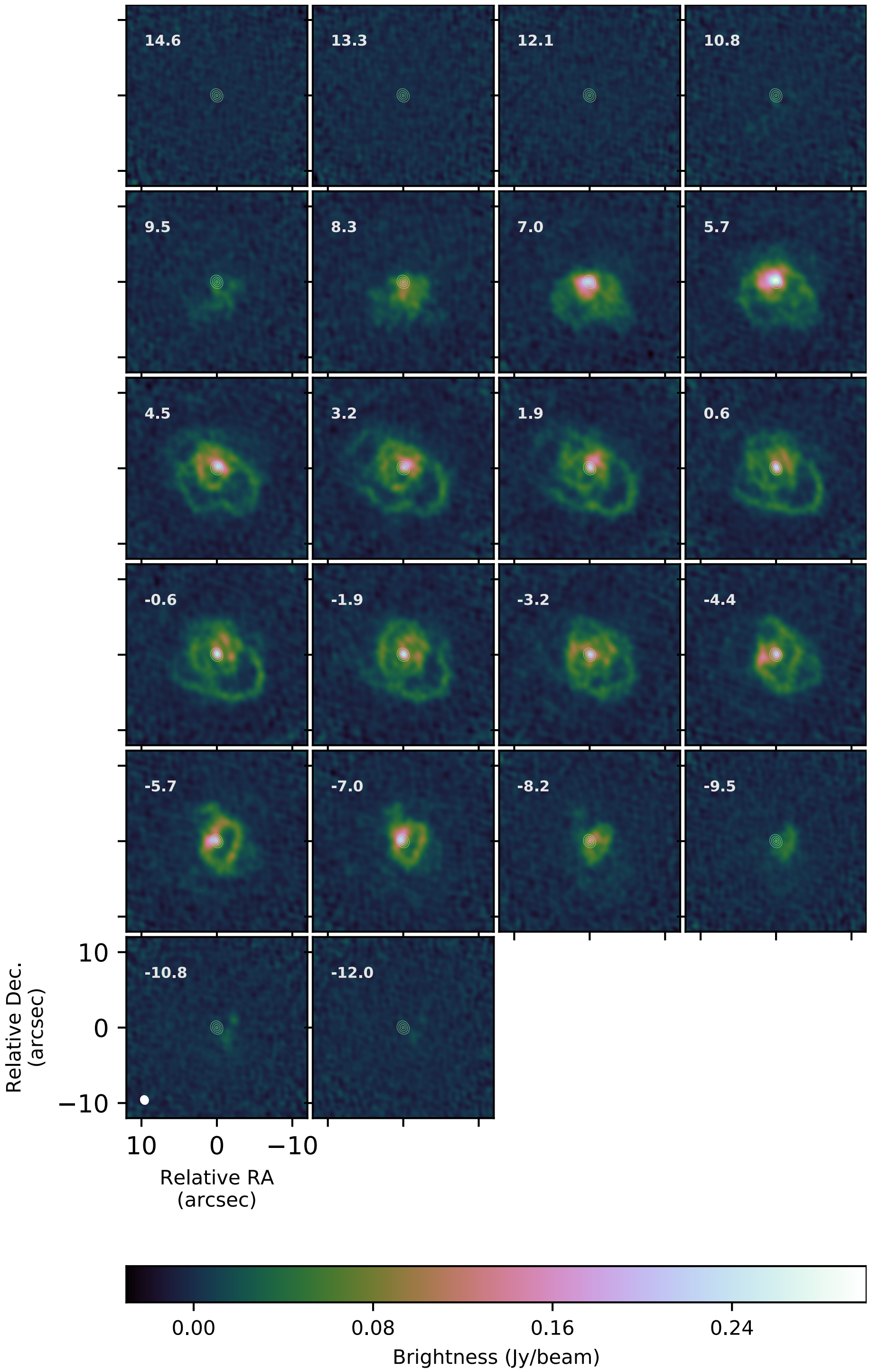

Fig. S38: **Low resolution $^{12}$CO J = 2 → 1 channel map of U Her.** Same as Fig. S9, but for U Her.

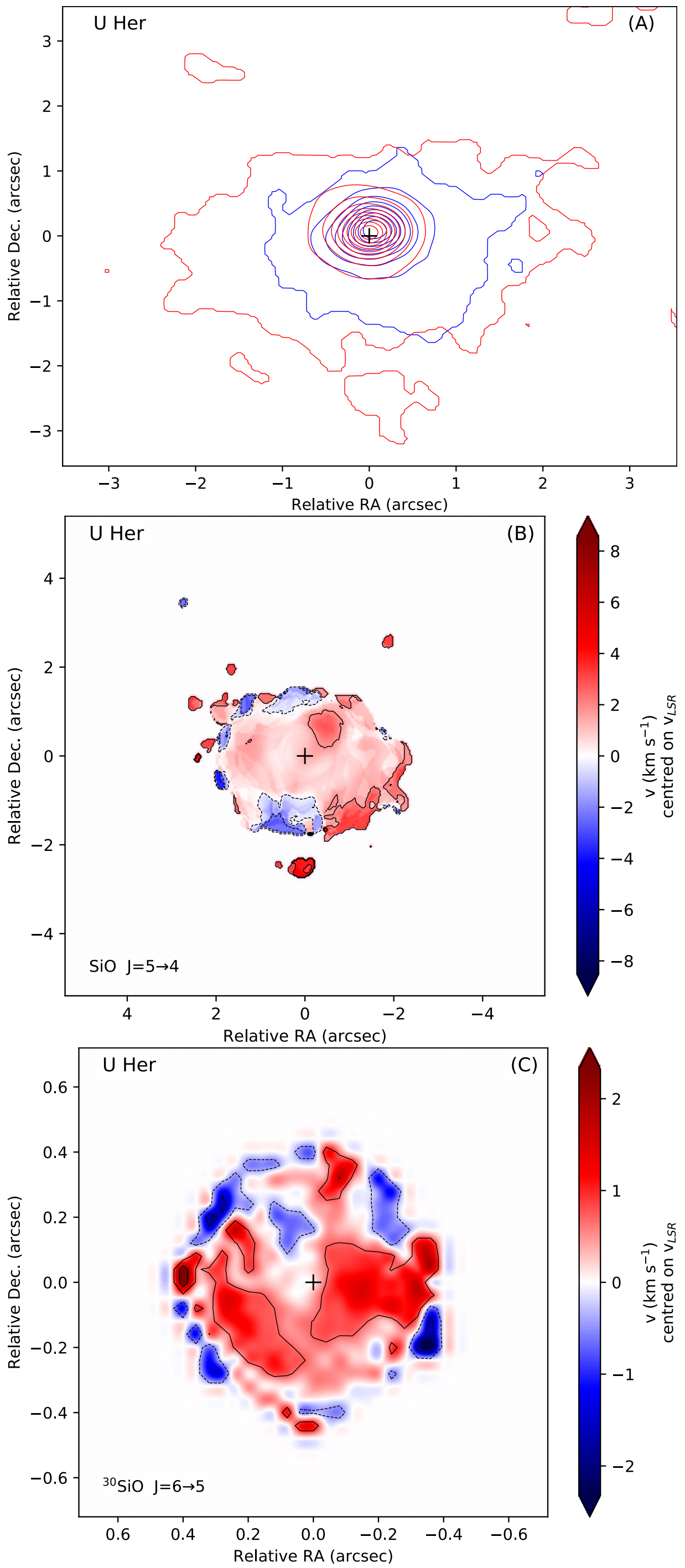

Fig. S39: **SiO stereogram and moment1-map of U Her.** Same as Fig. S10, but for U Her.

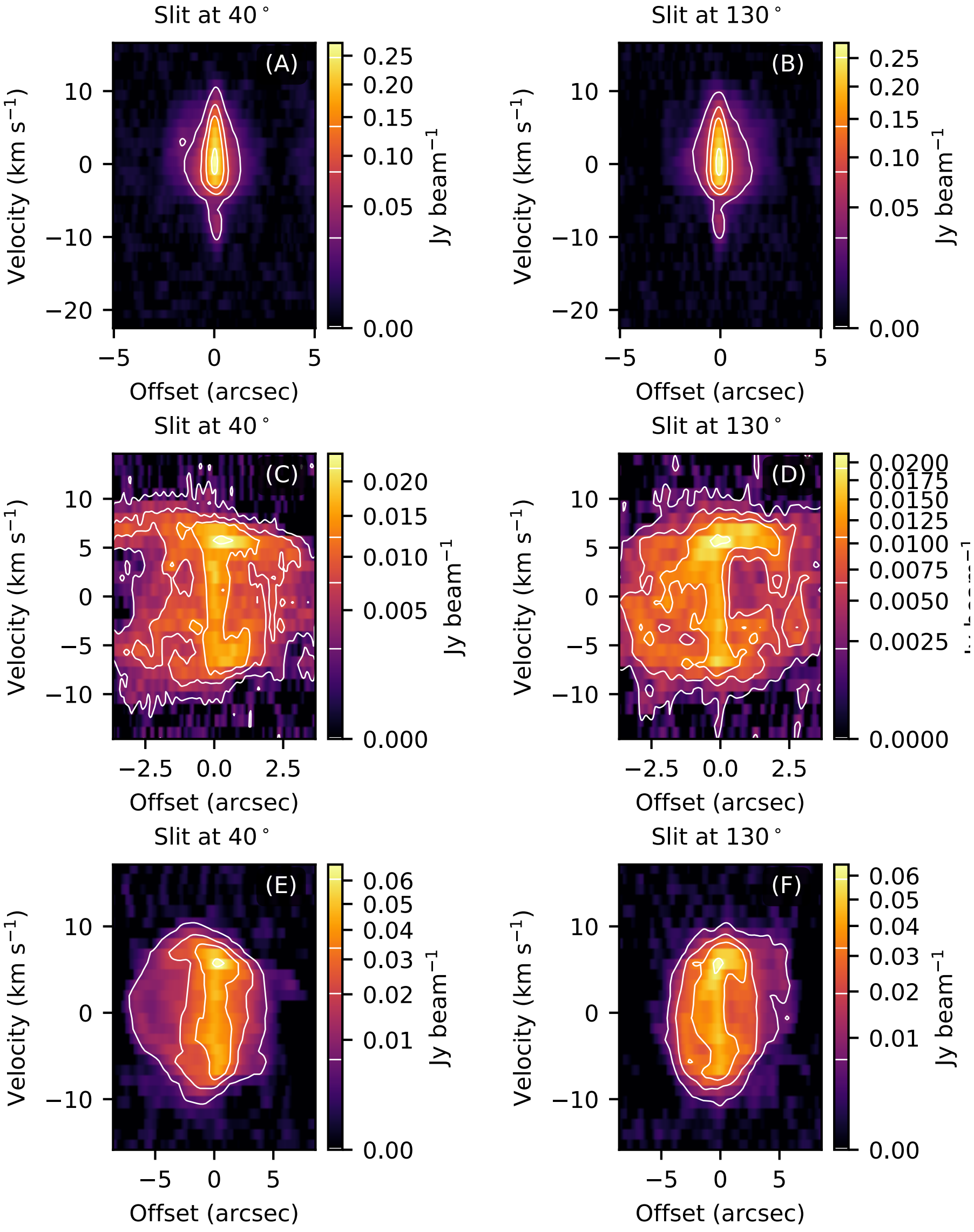

Fig. S40: **ALMA $^{12}$CO J = 2 → 1 and $^{28}$SiO J = 5 → 4 position-velocity (PV) diagram of U Her.** Same as Fig. S11, but for U Her.

## S7.9 $\pi^1$ Gru

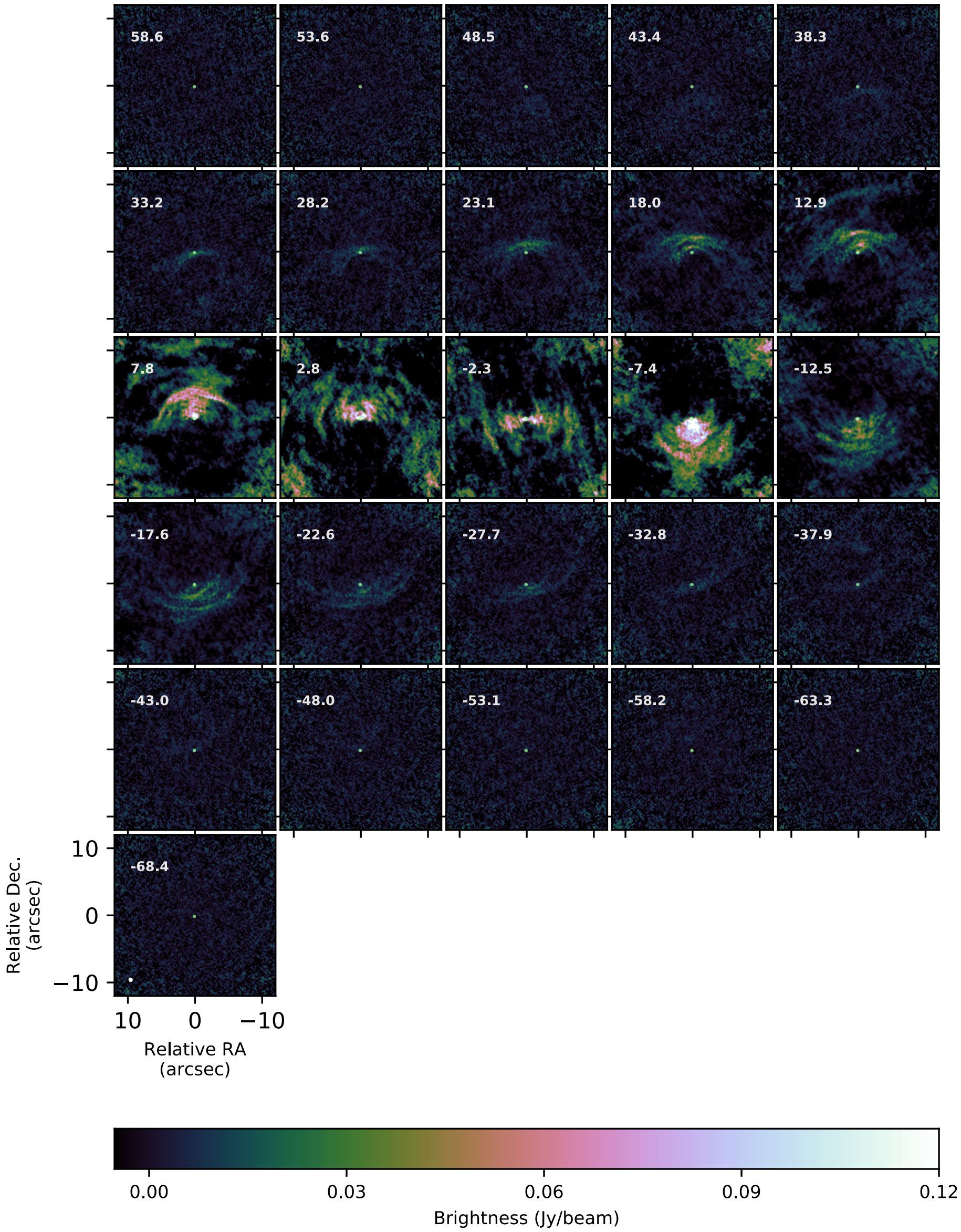

Fig. S41: **Medium resolution $^{12}$CO J = 2 → 1 channel map of $\pi^1$ Gru.** Same as Fig. S8, but for $\pi^1$ Gru. Only 1/4 of the channel map is shown, the velocity resolution is $\sim$1.25 km s$^{-1}$.

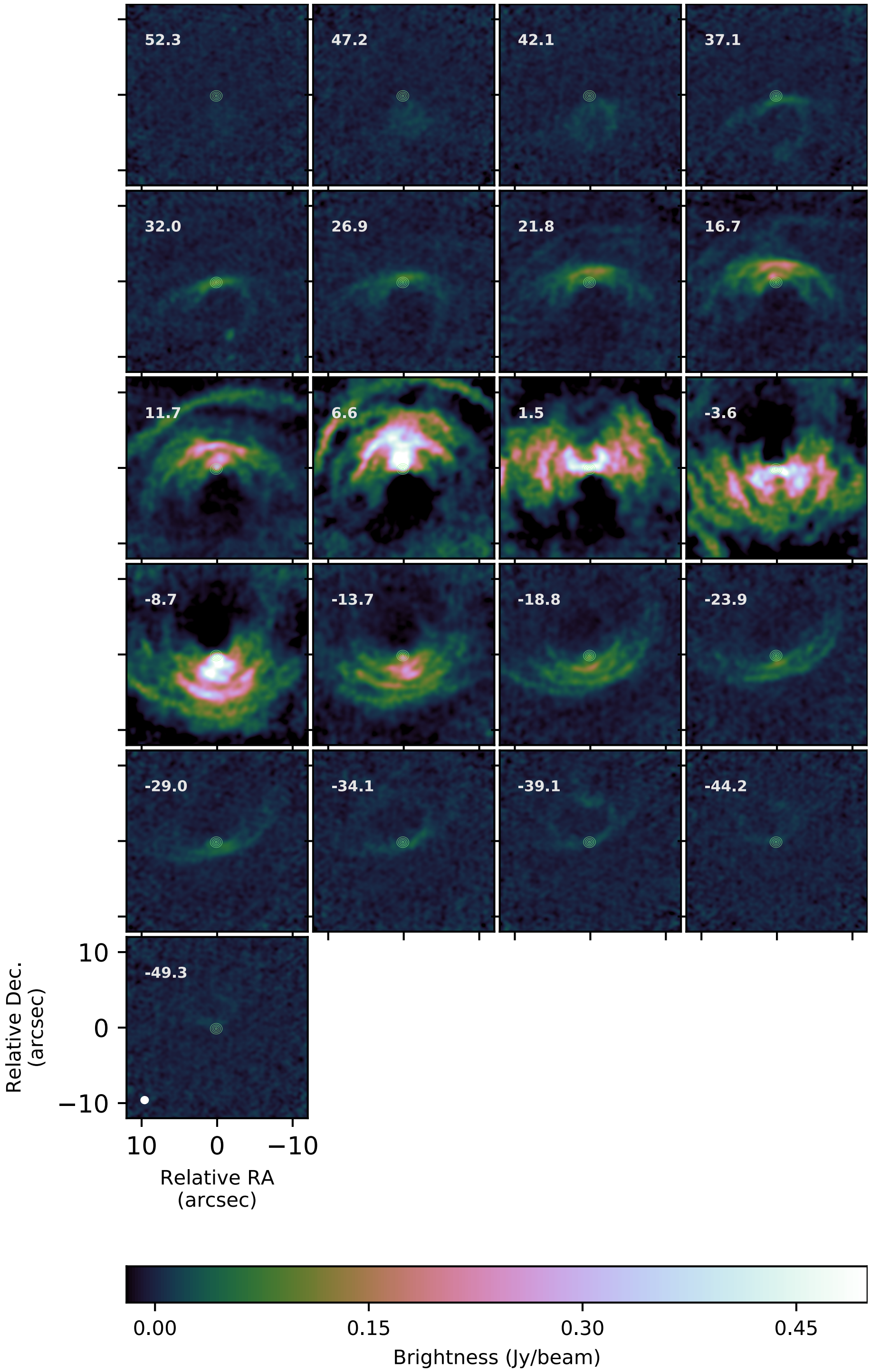

Fig. S42: **Low resolution $^{12}$CO J = 2 → 1 channel map of $\pi^1$ Gru.** Same as Fig. S9, but for $\pi^1$ Gru. Only 1/4 of the channel map is shown, the velocity resolution is $\sim 1.25\,\mathrm{km\,s^{-1}}$.

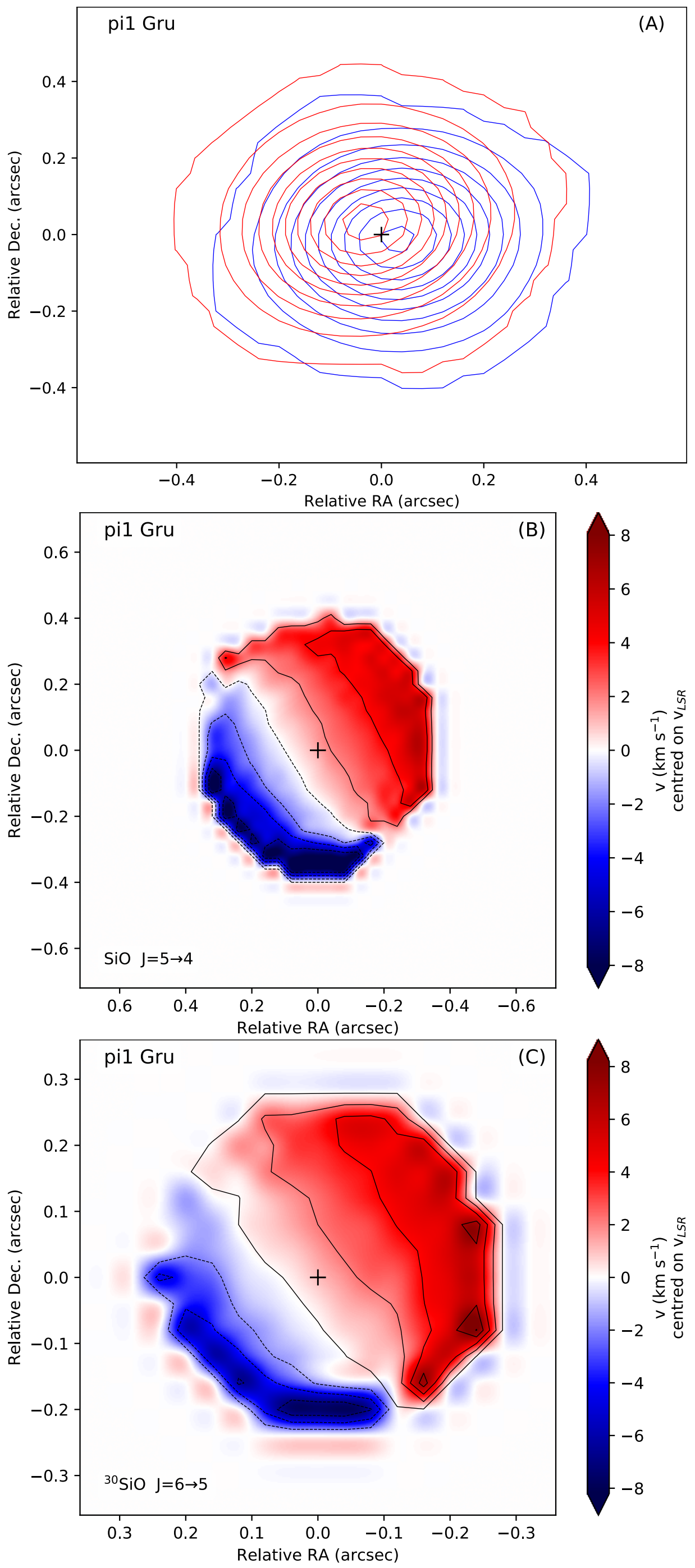

Fig. S43: **SiO stereogram and moment1-map of $\pi^1$ Gru.** Same as Fig. S10, but for $\pi^1$ Gru.

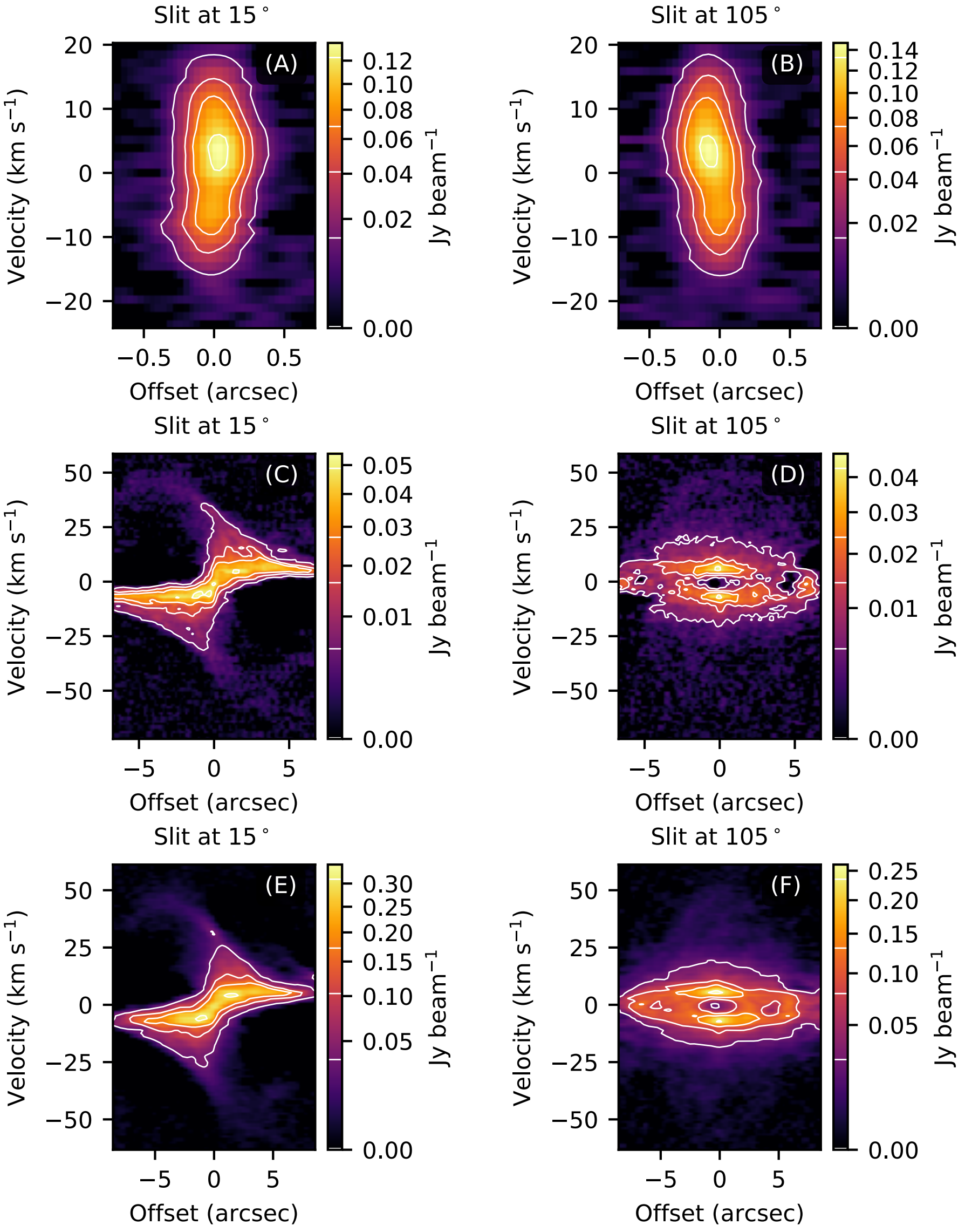

Fig. S44: **ALMA $^{12}$CO J = 2 → 1 and $^{28}$SiO J = 5 → 4 position-velocity (PV) diagram of $\pi^1$ Gru.** Same as Fig. S11, but for $\pi^1$ Gru.

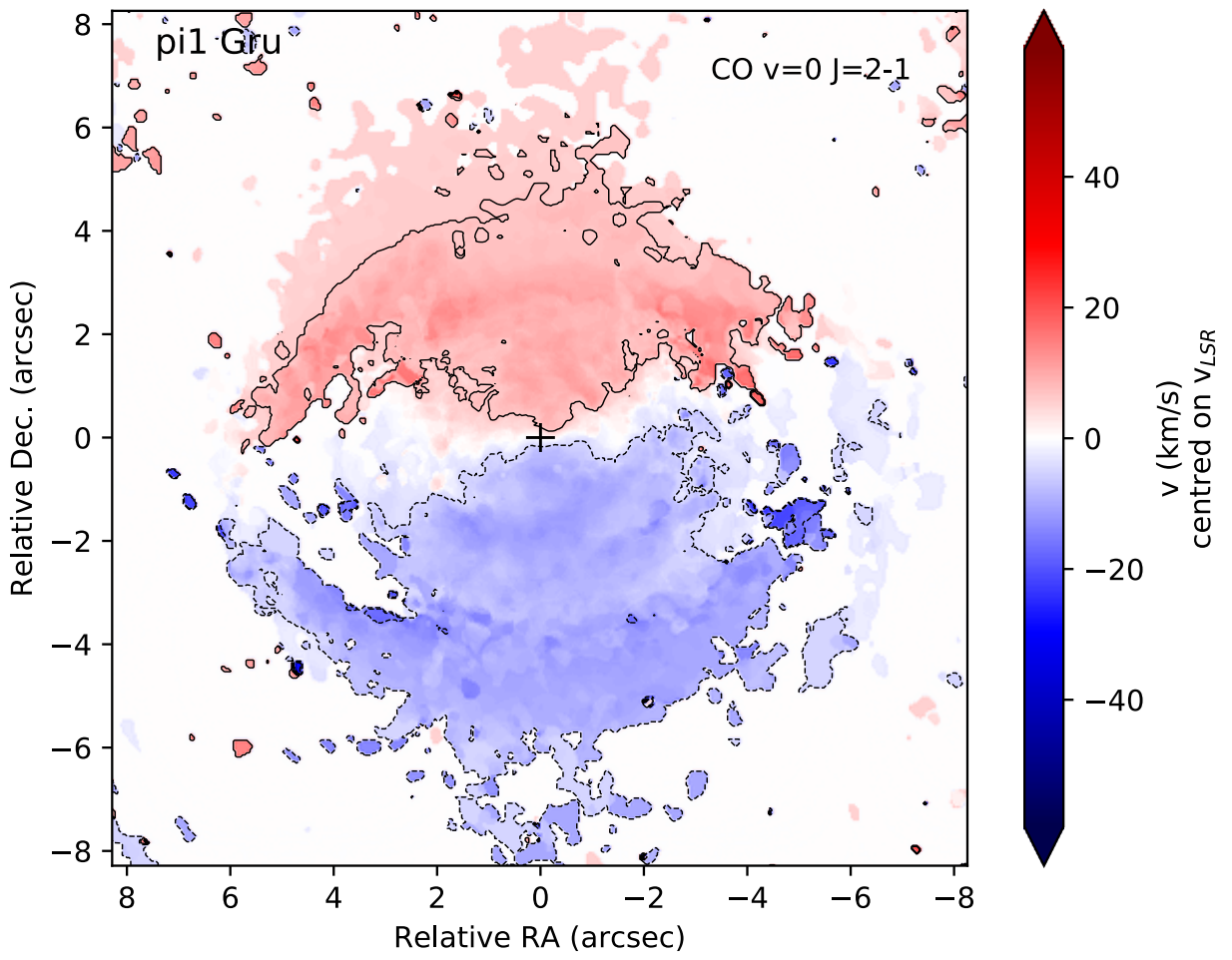

Fig. S45: **ALMA $^{12}$CO J = 2 → 1 moment1-maps of $\pi^1$ Gru.**

## S7.10 R Aql

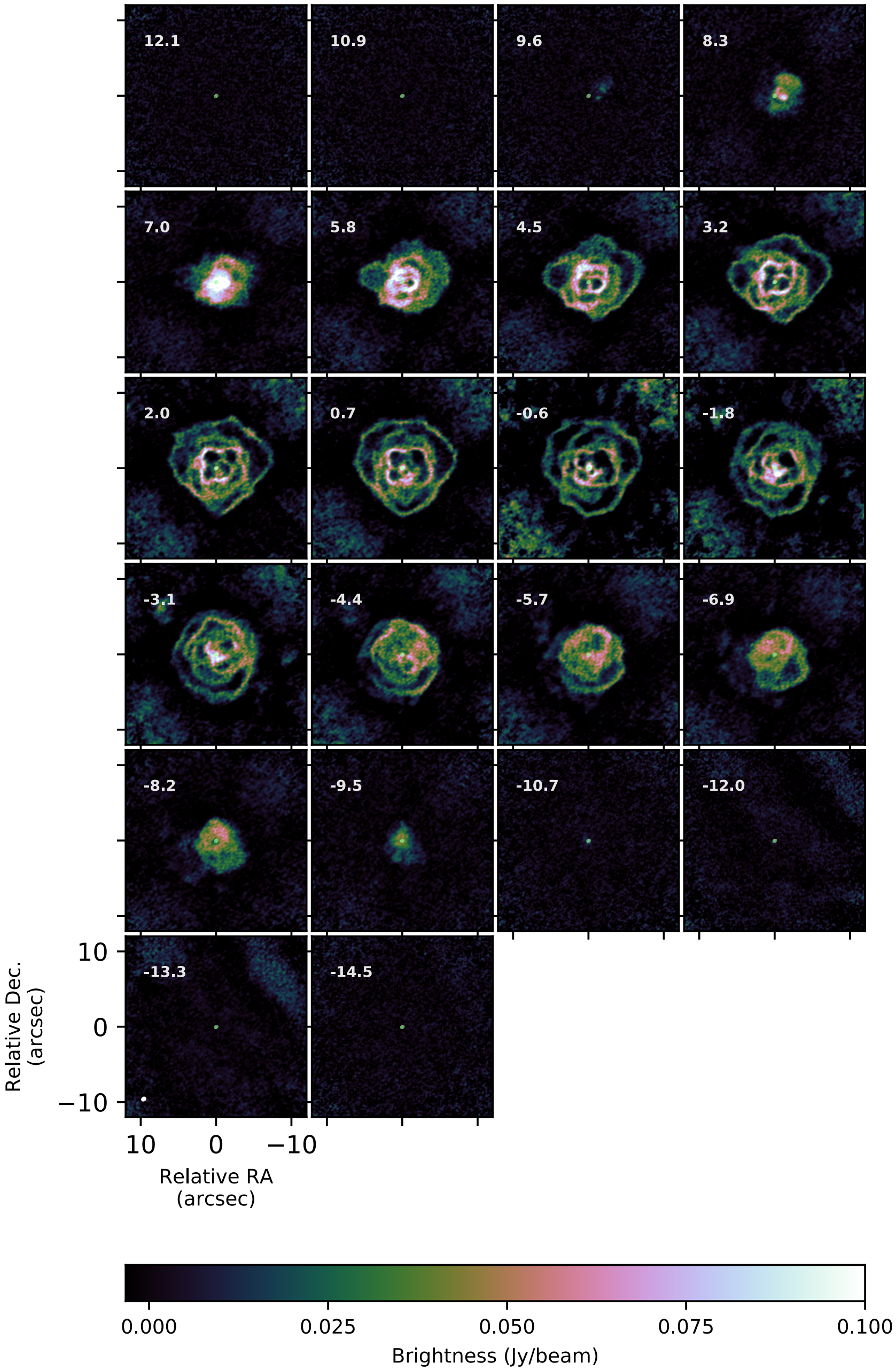

Fig. S46: **Medium resolution $^{12}$CO J = 2 → 1 channel map of R Aql.** Same as Fig. S8, but for R Aql.

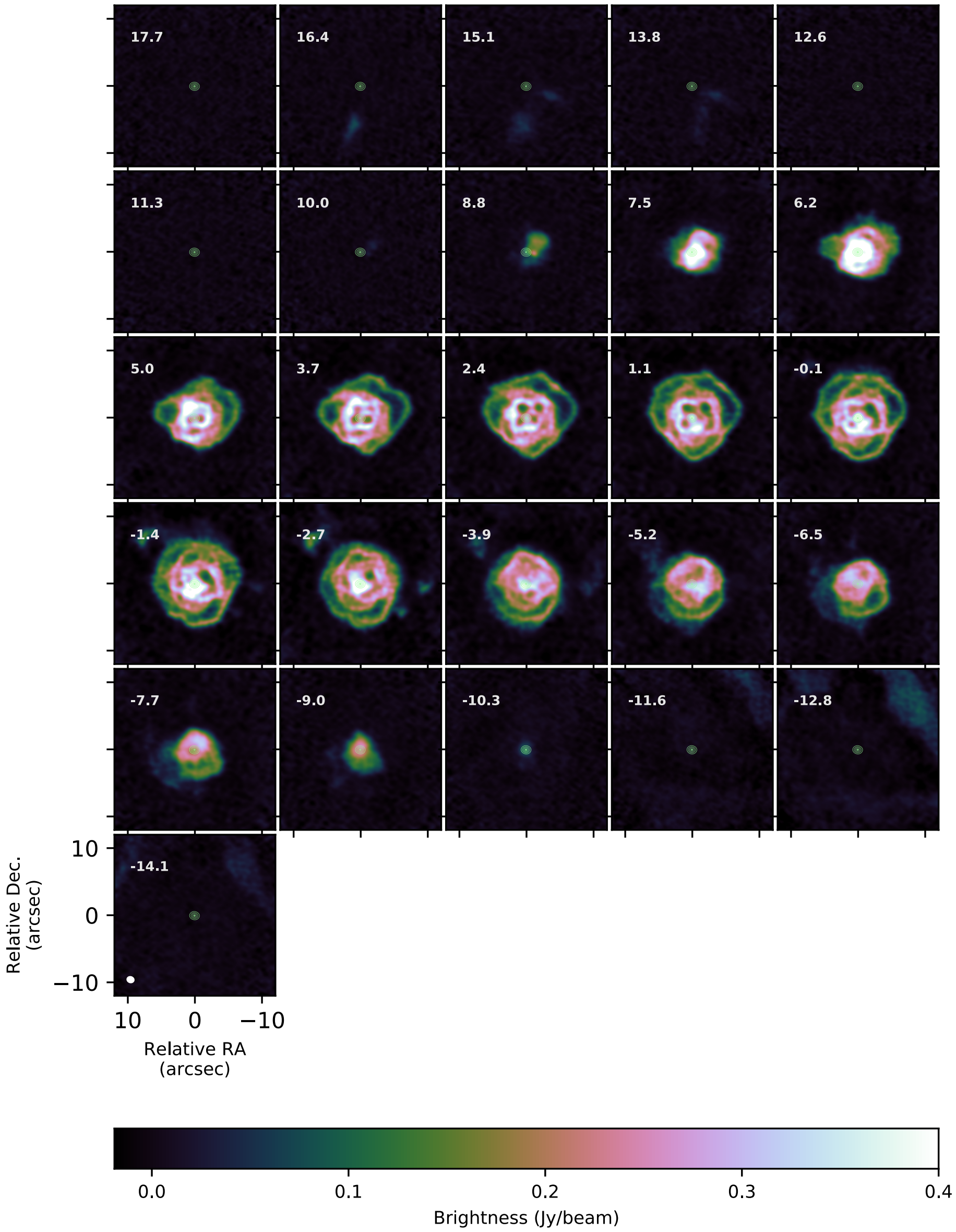

Fig. S47: **Low resolution $^{12}$CO J = 2 → 1 channel map of R Aql.** Same as Fig. S9, but for R Aql.

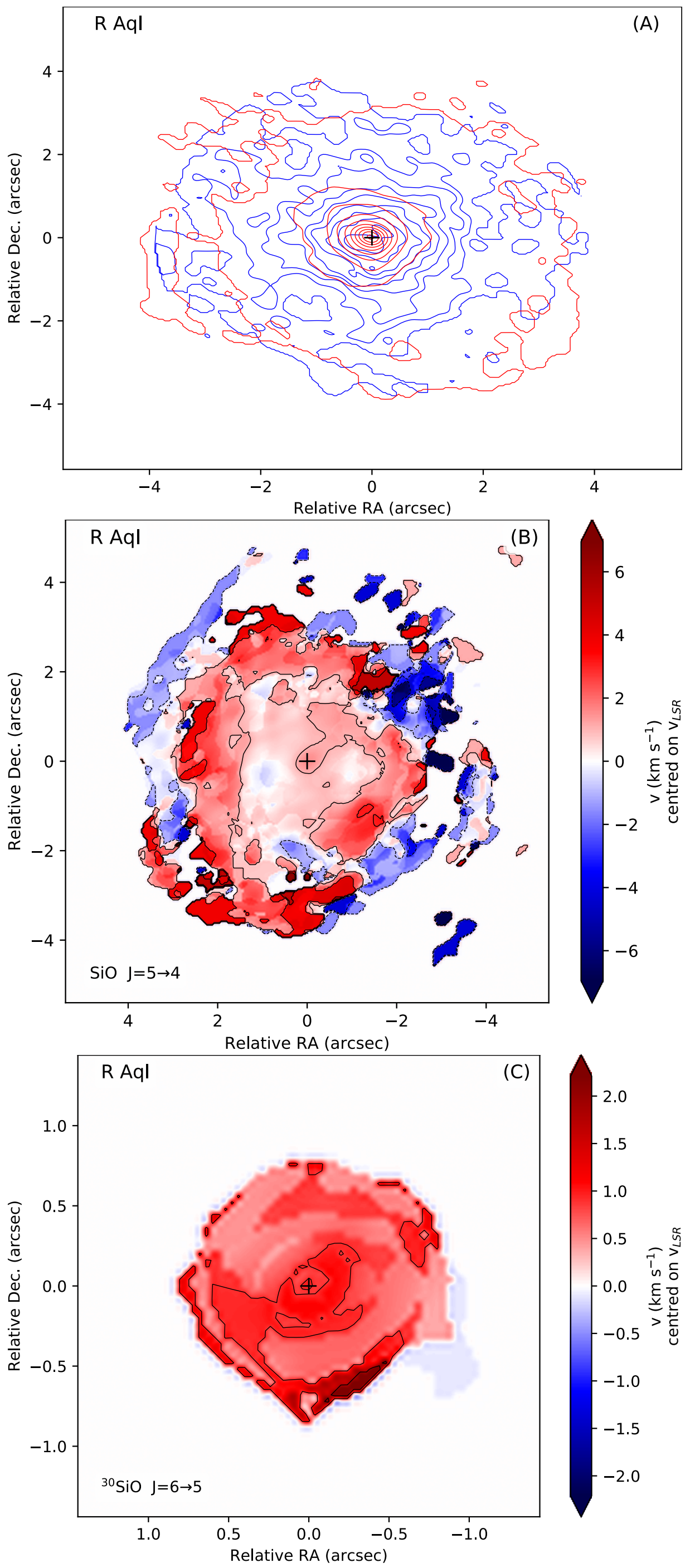

Fig. S48: **SiO stereogram and moment1-map of R Aql.** Same as Fig. S10, but for R Aql.

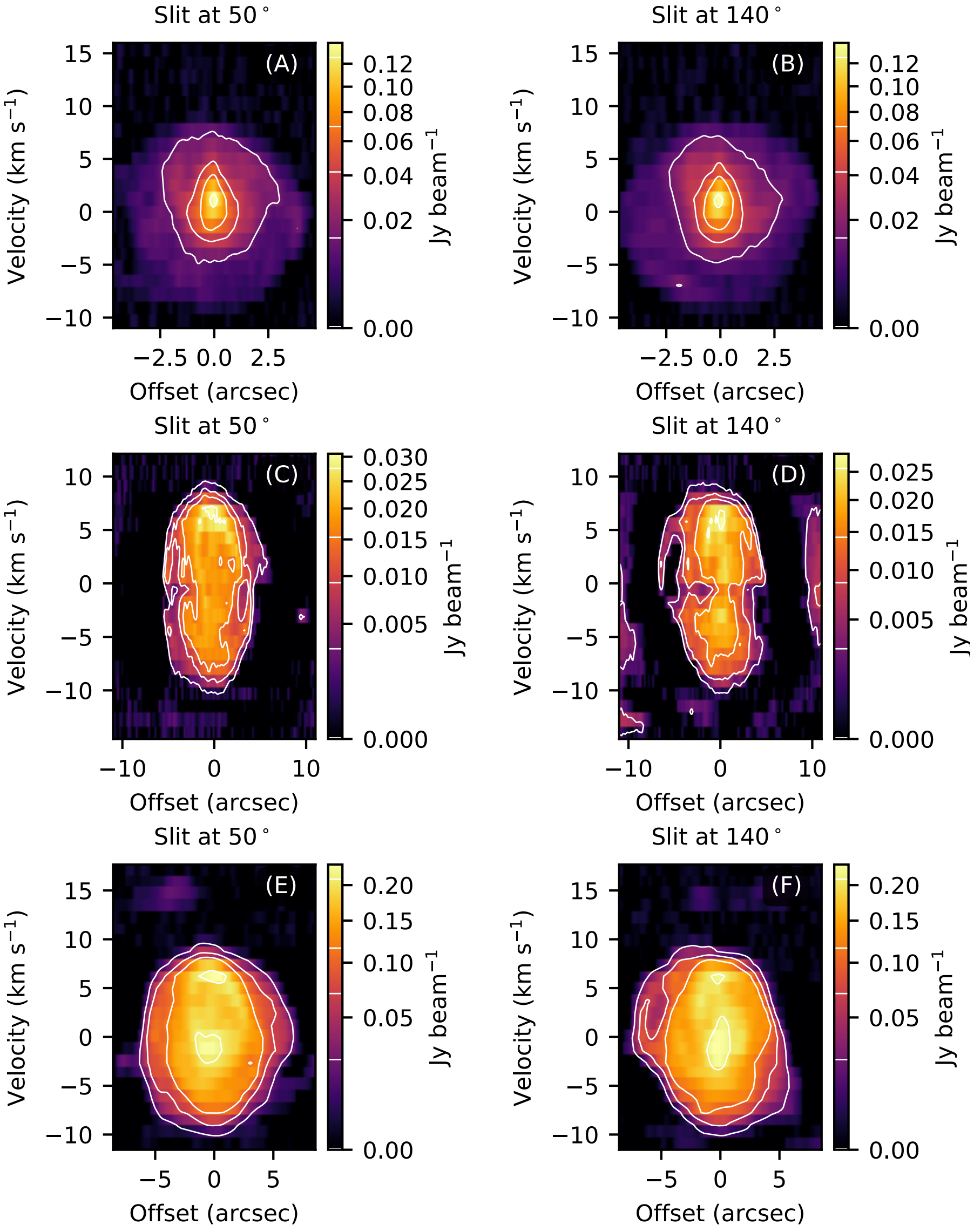

Fig. S49: **ALMA $^{12}$CO J$=2\rightarrow1$ and $^{28}$SiO J$=5\rightarrow4$ position-velocity (PV) diagram of R Aql.** Same as Fig. S11, but for R Aql.

## S7.11 W Aql

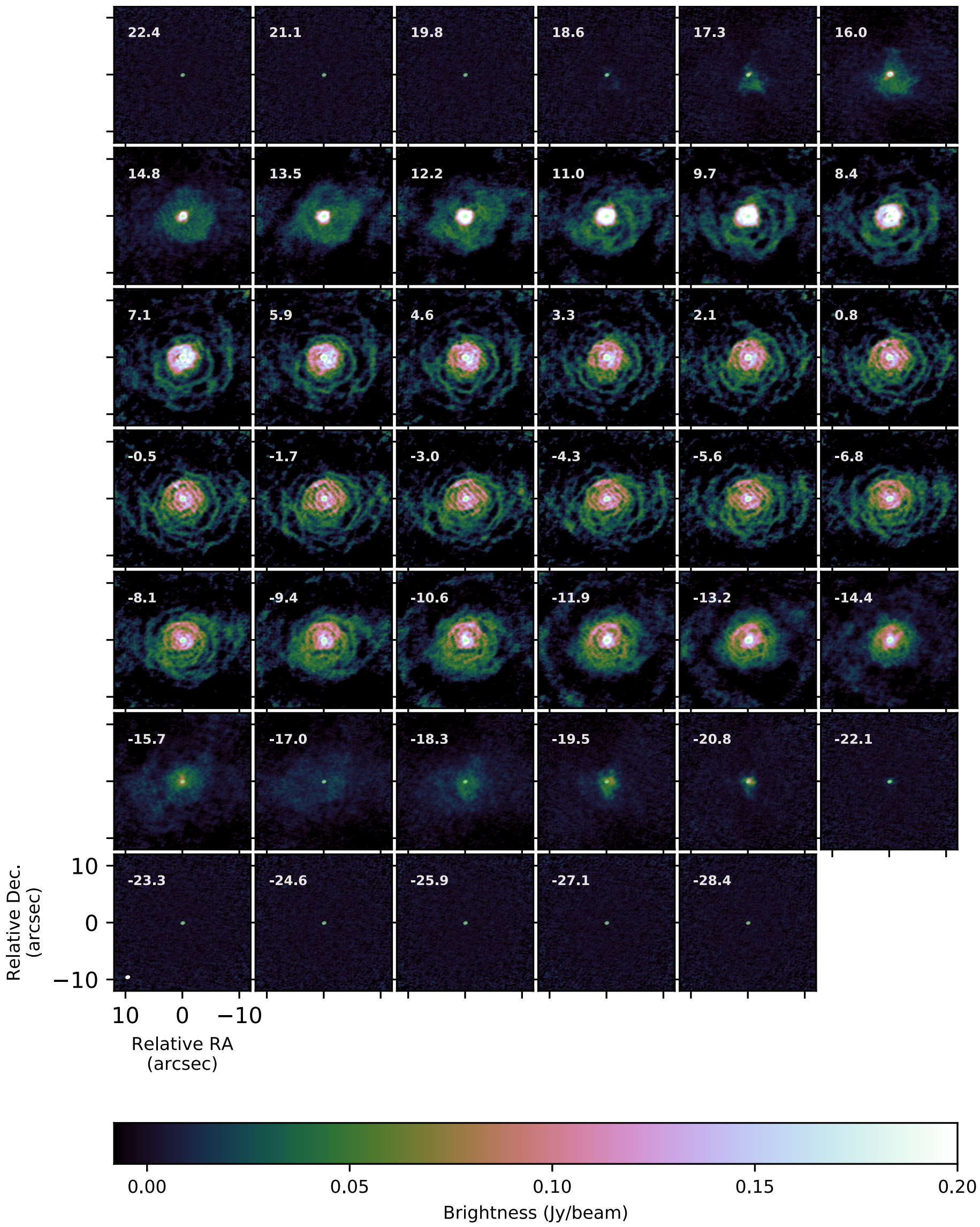

Fig. S50: **Medium resolution $^{12}$CO J = 2 → 1 channel map of W Aql.** Same as Fig. S8, but for W Aql.

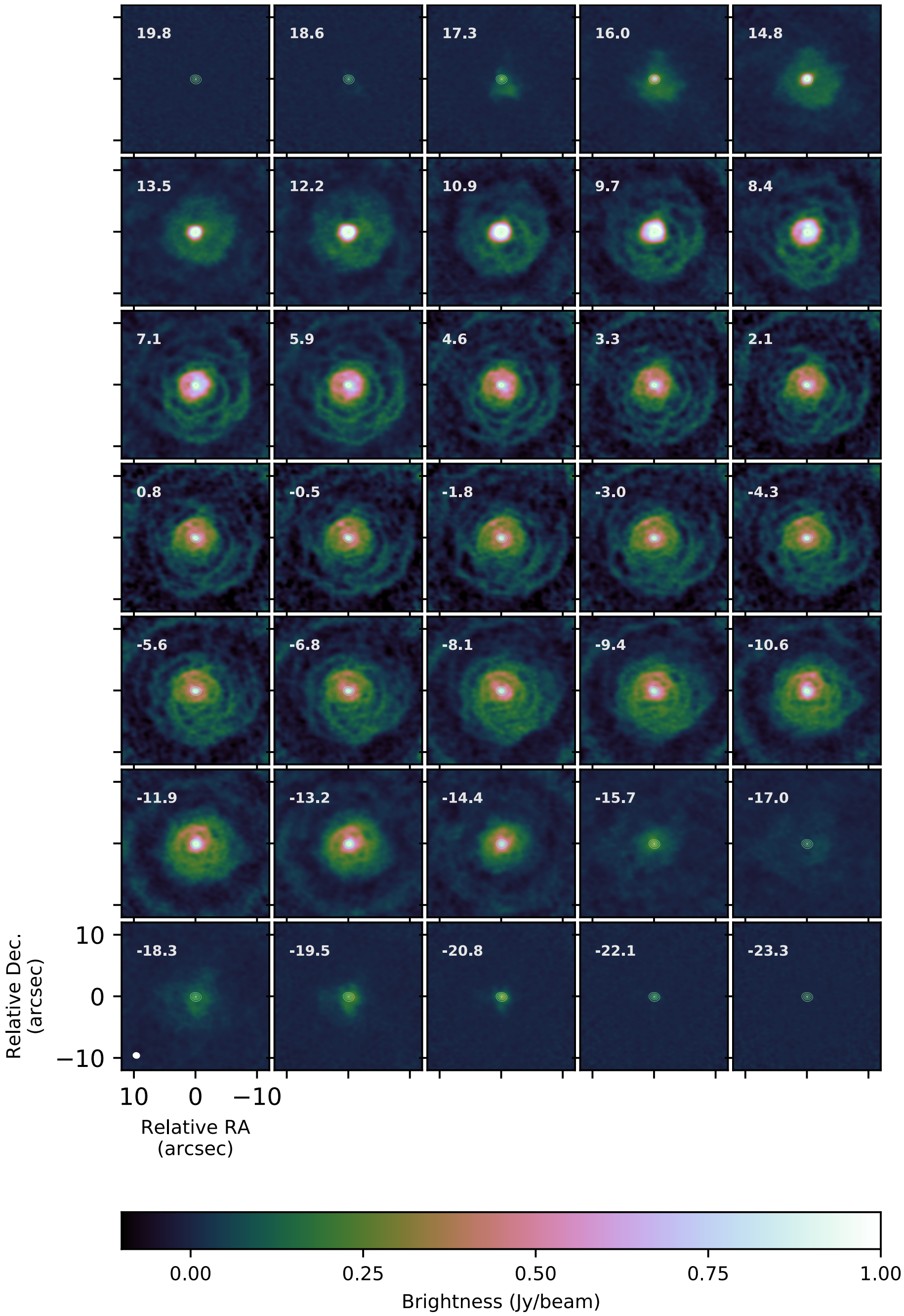

Fig. S51: **Low resolution $^{12}$CO J$=2\rightarrow1$ channel map of W Aql.** Same as Fig. S9, but for W Aql.

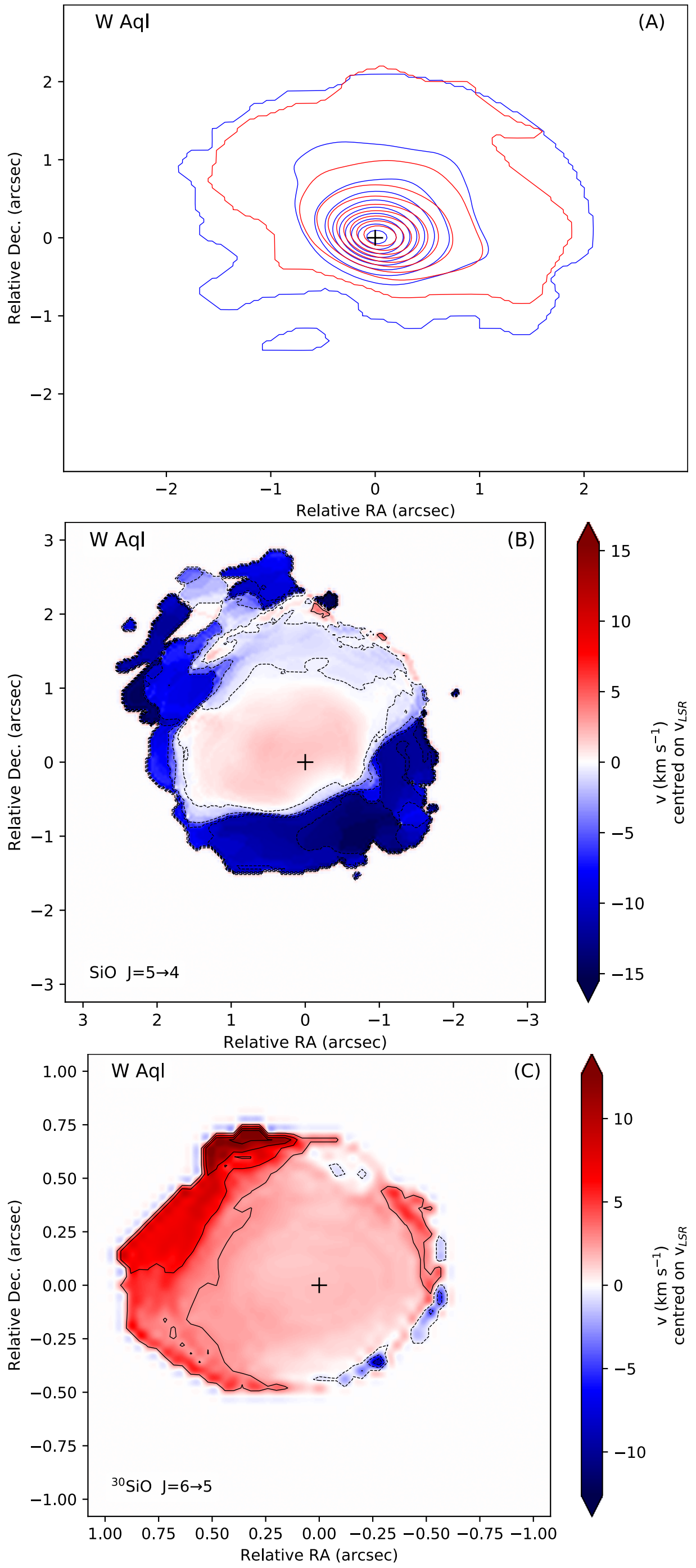

Fig. S52: **SiO stereogram and moment1-map of W Aql.** Same as Fig. S10, but for W Aql.

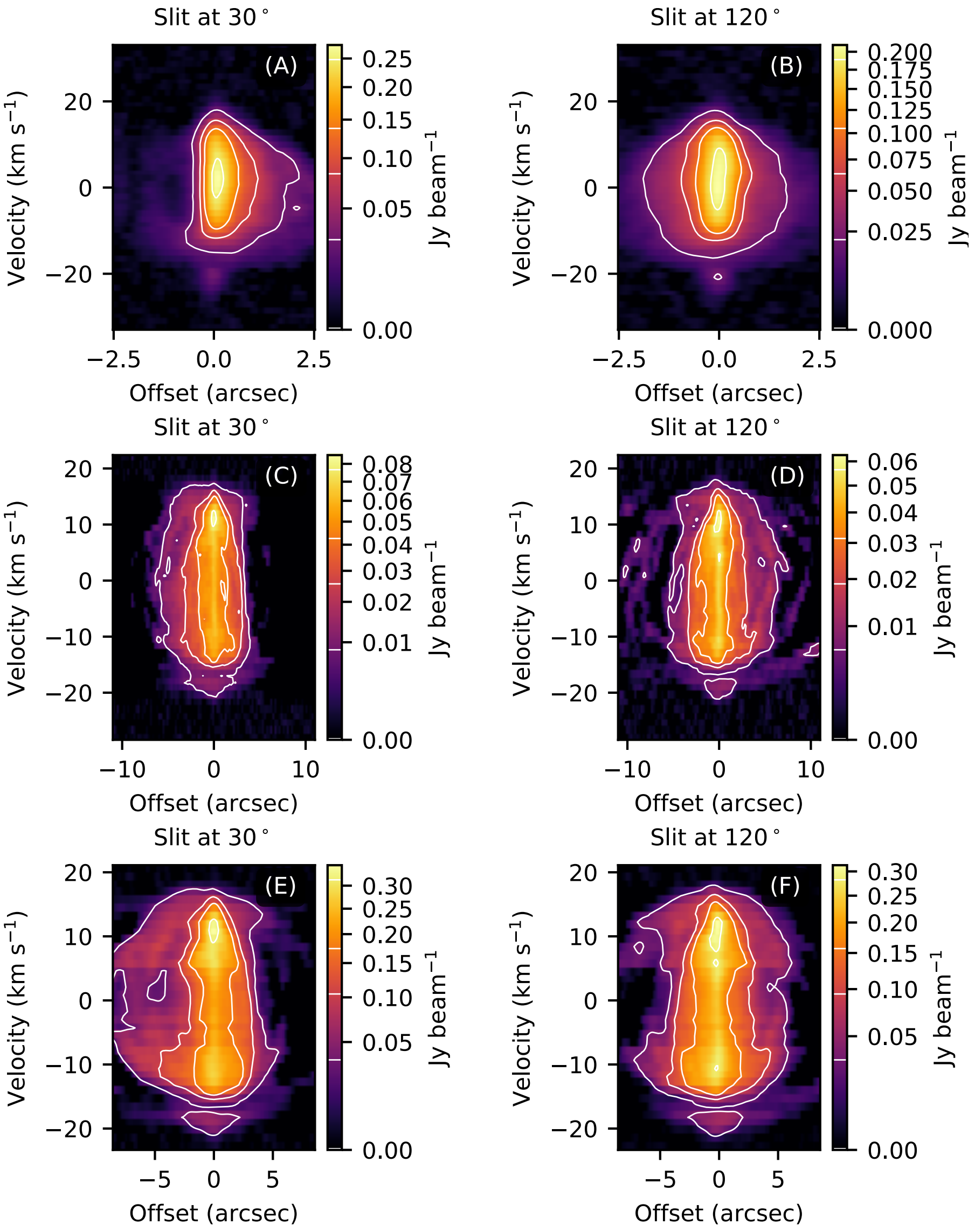

Fig. S53: **ALMA $^{12}$CO J = 2 → 1 and $^{28}$SiO J = 5 → 4 position-velocity (PV) diagram of W Aql.** Same as Fig. S11, but for W Aql.

## S7.12 GY Aql

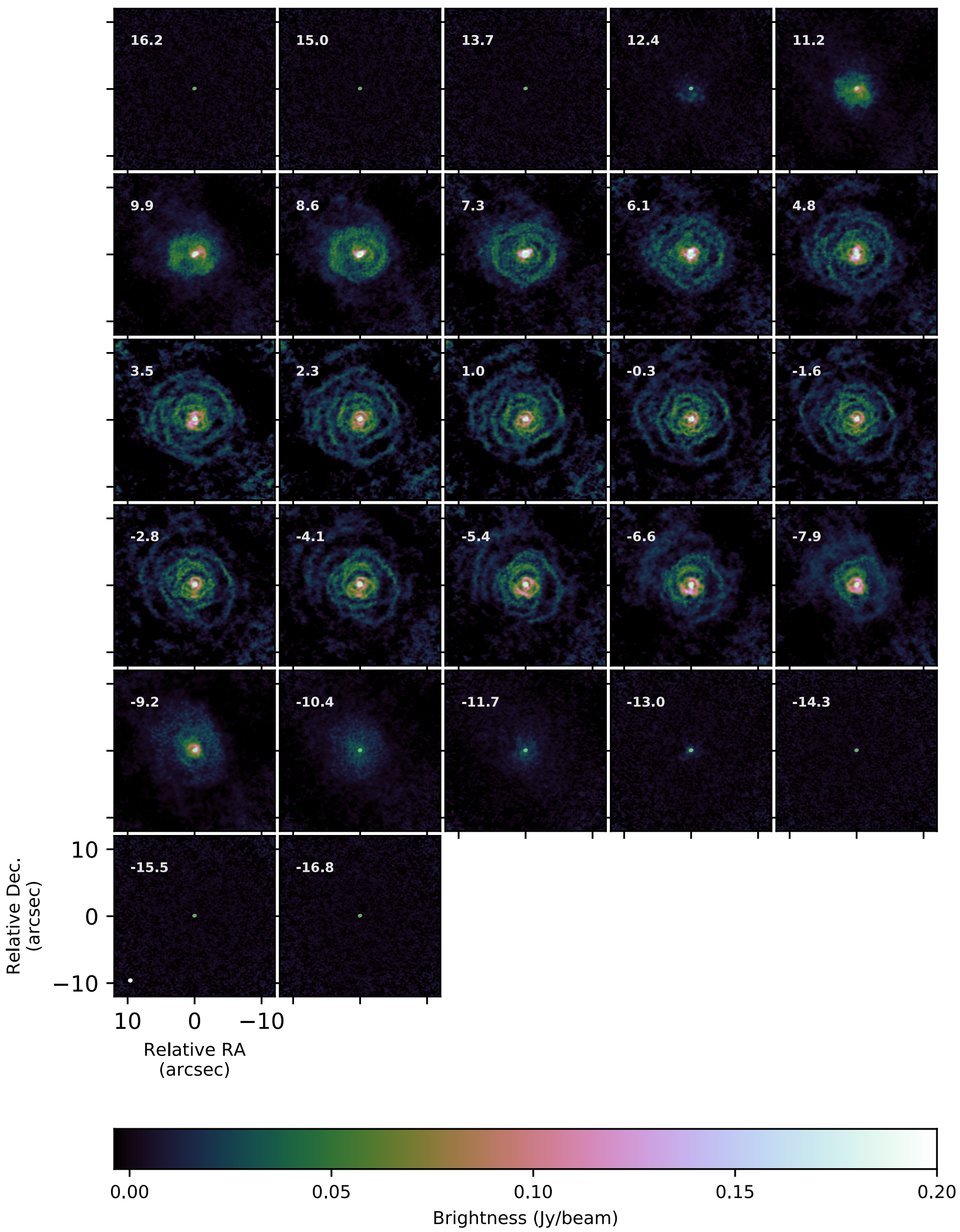

Fig. S54: **Medium resolution $^{12}$CO $J = 2 \rightarrow 1$ channel map of GY Aql.** Same as Fig. S8, but for GY Aql.

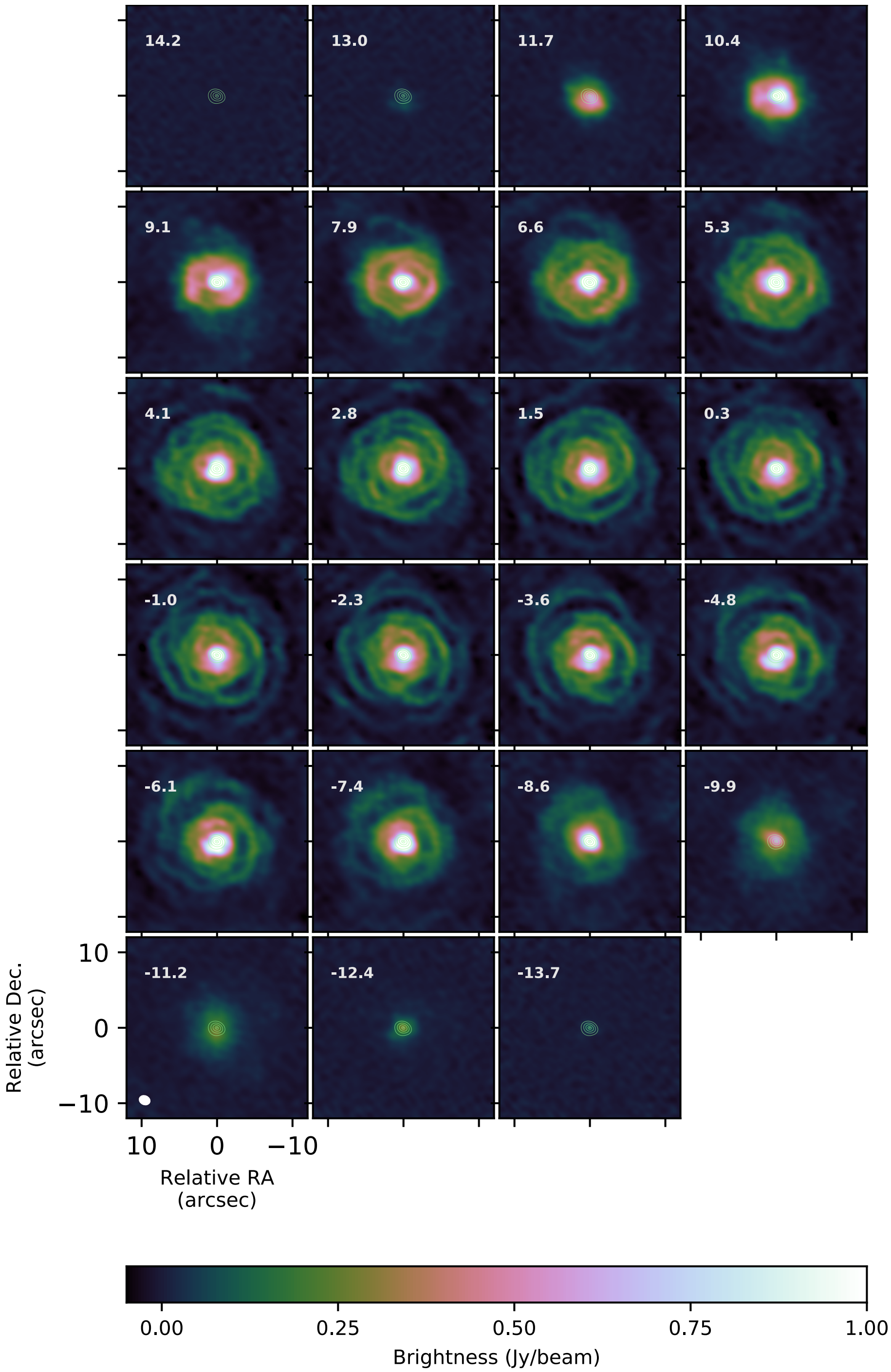

Fig. S55: **Low resolution $^{12}$CO J$=2\rightarrow1$ channel map of GY Aql.** Same as Fig. S9, but for GY Aql.

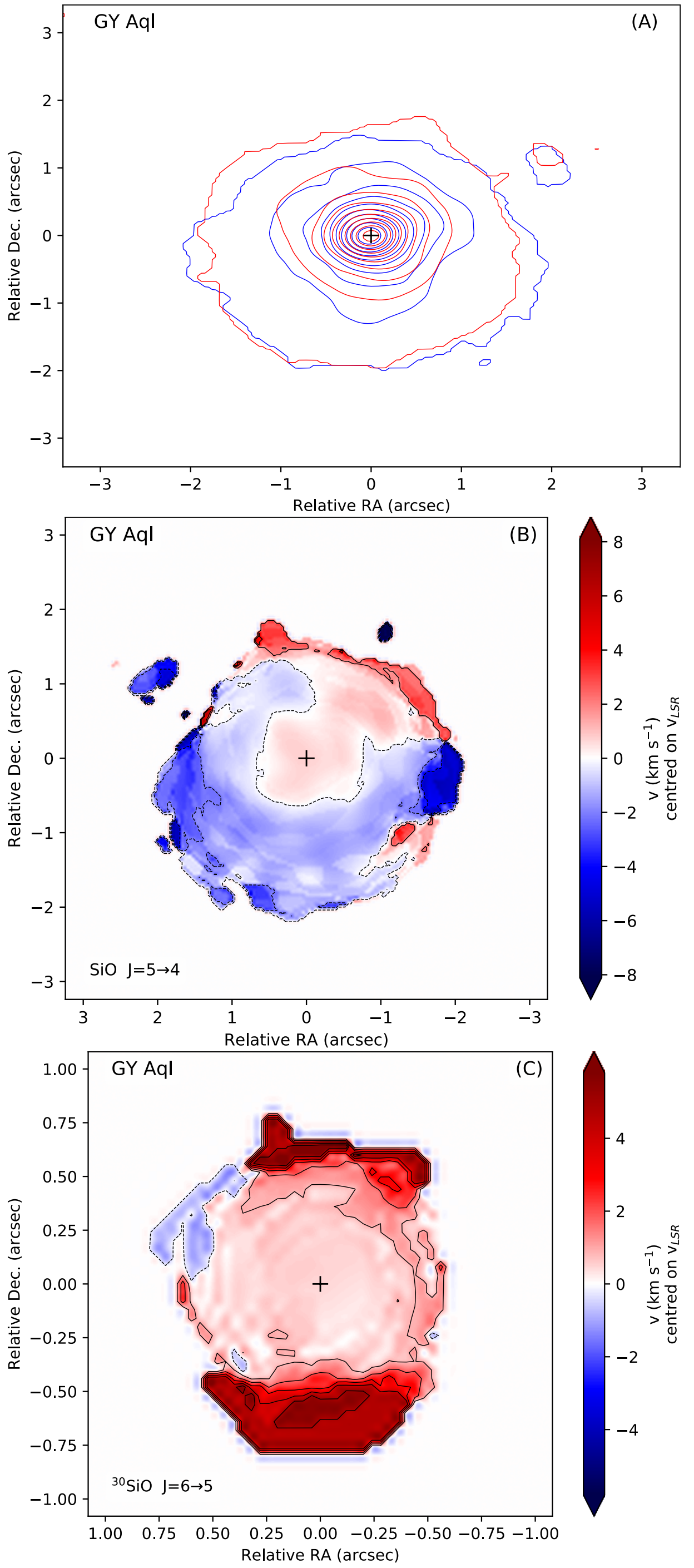

Fig. S56: **SiO stereogram and moment1-map of GY Aql.** Same as Fig. S10, but for GY Aql.

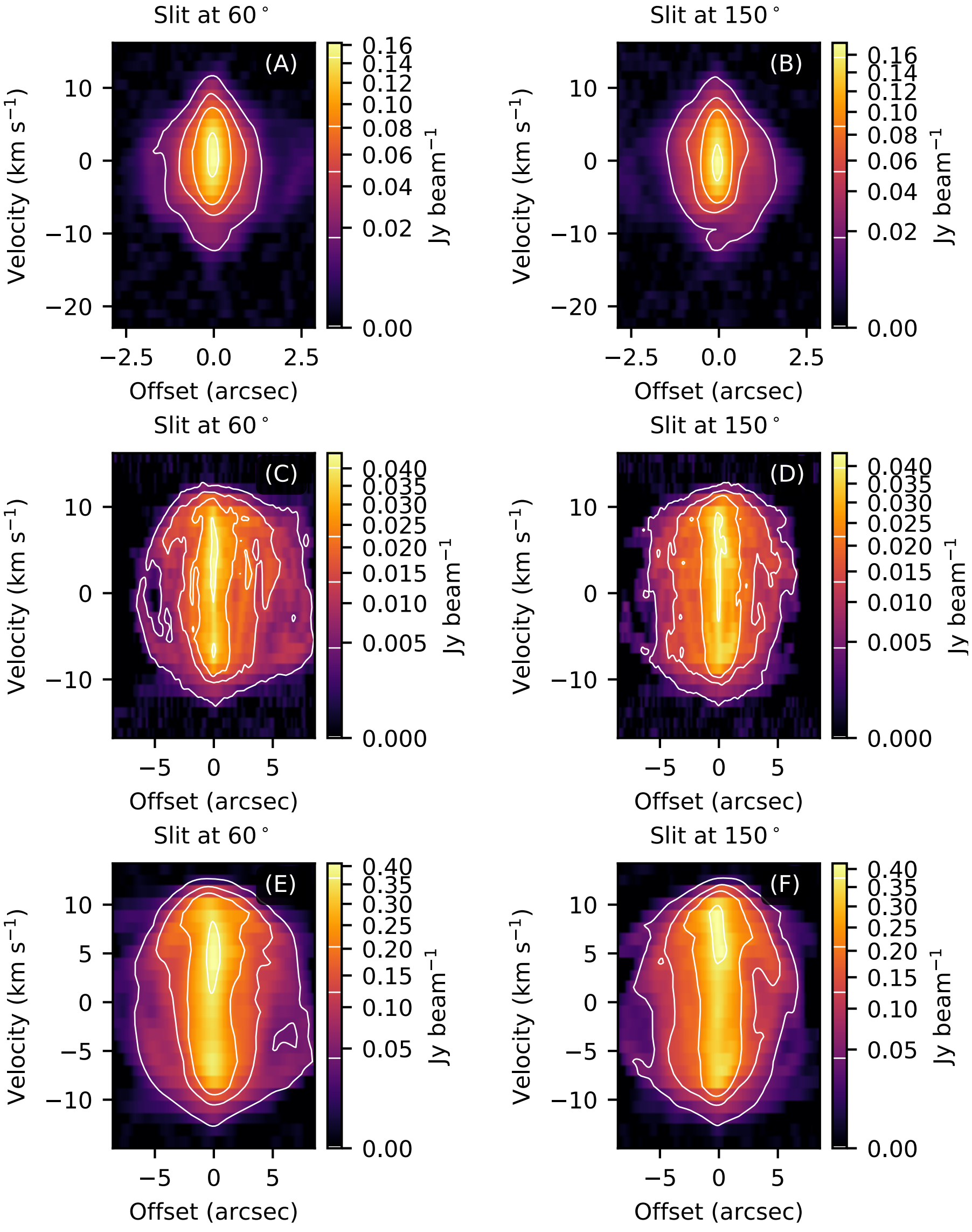

Fig. S57: **ALMA $^{12}$CO J = 2 → 1 and $^{28}$SiO J = 5 → 4 position-velocity (PV) diagram of GY Aql.** Same as Fig. S11, but for GY Aql.

## S7.13 IRC −10529

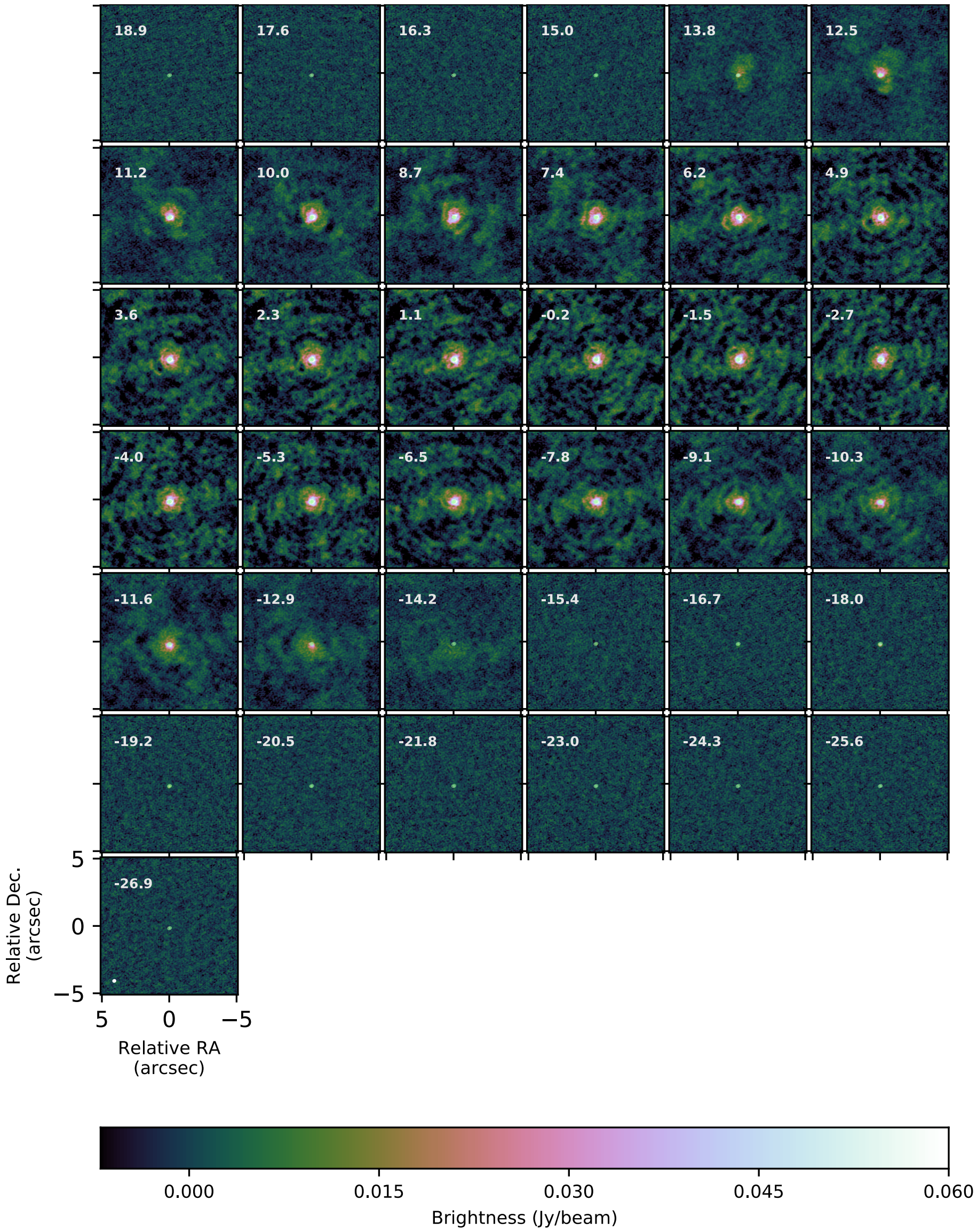

Fig. S58: **Medium resolution $^{12}$CO J = 2 → 1 channel map of IRC −10529.** Same as Fig. S8, but for IRC −10529.

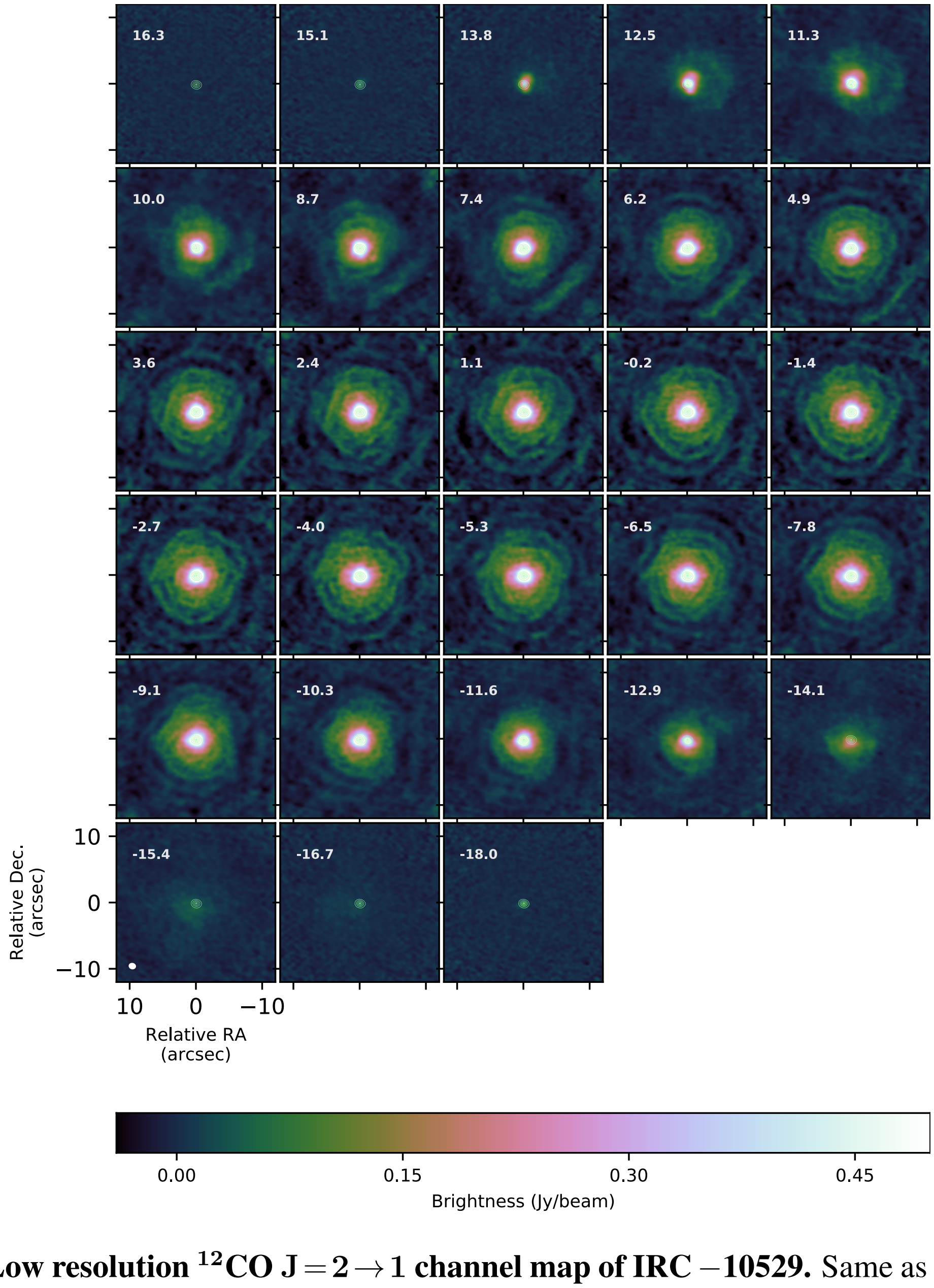

Fig. S59: **Low resolution $^{12}$CO J $= 2 \rightarrow 1$ channel map of IRC $-$10529.** Same as Fig. S9, but for IRC $-$10529.

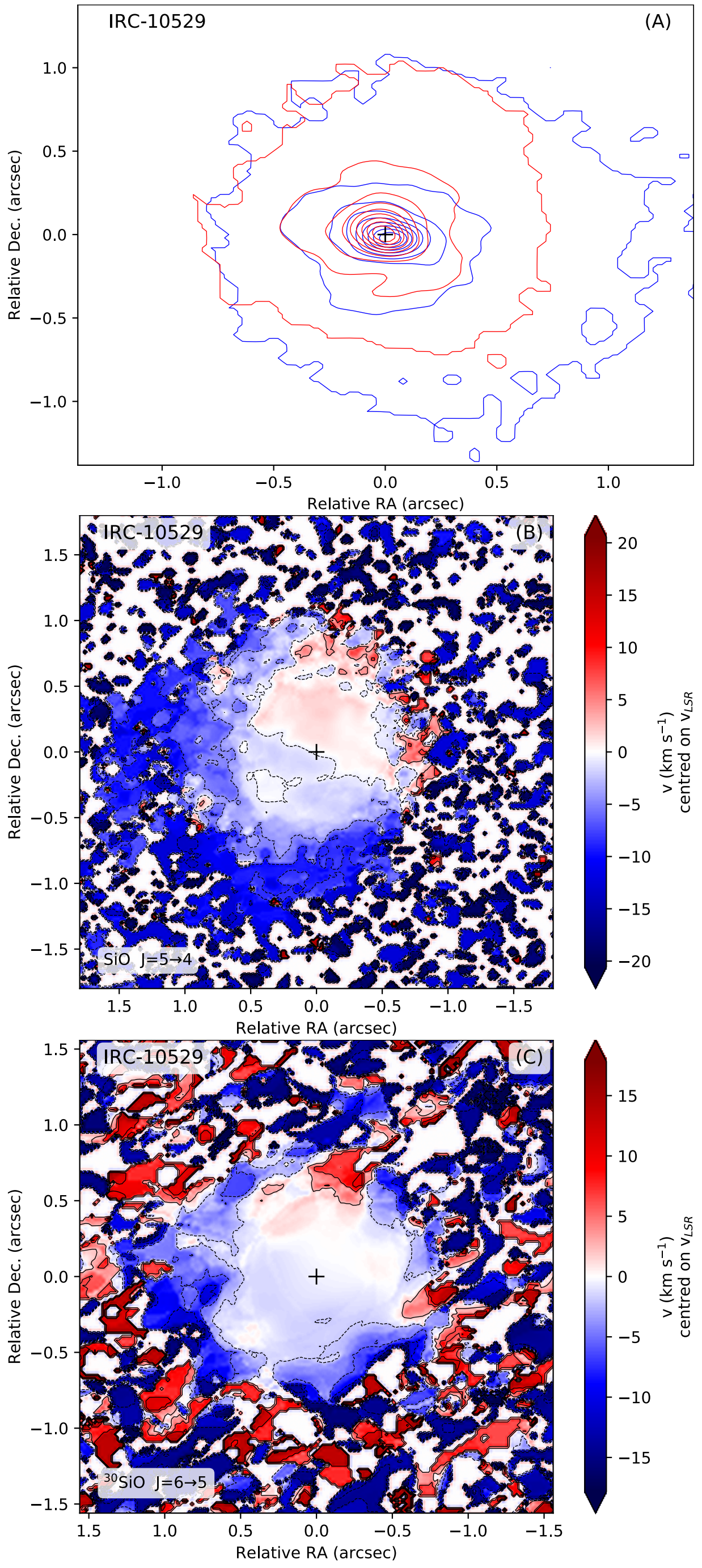

Fig. S60: **SiO stereogram and moment1-map of IRC −10529.** Same as Fig. S10, but for IRC −10529. Owing to the low signal-to-noise ratio of the SiO emission, these images should be interpreted with care (see also Sect. S3.2).

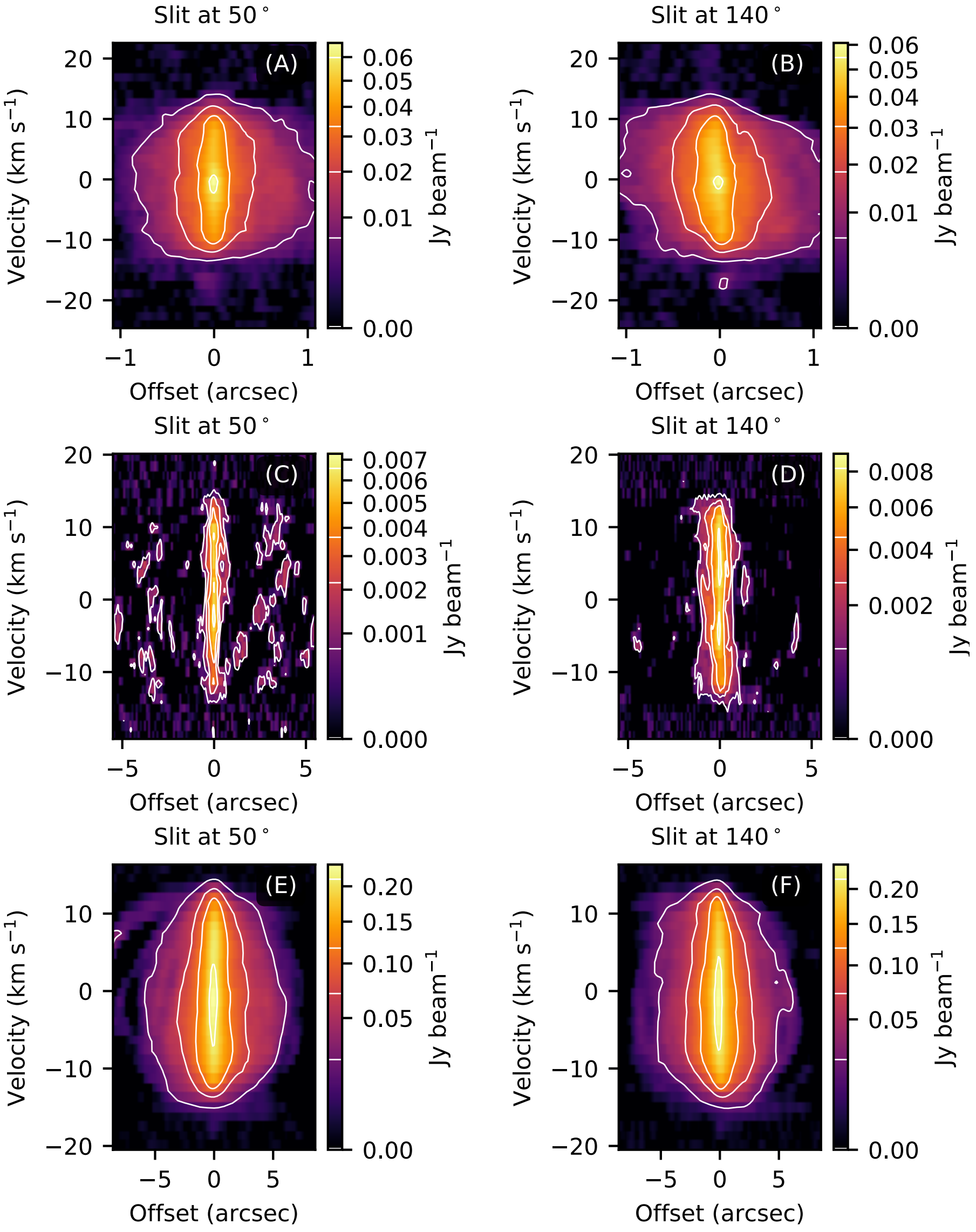

Fig. S61: **ALMA $^{12}$CO J = 2 → 1 and $^{28}$SiO J = 5 → 4 position-velocity (PV) diagram of IRC −10529.** Same as Fig. S11, but for IRC −10529.

## S7.14 IRC +10011

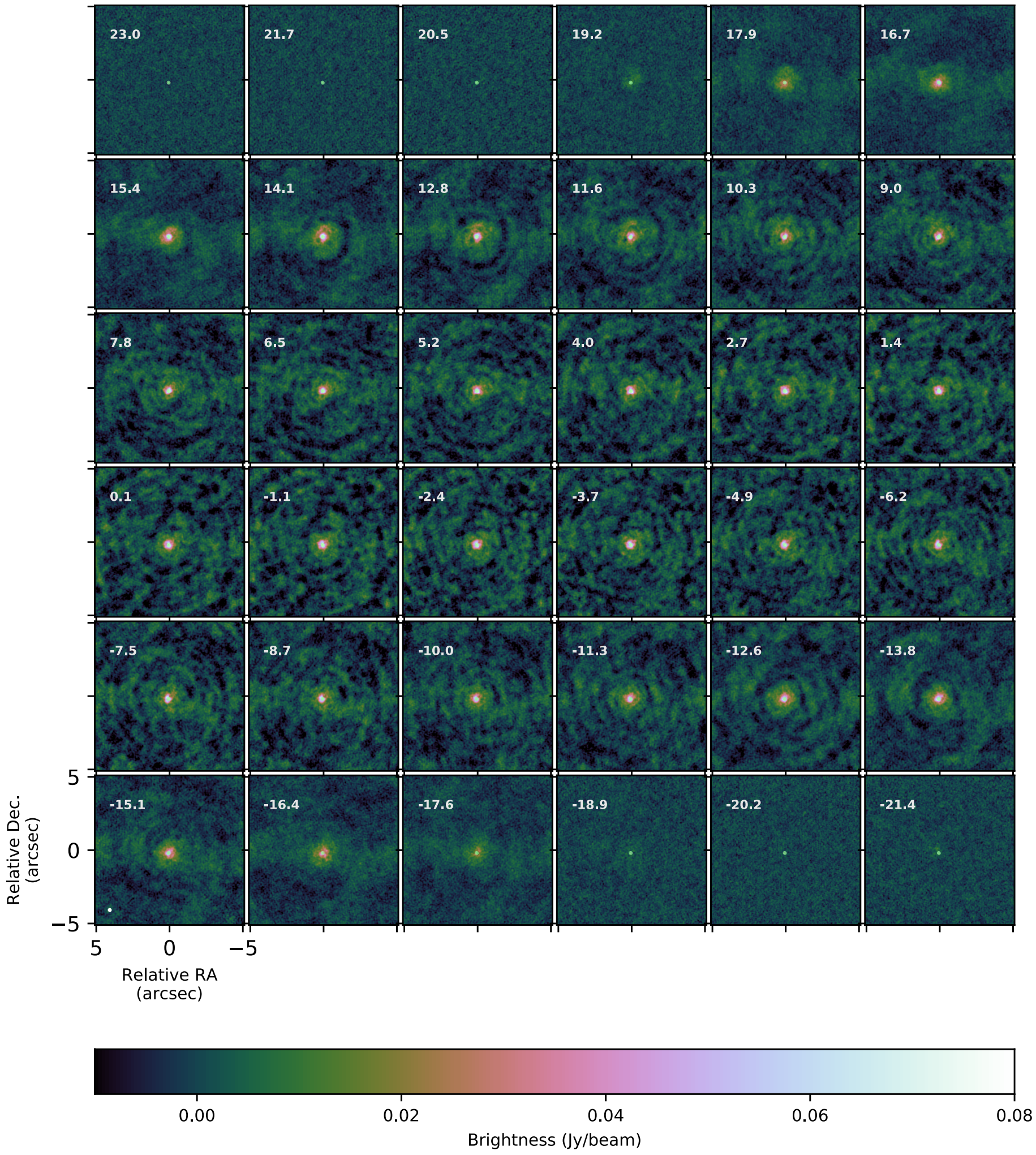

Fig. S62: **Medium resolution $^{12}$CO J = 2 → 1 channel map of IRC +10011.** Same as Fig. S8, but for IRC +10011.

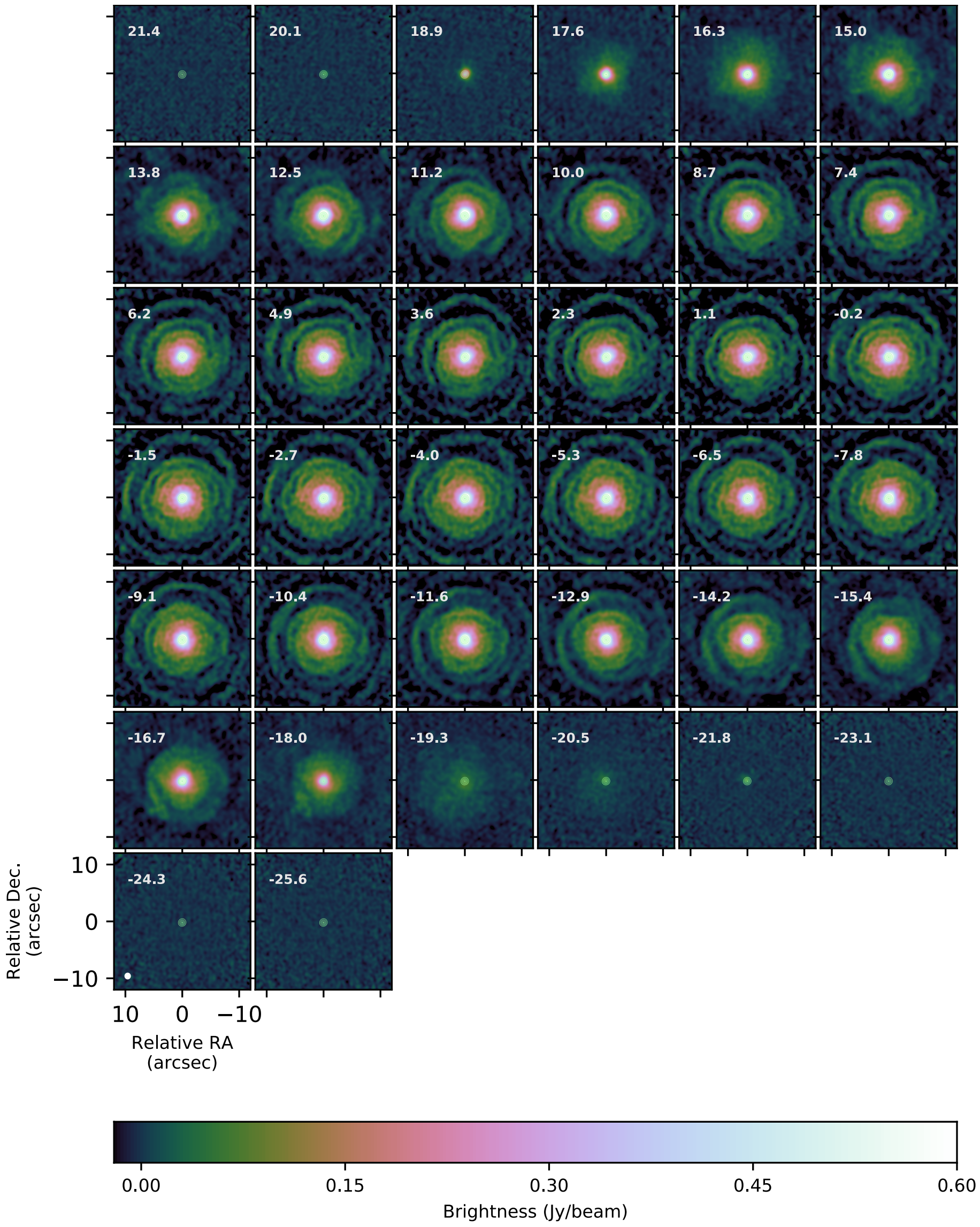

Fig. S63: **Low resolution $^{12}$CO J = 2 → 1 channel map of IRC +10011.** Same as Fig. S9, but for IRC +10011.

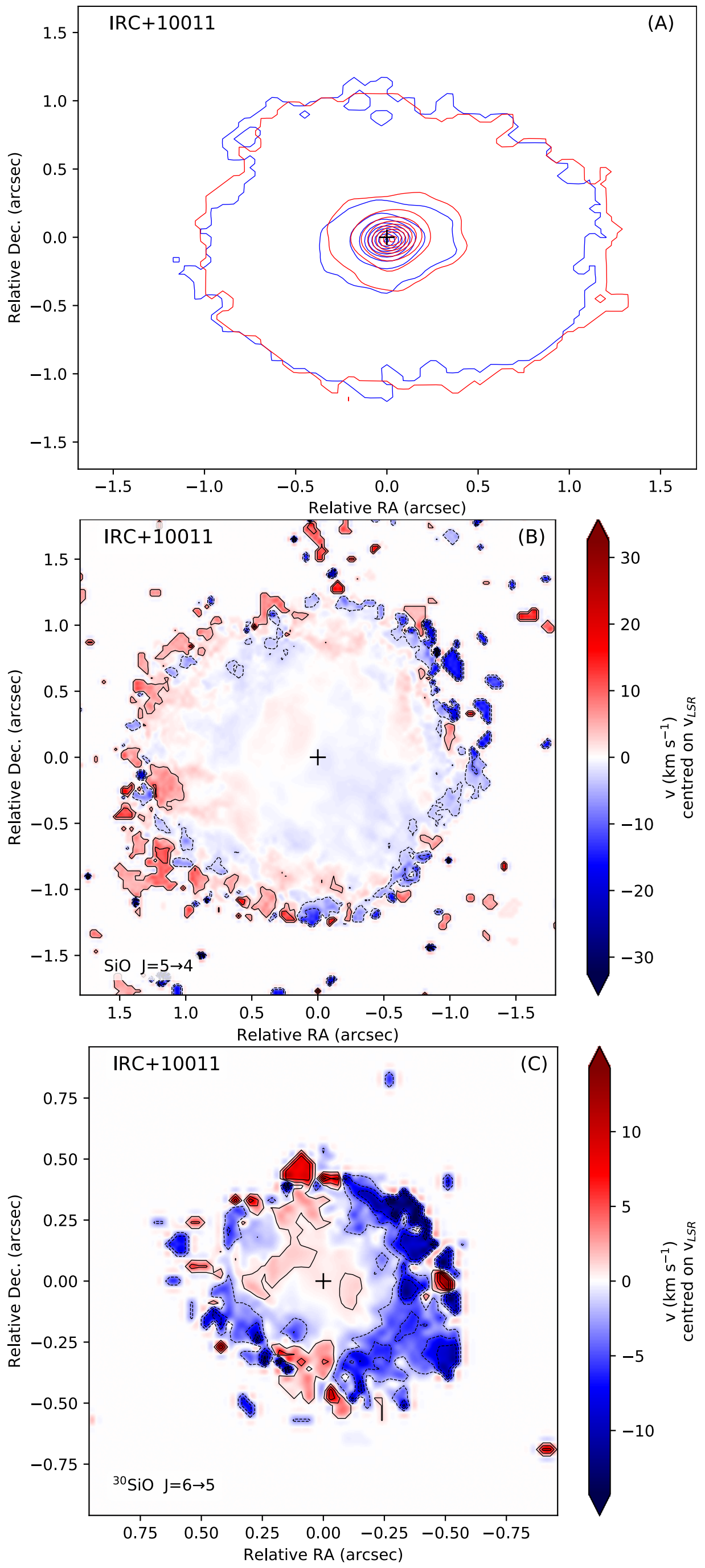

Fig. S64: **SiO stereogram and moment1-map of IRC +10011.** Same as Fig. S10, but for IRC +10011.

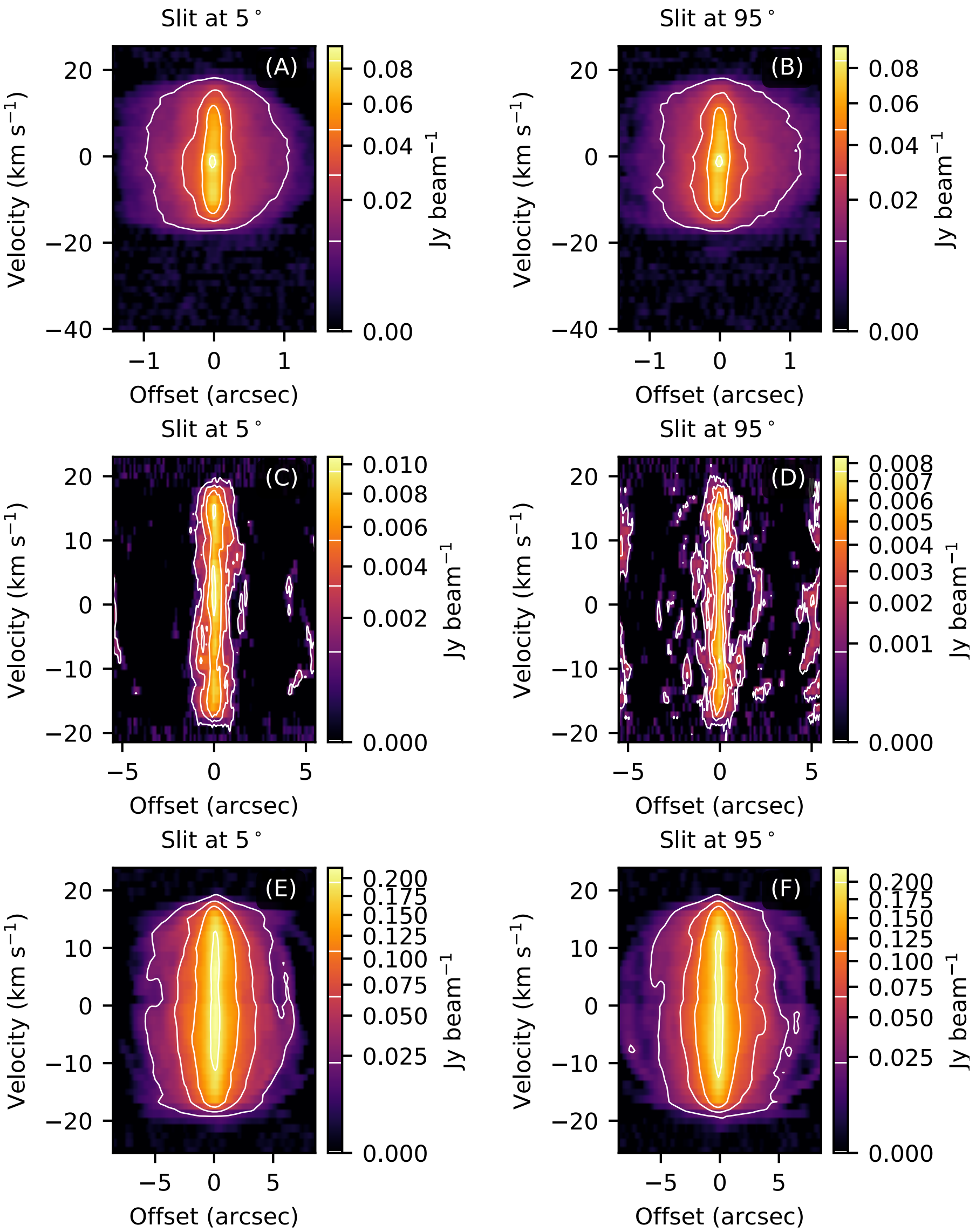

Fig. S65: **ALMA $^{12}$CO J = 2 → 1 and $^{28}$SiO J = 5 → 4 position-velocity (PV) diagram of IRC +10011.** Same as Fig. S11, but for IRC +10011.

# Supplementary Text

## S8 Hydrodynamical simulations of binary systems

Hydrodynamical simulations of the influence of a companion on a wide orbit around a mass-losing AGB star ($\dot{M} \sim 1 \times 10^{-5}\,\mathrm{M_\odot yr^{-1}}$, orbital separation between 2 – 50 au) have demonstrated that spiral patterns emerge *(92)*. Depending on the binary separation, wind velocity, and secondary mass the global large scale wind geometry can be described as bipolar, elliptical, or quasi-spherical *(92)*. The spiral structure in the stellar wind is caused by the orbital motion of the mass-losing AGB star around the common center-of-mass, or by the accretion wake of the companion owing to BHL flow *(19)*. The latter case gives rise to a spiral which is much more focussed toward the orbital plane. For binary systems with an eccentric orbit, a bifurcation of the spiral pattern might occur, including an asymmetry in the interarm density cavity and a spiral/ring appearance *(162, 163)*.

Simulations for companions with a mass of 0.1–0.5 $\mathrm{M_\odot}$ and separation of 3–10 au orbiting around an AGB star with mass-loss rate around $2 \times 10^{-7}\,\mathrm{M_\odot yr^{-1}}$ show the ejected AGB mass can form a circumbinary disk, or contribute to an accretion disk around the secondary *(20)* in addition to a spiral structure caused by BHL accretion. As exemplified by $\mathrm{L_2}$ Puppis *(17)*, a low mass secondary orbiting an AGB star can shape the AGB wind into a bipolar structure. When the effects of the AGB pulsations are included in simulations, the bipolar outflow displays ripple-like structures with lobes extending to 10 – 20 au *(20)*.

Focussing on the effect of the mass ratio between secondary and primary AGB star, it was shown that the mass-accretion efficiency by the secondary increases with the mass ratio of the system *(164)*. Those simulations were for a system with a primary star of 3 $\mathrm{M_\odot}$, orbital separation of 3 au, mass-loss rate of $10^{-6}\,\mathrm{M_\odot yr^{-1}}$, and a wind velocity of $25\,\mathrm{km\,s^{-1}}$. The higher the mass ratio, the more complex the structure of the outflow. The wind morphology was described as being ‘spider-like’ for $q = 0.33$ and as ‘rose-like’ for $q = 1.0$ *(164)*.

Simulations have been performed for a variety of binary mass ratios, initial wind velocities, orbital separations, and rotation rates of the donor star, and where the primary star loses mass at a rate of $1.5 \times 10^{-5}\,\mathrm{M_\odot yr^{-1}}$ and has a mass of 1.2 $\mathrm{M_\odot}$ *(135)*. That study showed that the strength of interaction depends on the binary mass ratio and the ratio of the orbital velocity to wind velocity. Smaller values for the orbital separation and larger companion masses induce stronger interaction. The authors also find that the corotation of the primary star modifies the outflow morphology if the initial wind velocity is below $5\,\mathrm{km\,s^{-1}}$.

Comparing these simulations to our ALMA observations, the morphologies can be explained by gravitational interaction with the companion causing the initially isotropic wind to be heavily disturbed. Other mechanisms, such as the stellar pulsations, magnetic fields, and rotation might contribute in an initial wind anisotropy thus adding complexities to the binary-induced morphological phenomena.

## S9 Observational evidence of binary-induced wind morphologies

In addition to our ATOMIUM survey, there are previously-published ALMA observations of 12 AGB stellar winds. Some have been described as being perturbed by a spiral-like signature. For carbon-rich AGB stars, examples include R Scl ($\dot{M}\sim 2\times10^{-5}$ $M_\odot yr^{-1}$) *(108, 109)*, CIT 6 ($\dot{M}\sim 8\times10^{-6}$ $M_\odot yr^{-1}$) *(105)*, and IRC +10216 ($\dot{M}\sim 1.5\times10^{-5}$ $M_\odot yr^{-1}$) *(106, 107)*, where all the three targets have terminal wind velocities of approximately 14–18 km $s^{-1}$. Two oxygen-rich high mass-loss rate targets, OH 26.5+06 and OH 30.1−0.7, have mass-loss rates about $0.3-1\times10^{-5}$ $M_\odot yr^{-1}$ and terminal wind speed of 15–18 km $s^{-1}$ *(55)*. The spiral patterns in these stars have been interpreted as being caused either by the orbital motion of the AGB star around the center-of-mass of a binary system, if the companion is relatively massive; or because the companion's gravity focusses a fraction of the wind material toward the equatorial plane, producing a (more flattened) spiral structure associated with the companion owing to the BHL flow it induces. Position-velocity diagrams can differentiate between both causes *(54)*. Bifurcated spiral-like structures in the circumstellar envelope of CIT 6, have been shown to be produced by companions on a wide, eccentric orbit *(105, 162)*.

The oxygen-rich AGB star $L_2$ Puppis ($\dot{M}\sim 5\times10^{-7}$ $M_\odot yr^{-1}$) is surrounded by a circumstellar disk in which the rotating gas has a Keplerian velocity profile within the central cavity of the dust disk (r <6 au), that changes to sub-Keplerian rotation ($6 < r <$20 au) beyond the inner dust rim *(17)*. The continuum map revealed a secondary source with an estimated mass of $\sim$10 Jupiter masses. Two more oxygen-rich examples are R Dor ($\dot{M}\sim 1\times10^{-7}$ $M_\odot yr^{-1}$) and EP Aqr ($\dot{M}\sim 1.2\times10^{-7}$), although the velocity field in the EDE/disk-like environments is more complex than Keplerian *(79, 101)*. In the case of EP Aqr, the CO emission exhibits the characteristic features of a nearly face-on spiral with a biconical outflow. The narrow width of the spiral signature in velocity space has been interpreted as an indication that the spiral is caused by a hydrodynamical perturbation in a face-on differentially rotating disk. For these three cases ($L_2$ Puppis, R Dor, and EP Aqr), the widths of the rotational SiO lines in the ground vibrational state are considerably larger than the widths of the ground-state CO lines obtained from the same observational setup (as is also the case for five of our sources, see Table S2). The SiO lines are radiatively excited, and hence are more diagnostic of the inner wind dynamics.

The Mira AB system is another well-known binary containing an oxygen-rich AGB star *(99, 100)*. When imaged at a spatial resolution of $\sim$0.5″, the map of $^{12}$CO $J=3\rightarrow2$ shows a lot of complexity. Spiral arcs oriented in the orbital plane are found around Mira A ($\dot{M}\sim 1\times10^{-7}$ $M_\odot yr^{-1}$). The accretion wake behind the companion (Mira B), which resides at an orbital separation of 0.″487, was visible, and several large ($\sim$5–10″) opposing arcs were also found. Some morphological resemblance between the $^{12}$CO $J=3\rightarrow2$ emission of the Mira AB system and the CO channel map of S Pav can be discerned [compare Fig. S8 with figure A1 of *(99)*].

R Aqr is a symbiotic stellar system, consisting of interacting AGB star and a white dwarf *(103, 104)*. In these systems, the strong binary interaction results in high mass transfer between both stars, equatorial flows, and ejection of fast bipolar jets. The two-arcminute-wide nebula of R Aqr is composed of an equatorial structure, and a precessing jet powered by the accretion on the white dwarf *(103)*. The orbital period is ∼44 yr with binary separation of ∼45 mas (or 10 au at 218 pc). $^{12}$CO data show that the CO structure traces mass ejection focused in the orbital plane, with two plumes in opposite directions probably corresponding to (the start of) a double spiral *(103)*. The mass-loss rate is $2 \times 10^{-7}\,\mathrm{M_{\odot}yr^{-1}}$ *(104)*.

Three of the fourteen ATOMIUM AGB sources are known to be part of a binary system: the two S-type AGB stars W Aql and $\pi^1$ Gru (whose C/O ratio is slighter lower than 1), and R Hya. W Aql ($\dot{M}\sim\, 3 \times 10^{-6}\,\mathrm{M_{\odot}yr^{-1}}$) has a known companion at a separation of 0.46″ (or ∼150 au), which is classified as an F8 to G0 main sequence star *(165)*. The CO $J = 3 \rightarrow 2$ line in W Aql in the inner 10″ of the circumstellar envelope is asymmetric with arc-like structures at separations of 2–3″ *(50)*. Farther out, weaker spiral structures are present at greater separations. The larger separations can be explained by the interaction with the known companion in an orbit with low eccentricity, but not the smaller separation pattern. Potential physical causes that have been suggested include a second, closer companion residing at an eccentric orbit; a recent change in the wind velocity; or variations in the mass-loss-rate *(50)*. $\pi^1$ Gru has a companion of spectral type G0V at a separation of 2.″7 (∼400 au) *(166)*. Observations of the $^{12}$CO $J = 3 \rightarrow 2$ and $^{13}$CO $J = 3 \rightarrow 2$ at a spatial resolution of ∼4″ confirmed the envelope structure includes a radially expanding equatorial torus (with a velocity of 8–13 km s$^{-1}$), and a fast bipolar outflow with a linear velocity increase of up to 100 km s$^{-1}$ *(49)*. Because wide binary systems are not expected to produce this kind of envelope structure, a second, closer companion (at 10–30 au) has been suggested *(167)*. R Hya is thought to be a wide binary system with an angular separation of 21″ and a very long orbital period *(168)*.

## S10 ATOMIUM observation properties

The ATOMIUM observation properties are listed in Table S5.

Table S5: **Observing properties of the ATOMIUM project.** Data for GY Aql only are shown as an example; full data for all targets are given in Data S1. Given are the Science Goal scheduling block (SG; in which a, b, c and d are observations in the medium resolution configuration, and e and f in the low resolution configuration), the observing date, the ALMA archival name (ASDM), the target name, the right ascension (R.A.) and declination (Dec.) of that specific observation, the time on source (ToS), the precipitable water vapour at the date of observations (PWV), the phase-reference source (Ph. Ref.), and the bandpass and flux density scale calibrator (Bandpass).

| SG | Date Obs. | ASDM | Target | R.A. (h:min:sec) | Dec. (deg:min:sec) | ToS (sec) | PWV (mm) | Ph. Ref. | ToS (sec) | Bandpass | ToS (sec) |
|---|---|---|---|---|---|---|---|---|---|---|---|
| GY_Aql_a_06_TM2 | 2018-11-13 22:46:33 | uid___A002_Xd51939_X6300 | GY Aql | 19:50:06.31 | –07:36:52.3 | 302 | 0.79 | J1951-0509 | 60 | J2000-1748 | 302 |
| GY_Aql_b_06_TM2 | 2018-11-11 22:31:26 | uid___A002_Xd50463_X9b77 | GY Aql | 19:50:06.31 | –07:36:52.3 | 302 | 2.68 | J1951-0509 | 60 | J1924-2914 | 302 |
| GY_Aql_b_06_TM2 | 2018-11-12 23:36:25 | uid___A002_Xd51939_X2e4 | GY Aql | 19:50:06.31 | –07:36:52.3 | 302 | 1.71 | J1951-0509 | 60 | J2000-1748 | 302 |
| GY_Aql_b_06_TM2 | 2018-11-13 22:28:25 | uid___A002_Xd51939_X6277 | GY Aql | 19:50:06.31 | –07:36:52.3 | 302 | 0.96 | J1951-0509 | 60 | J2000-1748 | 302 |
| GY_Aql_c_06_TM2 | 2018-11-14 23:37:44 | uid___A002_Xd52fc8_X876 | GY Aql | 19:50:06.31 | –07:36:52.3 | 302 | 0.75 | J1951-0509 | 60 | J2000-1748 | 302 |
| GY_Aql_d_06_TM2 | 2018-11-13 23:04:44 | uid___A002_Xd51939_X638d | GY Aql | 19:50:06.31 | –07:36:52.3 | 90 | 0.64 | J1951-0509 | 60 | J2000-1748 | 302 |
| GY_Aql_e_06_TM1 | 2019-01-06 16:19:01 | uid___A002_Xd7aa27_X7ceb | GY Aql | 19:50:06.31 | –07:36:52.3 | 302 | 1.84 | J1951-0509 | 60 | J1751+0939 | 302 |
| GY_Aql_e_06_TM1 | 2019-03-03 11:43:28 | uid___A002_Xd90607_X3948 | GY Aql | 19:50:06.31 | –07:36:52.3 | 302 | 1.65 | J1951-0509 | 60 | J2000-1748 | 302 |
| GY_Aql_f_06_TM1 | 2019-01-13 14:34:05 | uid___A002_Xd80784_X62d5 | GY Aql | 19:50:06.31 | –07:36:52.3 | 302 | 1.12 | J1951-0509 | 60 | J1751+0939 | 302 |

## S11 ATOMIUM image cube properties

The properties of each cube and continuum image are listed in Table S6 and Table S7, respectively.

Table S6: **Image cube properties.** Data for GY Aql only are shown as an example; full data for all targets are given in Data S2. Given are the source name, the configuration (low or medium spatial resolution), the cube number (see Fig. S1), the start and end frequency of each cube (in GHz), the size of the major and minor axis of the synthesized beam (bmaj and bmin, in arcsec), the angle of the synthesized beam (bpa, in degrees), the size of the image (imsize, in arcsec) and the noise $\sigma_{\rm rms}$ (in mJy) of the cube as measured in an emission-free channel of the cube.

| target | config | cubeNo | start (GHz) | end (GHz) | bmaj (arcsec) | bmin (arcsec) | bpa (deg) | imsize (arcsec) | $\sigma_{\rm rms}$ (mJy) |
|---|---|---|---|---|---|---|---|---|---|
| GY Aql | low | 00 | 213.838 | 215.711 | 1.343 | 1.045 | 64 | 24.0 | 2.8 |
| GY Aql | low | 01 | 216.038 | 217.910 | 1.360 | 1.047 | 66 | 24.0 | 3.3 |
| GY Aql | low | 04 | 227.234 | 229.107 | 1.286 | 1.013 | 69 | 24.0 | 2.9 |
| GY Aql | low | 05 | 229.589 | 231.462 | 1.265 | 0.986 | 67 | 24.0 | 3.1 |
| GY Aql | low | 08 | 244.051 | 244.987 | 1.294 | 0.929 | 65 | 24.0 | 3.0 |
| GY Aql | low | 09 | 245.350 | 247.223 | 1.278 | 0.925 | 66 | 24.0 | 2.6 |
| GY Aql | low | 12 | 258.623 | 260.496 | 1.223 | 0.885 | 66 | 24.0 | 2.9 |
| GY Aql | low | 13 | 262.104 | 263.039 | 1.207 | 0.876 | 66 | 24.0 | 3.5 |
| GY Aql | medium | 00 | 213.838 | 215.711 | 0.382 | 0.319 | −74 | 24.0 | 2.2 |
| GY Aql | medium | 01 | 216.037 | 217.911 | 0.375 | 0.318 | −76 | 24.0 | 2.3 |
| GY Aql | medium | 02 | 220.237 | 222.110 | 0.364 | 0.295 | −73 | 24.0 | 2.2 |
| GY Aql | medium | 03 | 223.631 | 225.504 | 0.358 | 0.290 | −76 | 24.0 | 1.9 |
| GY Aql | medium | 04 | 227.235 | 229.108 | 0.357 | 0.304 | −79 | 24.0 | 2.4 |
| GY Aql | medium | 05 | 229.589 | 231.462 | 0.358 | 0.298 | −78 | 24.0 | 2.4 |
| GY Aql | medium | 06 | 235.438 | 237.311 | 0.340 | 0.272 | −76 | 24.0 | 2.3 |
| GY Aql | medium | 07 | 239.157 | 240.092 | 0.340 | 0.275 | −74 | 24.0 | 2.4 |
| GY Aql | medium | 08 | 244.051 | 244.987 | 0.397 | 0.278 | −75 | 24.0 | 3.0 |
| GY Aql | medium | 09 | 245.350 | 247.223 | 0.395 | 0.273 | −76 | 24.0 | 2.8 |
| GY Aql | medium | 10 | 251.583 | 253.456 | 0.351 | 0.273 | −71 | 24.0 | 4.9 |
| GY Aql | medium | 11 | 253.947 | 255.820 | 0.349 | 0.271 | −71 | 24.0 | 4.7 |
| GY Aql | medium | 12 | 258.624 | 260.497 | 0.377 | 0.260 | −75 | 24.0 | 3.0 |
| GY Aql | medium | 13 | 262.104 | 263.039 | 0.371 | 0.262 | −75 | 24.0 | 3.4 |
| GY Aql | medium | 14 | 265.533 | 267.406 | 0.334 | 0.257 | −70 | 24.0 | 6.1 |
| GY Aql | medium | 15 | 267.783 | 269.657 | 0.333 | 0.257 | −73 | 24.0 | 5.8 |

Table S7: **Continuum image properties.** Given are the source name, the configuration (low or medium resolution), the size of the major and minor axis of the synthesized beam (bmaj and bmin, in arcsec), the angle of the synthesized beam (bpa, in degrees), the maximum recoverable scale (MRS, in arcsec), the size of the image (imsize, in arcsec), the noise $\sigma_{\rm rms}^{\rm cont}$ (in mJy), and in the last column the total line-free bandwidth (BW, in GHz), spread over the 56 GHz span of observations.

| target | config | bmaj (arcsec) | bmin (arcsec) | bpa (deg) | MRS (arcsec) | imsize (arcsec) | $\sigma_{\rm rms}^{\rm cont}$ (mJy) | BW (GHz) |
|---|---|---|---|---|---|---|---|---|
| GY Aql | low | 1.220 | 0.897 | 64 | 9.5 | 24.0 | 0.040 | 8.69 |
| GY Aql | medium | 0.324 | 0.247 | −70 | 4.0 | 24.0 | 0.026 | 18.03 |
| IRC +10011 | low | 0.722 | 0.686 | −59 | 7.4 | 24.0 | 0.051 | 8.53 |
| IRC +10011 | medium | 0.112 | 0.100 | 38 | 1.6 | 6.0 | 0.033 | 18.49 |
| IRC −10529 | low | 0.788 | 0.627 | 76 | 8.9 | 24.0 | 0.052 | 6.97 |
| IRC −10529 | medium | 0.146 | 0.113 | −63 | 2.0 | 4.0 | 0.027 | 15.33 |
| $\pi^1$ Gru | low | 0.866 | 0.774 | −86 | 9.3 | 24.0 | 0.036 | 10.36 |
| $\pi^1$ Gru | medium | 0.248 | 0.235 | 30 | 3.9 | 8.0 | 0.034 | 20.49 |
| RW Sco | low | 0.928 | 0.701 | 86 | 9.0 | 24.0 | 0.034 | 10.30 |
| RW Sco | medium | 0.147 | 0.120 | −86 | 1.9 | 4.0 | 0.040 | 6.75 |
| R Aql | low | 0.764 | 0.648 | 83 | 7.7 | 24.0 | 0.042 | 10.25 |
| R Aql | medium | 0.306 | 0.238 | −54 | 3.8 | 8.0 | 0.030 | 20.31 |
| R Hya | low | 0.830 | 0.600 | 79 | 8.7 | 24.0 | 0.051 | 10.09 |
| R Hya | medium | 0.256 | 0.223 | 70 | 3.5 | 8.0 | 0.028 | 19.02 |
| SV Aqr | low | 0.886 | 0.747 | 74 | 9.8 | 24.0 | 0.038 | 10.92 |
| SV Aqr | medium | 0.124 | 0.104 | −75 | 1.6 | 8.0 | 0.023 | 27.55 |
| S Pav | low | 1.026 | 0.983 | −56 | 8.7 | 24.0 | 0.051 | 10.22 |
| S Pav | medium | 0.304 | 0.234 | 56 | 3.3 | 8.0 | 0.022 | 20.20 |
| T Mic | low | 1.047 | 0.730 | −79 | 9.3 | 24.0 | 0.059 | 10.99 |
| T Mic | medium | 0.268 | 0.225 | −89 | 4.0 | 8.0 | 0.025 | 19.36 |
| U Del | low | 1.167 | 1.043 | 42 | 9.6 | 24.0 | 0.036 | 6.47 |
| U Del | medium | 0.316 | 0.235 | −33 | 3.3 | 8.0 | 0.028 | 14.84 |
| U Her | low | 0.997 | 0.843 | 26 | 9.7 | 24.0 | 0.054 | 9.75 |
| U Her | medium | 0.352 | 0.249 | −33 | 2.2 | 8.0 | 0.048 | 16.83 |
| V PsA | low | 0.995 | 0.753 | 87 | 9.0 | 24.0 | 0.030 | 11.28 |
| V PsA | medium | 0.283 | 0.229 | 85 | 4.0 | 8.0 | 0.020 | 20.99 |
| W Aql | low | 0.920 | 0.667 | 76 | 8.9 | 24.0 | 0.056 | 6.61 |
| W Aql | medium | 0.351 | 0.223 | −68 | 3.9 | 8.0 | 0.030 | 18.75 |